\documentclass[12pt,preprint,iop]{emulateapj}
\usepackage{amsfonts,amsmath,apjfonts}
\usepackage{natbib}
\usepackage{graphicx}
\usepackage[caption=false]{subfig}
\usepackage{booktabs}

\def\cfa{1}

\shorttitle{An \textit{HST} Study of the Locations of Long Gamma-Ray Bursts}
\shortauthors{Blanchard et al.}

\begin{document}

\title{The Offset and Host Light Distributions of Long Gamma-Ray Bursts: A New View From \textit{HST} Observations of \textit{Swift} Bursts}

\author{
Peter K. Blanchard\altaffilmark{\cfa}, 
Edo Berger\altaffilmark{\cfa}, and
Wen-fai Fong\altaffilmark{2,3}
}
\email{pblanchard@cfa.harvard.edu}

\altaffiltext{1}{Harvard-Smithsonian Center for Astrophysics, 60
Garden Street, Cambridge, MA 02138, USA}
\altaffiltext{2}{Einstein Fellow}
\altaffiltext{3}{Steward Observatory, University of Arizona, 933 North Cherry Avenue, Tucson, AZ 85721, USA}

\begin{abstract}
\medskip
We present the results of an extensive \textit{Hubble Space Telescope} (\textit{HST}) imaging study of $\sim$100 \textit{Swift} long-duration gamma-ray bursts (LGRBs) spanning $0.03 \lesssim z \lesssim 9.4$ using relative astrometry from ground- and space-based afterglow observations to locate the bursts within their host galaxies.  Using these data, we measure the distribution of LGRB offsets from their host centers, as well as their relation to the underlying host light distribution.  We find that the host-normalized offsets of LGRBs are more centrally concentrated than expected for an exponential disk profile, $\langle R/R_{h} \rangle$ = 0.67, and in particular they are more concentrated than the underlying surface brightness profiles of their host galaxies.  The distribution of offsets is inconsistent with the distribution for Type II supernovae (SNe) but consistent with the distribution for Type Ib/c SNe.  The fractional flux distribution, with a median value of 0.75, indicates that LGRBs prefer some of the brightest locations in their host galaxies but are not as strongly correlated as previous studies indicated.  More importantly, we find a clear correlation between the offset and fractional flux, where bursts at offsets $R/R_{h} \lesssim 0.5$ exclusively occur at fractional fluxes $\gtrsim 0.6$ while bursts at $R/R_{h} \gtrsim 0.5$ uniformly trace the light of their hosts.  This indicates that the spatial correlation of LGRB locations with bright star forming regions seen in the full sample is dominated by the contribution from bursts at small offset and that LGRBs in the outer parts of galaxies show no preference for unusually bright star forming regions.  Finally, we find no evidence for evolution  from $z \lesssim 1$ to $z \sim 3$ in the offset or fractional flux distributions.  We conclude that LGRBs strongly prefer the bright, inner regions of their hosts indicating that the star formation taking place there is more favorable for LGRB progenitor production.  This indicates that another environmental factor beyond metallicity, such as binary interactions or IMF differences, may be operating in the central regions of LGRB hosts.        
   
\smallskip
\end{abstract}

\keywords{gamma-ray burst: general}

\section{Introduction}

Long-duration gamma-ray bursts (LGRBs) are the most energetic explosions known in the universe, with a volumetric rate of $\lesssim 1\%$ of the core-collapse supernova rate.  A wide range of studies of their accompanying afterglow emission and host galaxies have been used to shed light on the properties of the bursts and their progenitors.  In particular, the association with broad-lined Type Ic supernovae (Type Ic-BL SNe) and exclusive locations in star forming galaxies firmly established that LGRBs result from the deaths of massive stars \citep[e.g.,][]{Christensen2004, WoosleyBloom2006, Wainwright2007}.  Detailed observations of the host galaxies have also indicated a preference for low metallicity environments \citep{Stanek2006,Levesque2010b,GrahamFruchter2013}, although some bursts occur in environments with solar metallicity \citep{Levesque2010a,Levesque2010b,Levesque2012}.  Similarly, afterglow observations established that LGRBs are powered by relativistic jets with an energy scale of $\sim 10^{51}$ erg \citep[e.g.,][]{Frail2001}.  These observations support the idea that the progenitors of LGRBs are rapidly-rotating massive stars that undergo core-collapse to form a hyper-accreting black hole \citep[collapsars:][]{Woosley1993, MacFadyenWoosley1999}.  However, it is unclear at the present whether these massive stars are single or whether they are subject to a wide range of possible binary interaction scenarios \citep{Podsiadlowski2004,vanYoon2007}.  

Since direct observations of LGRB progenitors are unlikely due to their low volumetric rate, insight into the nature of the progenitors has to rely on studies of the environments in which LGRBs occur.  On a large scale, LGRB hosts have been found to generally be blue, compact, low luminosity, low mass, low metallicity galaxies with high specific star formation rates \citep[e.g.,][]{LF2003, Christensen2004, Wainwright2007, Savaglio2009}, especially when compared to core-collapse SN hosts \citep{Svensson2010}.  These results have led to the notion of a preference for low metallicity progenitors.

The sub-galactic environments of LGRBs can provide additional clues about the nature of the progenitors.  To date, two major approaches have been employed in the literature: (i) measuring the offset of LGRBs relative to the centers of their hosts, and (ii) measuring the fractional brightness at the LGRB positions relative to the overall distribution of light within their hosts.  In this context, \citet{Bloom2002} combined ground-based afterglow observations and \textit{Hubble Space Telescope} (\textit{HST}) host observations to measure the offsets of 20 LGRBs, and found that they were consistent with being drawn from an exponential disk profile.  This led them to conclude that LGRBs were associated with star formation, and to rule out the compact object merger progenitor scenario.  However, the small sample size and the large uncertainties in many of the measured offsets prevented a more  detailed analysis.  \citet{Fruchter2006}, on the other hand, argued that the irregular morphologies of at least some LGRB hosts necessitate the use of the fractional brightness technique (hereafter, fractional flux).  Using this analysis for 30 LGRB hosts with \textit{HST} observations, in comparison to 16 core-collapse SNe from the GOODS survey, these authors found that LGRBs are more highly concentrated in the brightest regions of their hosts.  \citet{Svensson2010} expanded the sample and reached a similar conclusion, including that LGRBs occur in regions with a higher surface luminosity compared to core-collapse SNe.  These results are in tension with the conclusion of \citet{Bloom2002} that LGRBs track an exponential light distribution.

In addition to the tension between the results of previous studies, and the relatively small samples used in these studies, it is important to note other potential challenges.  First, the aforementioned studies were based on LGRB samples collected from multiple satellites using different trigger criteria, potentially introducing selection effects that are difficult to quantify.  Second, the redshifts were mainly obtained from direct host galaxy spectroscopy (emission lines), leading to a bias toward lower redshifts and more luminous hosts; in conjunction with the small samples this also limited the ability to probe any redshift evolution in the LGRB locations.  Finally, these studies did not include ``dark'' bursts (i.e., bursts that suffer from large rest-frame extinction), leading to a potential bias against dusty environments; at least on the global scale the hosts of dark bursts appear to be more luminous and massive than the general LGRB host sample \citep{Kruhler2011,Perley2013}.  These shortcomings can now be overcome with the much larger and uniform sample of {\it Swift} LGRBs.

Beyond the direct study of LGRB progenitors, the fractional flux distribution of LGRBs has also been used as a point of comparison with core-collapse SNe.  \citet{Fruchter2006} found that the core-collapse SNe from GOODS were consistent with a uniform distribution.  \citet{Svensson2010} reached a similar conclusion using a larger sample of GOODS and PANS core-collapse SNe.  For more local SN samples, \citet{Kelly2008} measured the fractional flux distributions for Type II and Ib/c SNe and found that Type Ic SNe are consistent with being drawn from the same fractional flux distribution as LGRBs, whereas Type II and Ib SNe uniformly track the light of their hosts.  They suggest that LGRBs and Type Ic SNe share a common progenitor, and that a factor such as metallicity may determine whether core-collapse results in a LGRB with an associated Ic-BL SNe or a normal Type Ic SNe with no associated LGRB.  However, recent work has found that LGRBs and Ic-BL SNe occur in host galaxies with high star formation rate density and that this preference cannot be due to metallicity \citep{Kelly2014}.  Finally, the LGRB fractional flux distribution has also been compared to that for super-luminous supernovae yielding a result that is suggestive of agreement \citep{Lunnan2015}.  With the growing samples of SNe, comparative studies of SN and LGRB environments have become limited by the small sample of LGRBs from the studies carried out a decade ago \citep{Bloom2002, Fruchter2006}.  This situation can now be remedied with the much larger sample of {\it Swift} LGRBs.

Here we present the first uniform analysis of \textit{HST} follow-up observations of the host galaxies of about 100 {\it Swift} LGRBs collected over the last decade.  The goal of this analysis is twofold.  First, to investigate the offset and fractional flux distributions for a much larger and more uniform sample of LGRBs than previously possible.  Second, to provide a much larger comparison sample for future studies of the locations of other astrophysical transients, such as various SN types.  The structure of our paper is as follows.  In Section 2 we present the details of the host galaxy and afterglow observations, data analysis, and astrometric matching.  In Section 3 we discuss our offset, fractional flux, and host assignment methodologies, and present our measurements.  In Section 4 we present our resulting offset and fractional flux distributions, and in Section 5 we discuss potential 
trends with redshift, the relationship between fractional flux and offset, and implications for LGRB progenitors.  We conclude with a summary in Section 6.

In this paper we use $H_{0} = 67$ km s$^{-1}$ Mpc$^{-1}$, $\Omega_{m} = 0.32$, and $\Omega_{\Lambda} = 0.68$ \citep{Planck2013}, as well as AB magnitudes \citep{OkeGunn1983}, corrected for Galactic extinction \citep{SF2011}.        

\section{Observations and Astrometry}
\subsection{\textit{HST} Observations and Reduction}
\begin{figure*}[ht!]
\begin{center}
\includegraphics[scale=0.55]{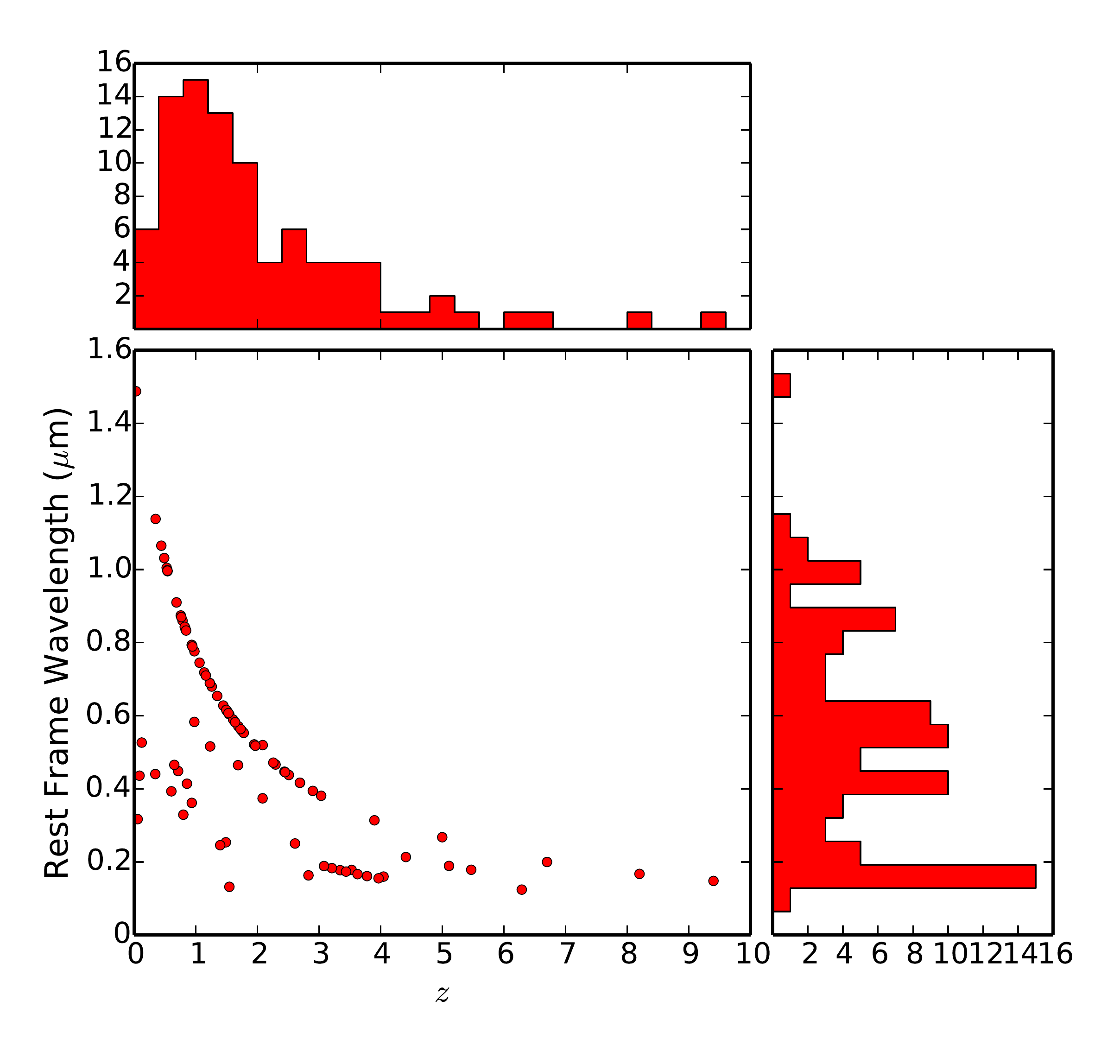}
\end{center}
\caption{Rest-frame wavelength versus redshift of the observations used to measure offsets and fractional flux for each LGRB.  Only bursts with measured redshifts are included in this plot.  Also shown are projected histograms of the redshift distribution and the distribution of rest-frame wavelengths probed by the \textit{HST} observations.}
\label{fig:red}
\end{figure*}

To locate the LGRBs within their host galaxies we utilize high resolution \textit{HST} images.  We primarily use Wide Field Camera 3 (WFC3) and Advanced Camera for Surveys (ACS) data, but also Wide Field Planetary Camera 2 (WFPC2) and Near Infrared Camera and Multi Object Spectrometer (NICMOS) data when the former are not available.  Our sample is composed of 100 LGRBs observed over the last decade (August 2004 to July 2013) under multiple programs (see Table \ref{tab:obs} for program ID numbers).  We perform the first uniform analysis of these data and utilize them for the purpose of investigating the locations of LGRBs within their hosts and their relation to the underlying host light distribution.  This sample has the significant advantage of largely consisting of bursts discovered by the \textit{Swift} satellite and thus does not suffer from any potential biases that may be associated with combining data from multiple satellites as was done in previous studies.  About 86\% of the bursts in the sample have measured redshifts ($0.03 \lesssim z \lesssim 9.4$) and the \textit{HST} imaging for these bursts span rest-frame UV and optical wavelengths (Figure \ref{fig:red}).  The wide redshift range allows us to investigate trends as a function of cosmic time.  Unlike previous studies, the sample also includes a substantial number of dark LGRBs, which either lack optical afterglows or are much fainter than expected (GRBs 051022, 060719, 060923A, 061222A, 070306, 070521, 070802, 080207, 080325, 080607, 081109, 081221, 090404, 090407, 090417B, 090709A, 100413A, 100615A, and 110709B).  Some are detected in the NIR while others are exclusively detected in the radio and X-ray bands.    

We retrieved the \textit{HST} observations from the MAST archive; for the ACS images we obtained charge transfer efficiency (CTE) corrected images, while for WFC3/UVIS data we used software from STScI to apply CTE corrections.  We reduced the data using the {\tt astrodrizzle} \citep{Gonzaga2012} task in the STSDAS IRAF package utilizing recommended parameter settings for each instrument.  By combining dithered exposures, this task enables the reconstruction of a higher resolution image than sampled by the instrumental point spread function.  The task also applies distortion corrections to the images, which are critical for precise astrometric alignment.  We use {\tt final\_pixfrac} = 0.8 and {\tt final\_scale} = 0.065", 0.02", 0.03", 0.05", and 0.15" per pixel for WFC3/IR, WFC3/UVIS, ACS, WFPC2, and NICMOS, respectively.  In Table \ref{tab:obs} we list for each LGRB the program ID, instrument, filter, and exposure time of the final drizzled image.  In cases where there are multiple epochs of imaging we list only the final epoch, unless an earlier epoch was chosen as the best host image.  In Figure \ref{fig:mosaic} we show the final drizzled image for each burst, with the location of the afterglow shown as determined by relative or absolute astrometry (Sections 2.3 and 2.4).

\begin{figure*}[h!]
\begin{center}
\subfloat[Panel One]{\includegraphics[scale=0.7]{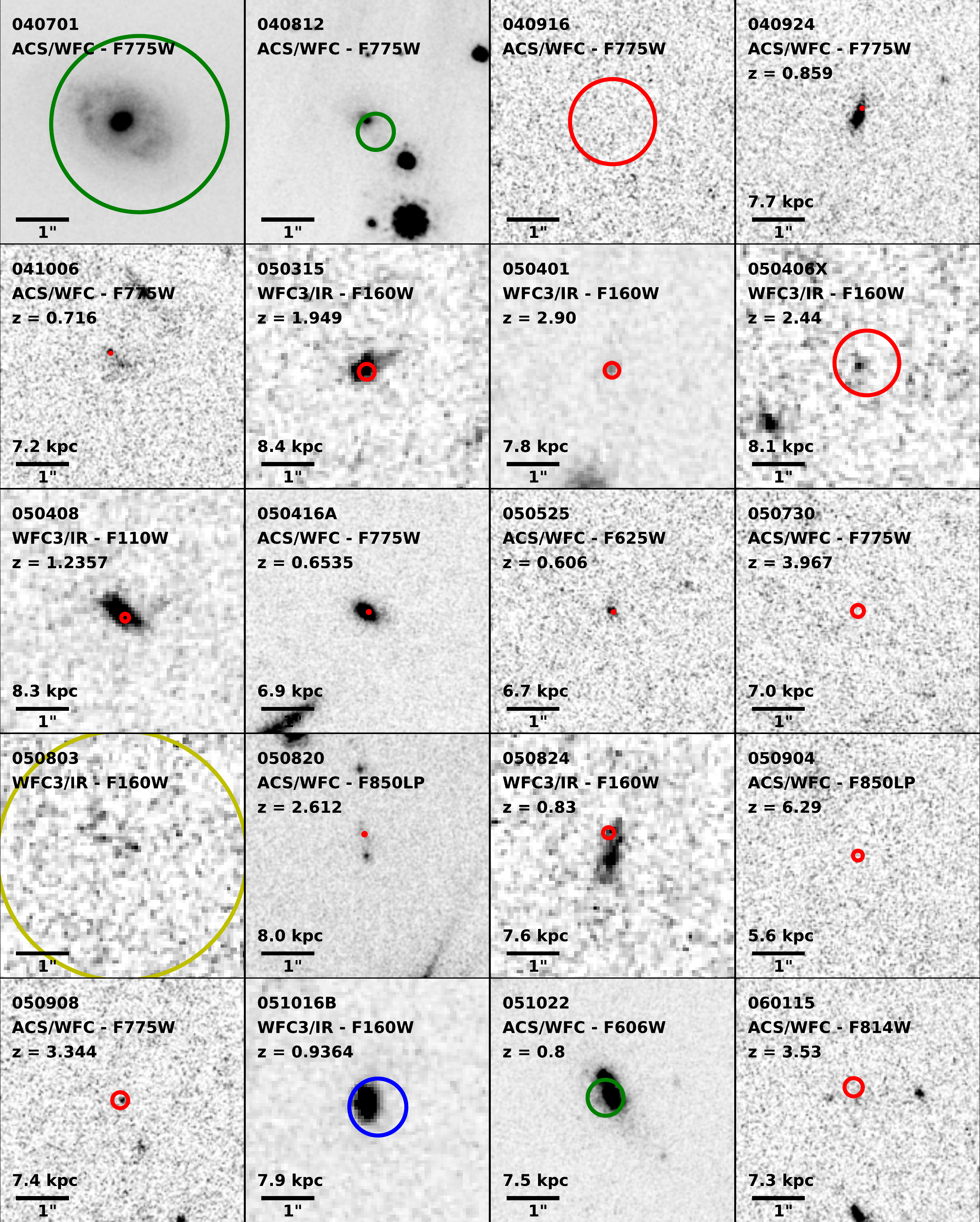}}

\end{center}
\caption{\textit{Hubble Space Telescope} drizzled images of 100 LGRB host galaxies studied in this work.  The circles (3$\sigma$) indicate the location of each GRB.  North is up and East is to the left.  Red, blue, green, cyan, and yellow circles indicate that  afterglow positions were determined from optical/NIR, \textit{Swift}/UVOT, \textit{Chandra}, radio, and XRT images, respectively.  Physical scales are given when the LGRB redshift is known.}
\label{fig:mosaic}
\end{figure*}

\begin{figure*}
\begin{center}
\ContinuedFloat
\subfloat[Panel Two]{\includegraphics[scale=0.7]{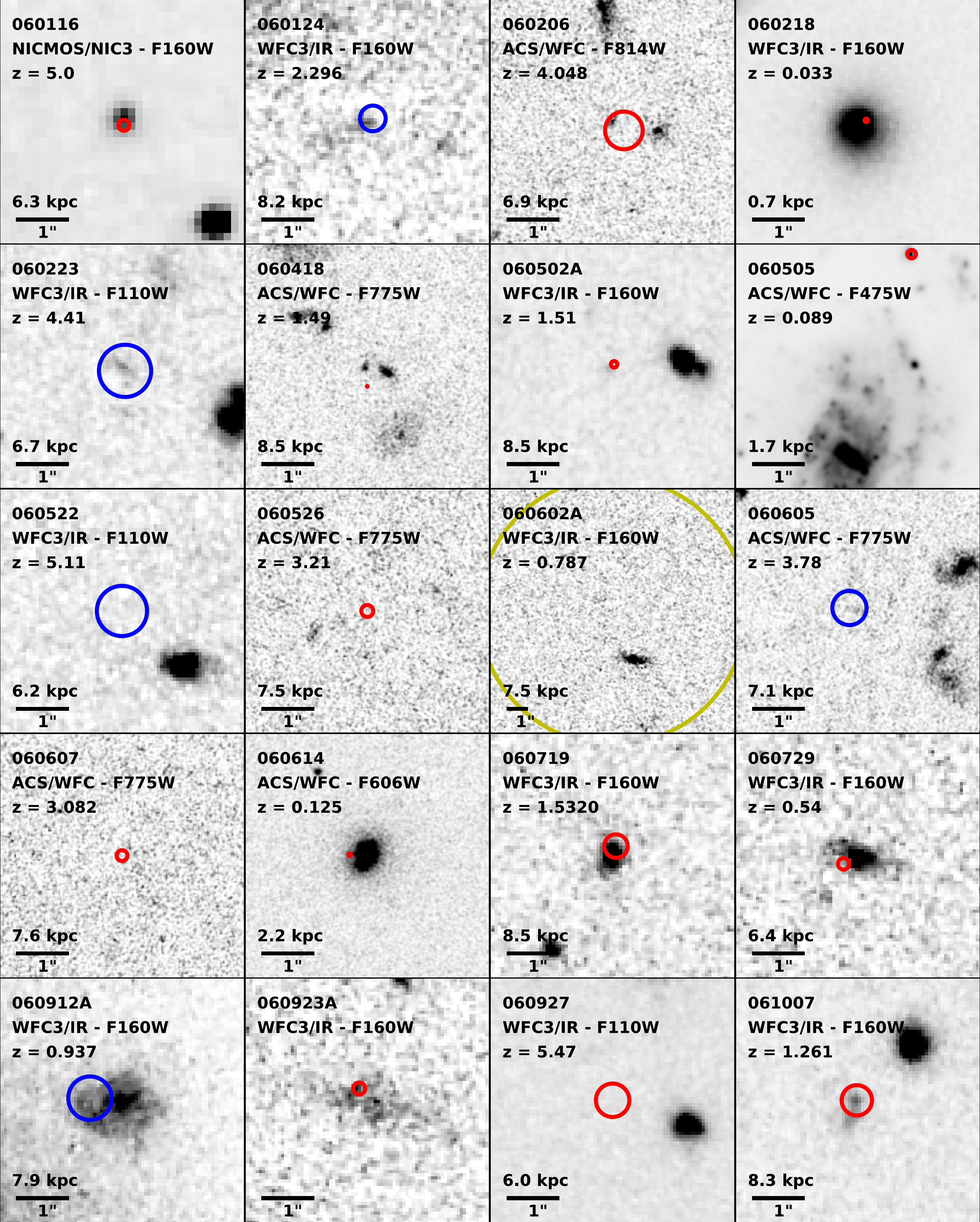}}
\end{center}
\caption{Continued}
\end{figure*}

\begin{figure*}
\begin{center}
\ContinuedFloat
\subfloat[Panel Three]{\includegraphics[scale=0.7]{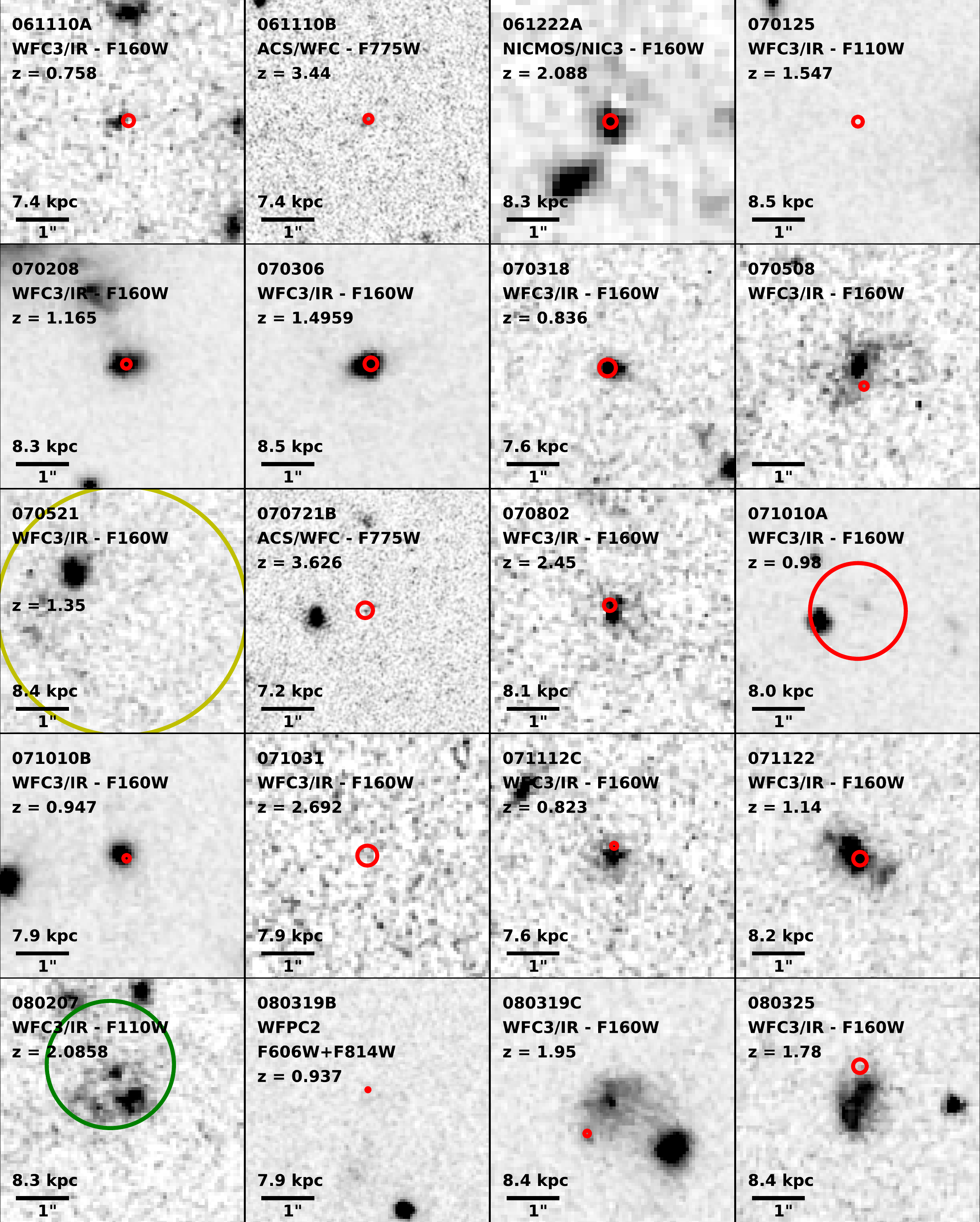}}
\end{center}
\caption{Continued}
\end{figure*}

\begin{figure*}
\begin{center}
\ContinuedFloat
\subfloat[Panel Four]{\includegraphics[scale=0.7]{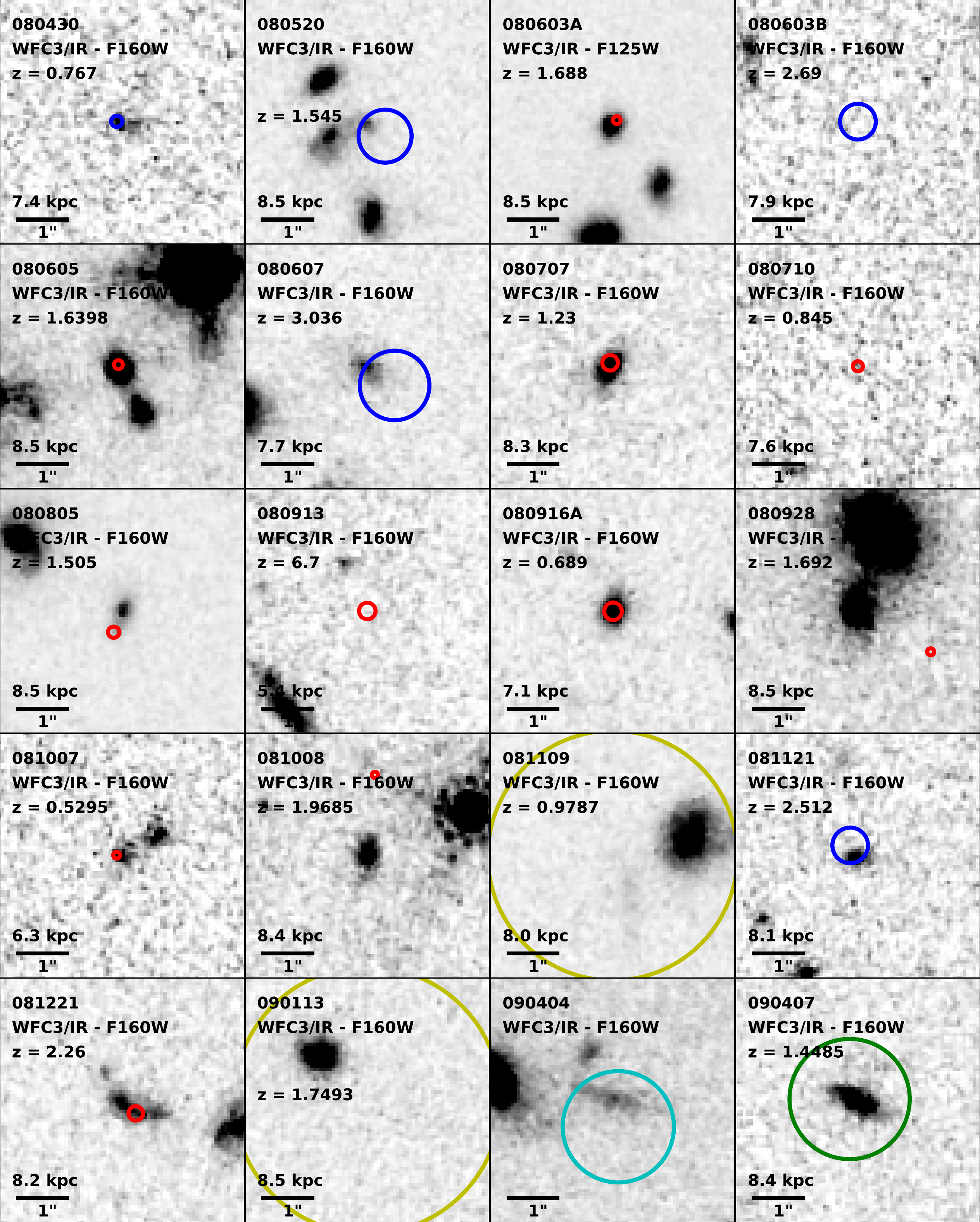}}
\end{center}
\caption{Continued}
\end{figure*}

\begin{figure*}
\begin{center}
\ContinuedFloat
\subfloat[Panel Five]{\includegraphics[scale=0.7]{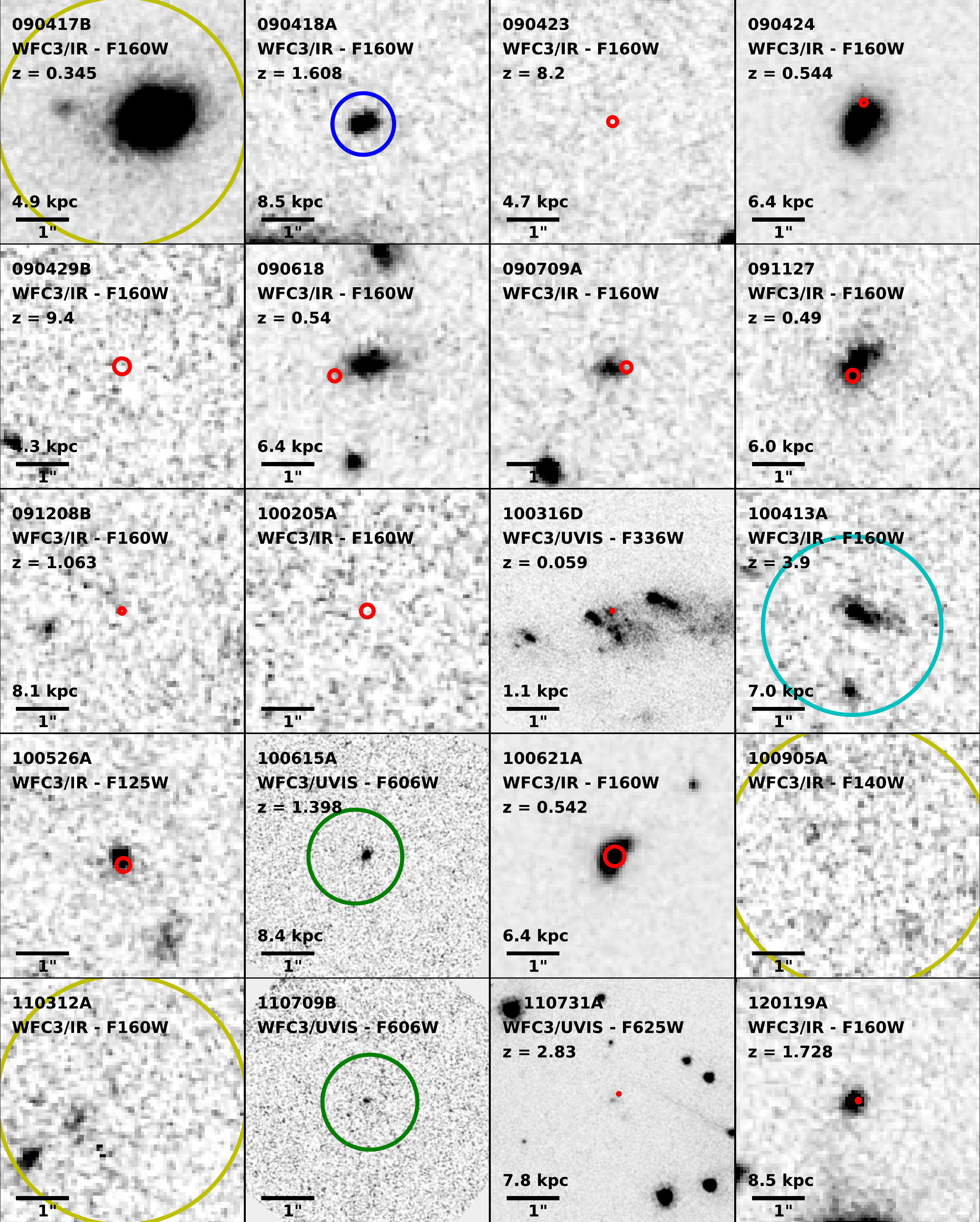}}
\end{center}
\caption{Continued}
\end{figure*}

\subsection{Afterglow Imaging}
To precisely locate the LGRBs within their host galaxies requires the detection of an afterglow.  About 85\% of bursts in the sample have detected optical or NIR afterglows.  We utilize the deepest, highest resolution afterglow images available to best match the quality of \textit{HST} imaging.  We primarily rely on publicly available afterglow images from the 8-m Very Large Telescope (VLT) and 8-m Gemini North and South telescopes.  In 13 cases, an afterglow image is available from \textit{HST}.  We also use images from the \textit{Swift} UV/Optical Telescope (UVOT), 6.5-m Magellan telescopes, Palomar 60-inch Telescope, Swope 40-inch Telescope, and the 8.2-m Subaru Telescope.  

We use standard IRAF packages to analyze the afterglow data.  We use the task {\tt uvotimsum} in the HEASoft package to co-add \textit{Swift}/UVOT images, combining multiple filters if necessary to obtain a high signal-to-noise ratio (S/N) afterglow detection.  For \textit{HST} afterglow images we use {\tt astrodrizzle} as described in Section 2.1.  For GRB 070306 we use the {\tt eclipse} pipeline to reduce the NIR data from VLT/ISAAC.  For GRBs 080325 and 090709A we use the MCSRED IRAF package to reduce the NIR data from Subaru/MOIRCS.  For the remaining bursts without optical or NIR afterglows, we rely on radio or X-ray detections.  We obtain the reported radio positions from the GCN circular archive.  When an X-ray afterglow image is available from \textit{Chandra}, we retrieve the image from the \textit{Chandra} Data Archive.  For each burst, we list in Table \ref{tab:obs} the telescope and filter utilized to obtain the afterglow image that was used for astrometry.

\subsection{Relative Astrometry}
\label{sec:rel}
To precisely locate the afterglows on the \textit{HST} images we use relative astrometry.  We first identify  common point-like sources between the afterglow and \textit{HST} images, using {\tt SExtractor} \citep{BertinArnouts1996} to measure the source positions.  We then match the coordinates with the IRAF task {\tt ccmap}, using a second or third order polynomial to compute the astrometric solution.  Using the afterglow images from the telescopes listed in Section 2.2 we generally obtain the following astrometric tie uncertainties, $\sigma_{\rm tie}$ ($1\sigma$):

        \begin{itemize}
          \item 6 mas using an \textit{HST} detection of an optical or NIR afterglow 
          \item 30 mas using an optical or NIR afterglow detection from ground-based telescopes 
          \item 100 mas using a UV or optical afterglow detection from \textit{Swift}/UVOT 
          \item 300 mas using an X-ray afterglow detection from \textit{Chandra} 
        \end{itemize}

In a few cases, the lack of sufficient common sources for a robust astrometric tie necessitates the use of an intermediate image to locate the afterglow position on the \textit{HST} image.  This is the case for bursts detected with \textit{Chandra}.  We use the CIAO task {\tt wavdetect} to measure the positions of X-ray sources.  In these cases, the uncertainty associated with the match between the \textit{Chandra} and intermediate ground-based images dominates the total astrometric tie uncertainty.  Table \ref{tab:data} lists the number of common sources and astrometric tie uncertainty for each burst.  Bursts with no listed astrometric tie uncertainty are those for which the best afterglow position is from XRT or for which we were not able to obtain an afterglow image.  GRB 060418 is a unique case where the afterglow and host are well-detected in a single \textit{HST} image and are sufficiently offset from each other such that their positions are not biased by light contribution from each other (later epoch imaging revealed no host emission at the afterglow position).  Therefore astrometric matching is not necessary for this burst.

\subsection{Absolute Astrometry}
In cases where the afterglow is detected only in the radio band, or when the X-ray observations do not contain any common sources for relative astrometry, absolute astrometry is performed by matching the late-time \textit{HST} images to the 2MASS astrometric system.  This is mainly the case for dark GRBs.  For the two GRBs where a radio position was used (090404 and 100413A), the uncertainty in the afterglow position with respect to the \textit{HST} image is dominated by the astrometric accuracy of those radio positions, $\sim$ 0.5".  Similarly, when using absolute astrometry to locate the X-ray afterglow of GRB 090407, the error is dominated by the astrometric accuracy of \textit{Chandra}, $\sim$ 0.4".     

\section{Measurement Methodology}
\subsection{Offset Measurements}
\label{sec:off}
Once the afterglow and \textit{HST} images are registered, we identify the host galaxy in the \textit{HST} image.  First we measure the afterglow centroid using {\tt SExtractor} to determine the afterglow position relative to the \textit{HST} image.  The afterglow positional uncertainty ($\sigma_{\rm AG}$) is estimated as $\frac{FWHM}{2(S/N)}$.  In some cases, it is apparent in the afterglow image that there is an underlying extended object, presumably the host galaxy.  To reliably measure the centroid of the afterglow we perform image subtraction, using the ISIS software package \citep{Alard2000}, relative to a late-time image of the host ideally from the same telescope and in the same filter.  Once the afterglow position is determined, we use a uniform host assignment procedure discussed in Section \ref{sec:pcc}.  Here we continue our discussion of the offset measurement procedure assuming a host galaxy has been identified.    

We measure the host galaxy center and corresponding uncertainty using {\tt SExtractor}.  We use a uniform definition of the host center as the brightest region of the galaxy (or peak of the light distribution) determined by sequentially increasing the threshold for detection in {\tt SExtractor}.    The statistical uncertainty on the centroid determined by {\tt SExtractor} does not take into account the larger systematic uncertainties caused by the fact that LGRB hosts tend to have irregular morphologies.  Instead, the uncertainty on the host center ($\sigma_{\rm host}$) is estimated by taking the quadrature sum of the standard deviation in both directions of the positions measured by multiple runs of {\tt SExtractor} using varying thresholds for detection from $3\sigma$ to the highest threshold that yields a detection. 

Using the afterglow position and host center we calculate the projected angular offset ($R$) between the afterglow position and host center.  The total uncertainty on the angular offset ($\sigma_{R}$) is taken to be the quadrature sum of $\sigma_{\rm tie}$, $\sigma_{\rm AG}$, and $\sigma_{\rm host}$.  We also calculate the projected physical offsets using the measured redshift for each burst.  For bursts with no measured redshift we use the fact that at $z \gtrsim 0.5$, typical of most LGRBs, the angular diameter distance is approximately constant so that we can use a conversion factor of about 8 kpc/".  The projected angular and physical offsets and corresponding uncertainties are listed in Table \ref{tab:data}.  Bursts for which no offsets are listed indicate either bursts with well-localized afterglows but no detected host (see Section \ref{sec:pcc} for a discussion of distinguishing between no host detection and a large offset) or bursts for which we only have an XRT position or were unable to access the afterglow image.  The uncertainties on the XRT positions are typically $\gtrsim 1\arcsec$ in radius, too large to enable a reliable host association.  

\subsection{Host Galaxy Sizes, Normalized Offsets, and Magnitudes}
As our interest lies in the positions of LGRBs relative to the light of their host galaxies, we normalize the offsets by the sizes of the hosts.  Here we use the half-light radius ($R_{h}$) as measured by {\tt SExtractor} (with {\tt FLUX$\_$RADIUS} = 0.5).  We also measure the 80\% light radius ($R_{\rm 80}$) for comparison with previous studies of LGRBs and SNe.  In Table \ref{tab:data} we list the half-light radii and host-normalized offsets ($R_{\rm norm}$) for each burst.  

We also measure the apparent magnitudes of the host galaxies, and corresponding uncertainties, using the MAG$\_$AUTO estimate from {\tt SExtractor} which uses Kron apertures.  We obtain the magnitude zeropoints from STScI where tabulated, or from the photometry keywords in the \textit{HST} image headers.  The magnitudes corrected for Galactic extinction \citep{SF2011} and corresponding uncertainties are listed in Table \ref{tab:data} for each burst with an associated host galaxy.

\subsection{Host Galaxy Assignment}
\label{sec:pcc}
To assign host galaxies to each LGRB we calculate the probability of chance coincidence, for several candidate extended objects surrounding the afterglow position, given the observed number density of field galaxies from deep optical and NIR surveys \citep{Hogg1997,Beckwith2006,Metcalfe2006}.  Nominally we assign the candidate with the lowest probability of chance coincidence as the host.  For candidates at small offsets from the afterglow position, such that the afterglow position is essentially coincident with the candidate, this candidate will in most cases have the lowest probability of chance coincidence.  Bursts for which there is no coincident candidate require further investigation to distinguish between the scenario where the host was not detected and the scenario where the burst occurred at a large offset from a neighboring galaxy.  We discuss these cases in detail in Section \ref{sec:lum}. Following the methodology of \citet{Bloom2002} we determine the probability of chance coincidence ($P_{\rm cc}$) using the following equation:
\begin{equation}
P_{\rm cc} = 1 - e^{-\pi R_{e}^{2} \sigma(\leq m)}
\end{equation}
where $\sigma(\leq m)$ is the observed number density of galaxies brighter than magnitude $m$, 
and $R_{e}$ is the effective radius.  For optical observations we use the galaxy number densities given by Equation 1 of \citet{Berger2010}, whereas for NIR observations we use the $H$-band number counts measured by \citet{Metcalfe2006}.  As discussed by \citet{Bloom2002}, the appropriate value of $R_{e}$ depends on the offset, size of the galaxy, and uncertainty on the afterglow position.  Following \citet{Bloom2002} we take $R_{e}$ to be the maximum of ($3\sqrt{\sigma_{\rm tie}^{2} + \sigma_{\rm OT}^{2}}, \sqrt{R^{2} + 4R_{h}^{2}}$).  In other words, if the localization is precise, such that the first term is much smaller than the offset and galaxy size, then the second term dominates.  On the other hand, if the GRB localization is poor the positional uncertainty dominates over the offset and galaxy size, and its radius becomes the appropriate value of the effective radius for determining $P_{\rm cc}$.  In the majority of cases ($\sim 85\%$) the localizations are sufficiently precise such that the second term dominates over the first.  As discussed in Section \ref{sec:rel}, $\sigma_{\rm tie}$ is typically $\gtrsim 100$ mas for bursts where we use afterglow detections from \textit{Swift}/UVOT or \textit{Chandra}.  In these cases, the GRB positional uncertainty can dominate over the offset and galaxy size.          

We list the values of $P_{\rm cc}$ in Table \ref{tab:data}.   For the bursts with coincident extended objects, $P_{\rm cc}$ tends to fall in the range $10^{-3}$ to $5\times10^{-2}$.  Therefore we do not expect significant contamination of our host galaxy sample by unrelated galaxies along the line of sight. 

\subsubsection{Bursts With No Coincident Host Candidate} 
\label{sec:lum}
 The subset of bursts that lack a coincident extended object must be examined further to determine if the object with the lowest $P_{\rm cc}$ can realistically be considered the host galaxy.  Relying simply on $P_{\rm cc}$ may result in the inclusion of unrealistically luminous objects in the sample.  We identify as possible host candidates the nearest extended objects with $P_{\rm cc}$ $\lesssim$ 0.1.  Given the measured redshift of the burst, we can determine whether the luminosity of the host candidate is consistent with the range of LGRB host luminosities observed at that redshift.  It may be more likely that the true host lies beyond the magnitude limit of our observations and was not detected.  However, it is important to avoid biasing our results by eliminating potential large offsets.  Here we discuss our assessment of bursts whose host candidates satisfy $P_{\rm cc}$ $\lesssim$ 0.1 but are offset by significantly more than the half-light radius of the host candidate.  
  
To carry out the assessment of non-coincident host candidates with $P_{\rm cc}$ $\lesssim$ 0.1 we compare their luminosities to the luminosities of the coincident host galaxies.  We convert our measured apparent magnitudes to rest-frame  luminosities in units of $L^{*}$.  Given the redshift, we determine the absolute magnitude and using measured values of $L^{*}$ from galaxy surveys \citep{Faber2007,Marchesini2007,Bouwens2015} we calculate $L/L^{*}$ for each host galaxy.  For bursts with $z \lesssim 1.7$ we use the measured value of $L^{*}$ in the $B$-band for the blue galaxy sample analyzed by \citet{Faber2007}.  In the range $1.7 \lesssim z \lesssim 3.2$ we use $L^{*}$ values in the $B$-band from \citet{Marchesini2007} using their $U - V < 0.25$ mag galaxy subsample.  For bursts with $z \gtrsim 3.2$ we use $L^{*}$ corresponding to UV bands as measured by surveys of Lyman-break galaxies \citep{Bouwens2015}.  These were chosen so that the wavelength regime in which the $L^{*}$ values are appropriate  correspond roughly to the rest-frame wavelengths probed by our observations in these redshift bins.  We do not perform a $K$-correction to a common band.  

\begin{figure*}[ht!]
\begin{center}
\includegraphics[scale=0.5]{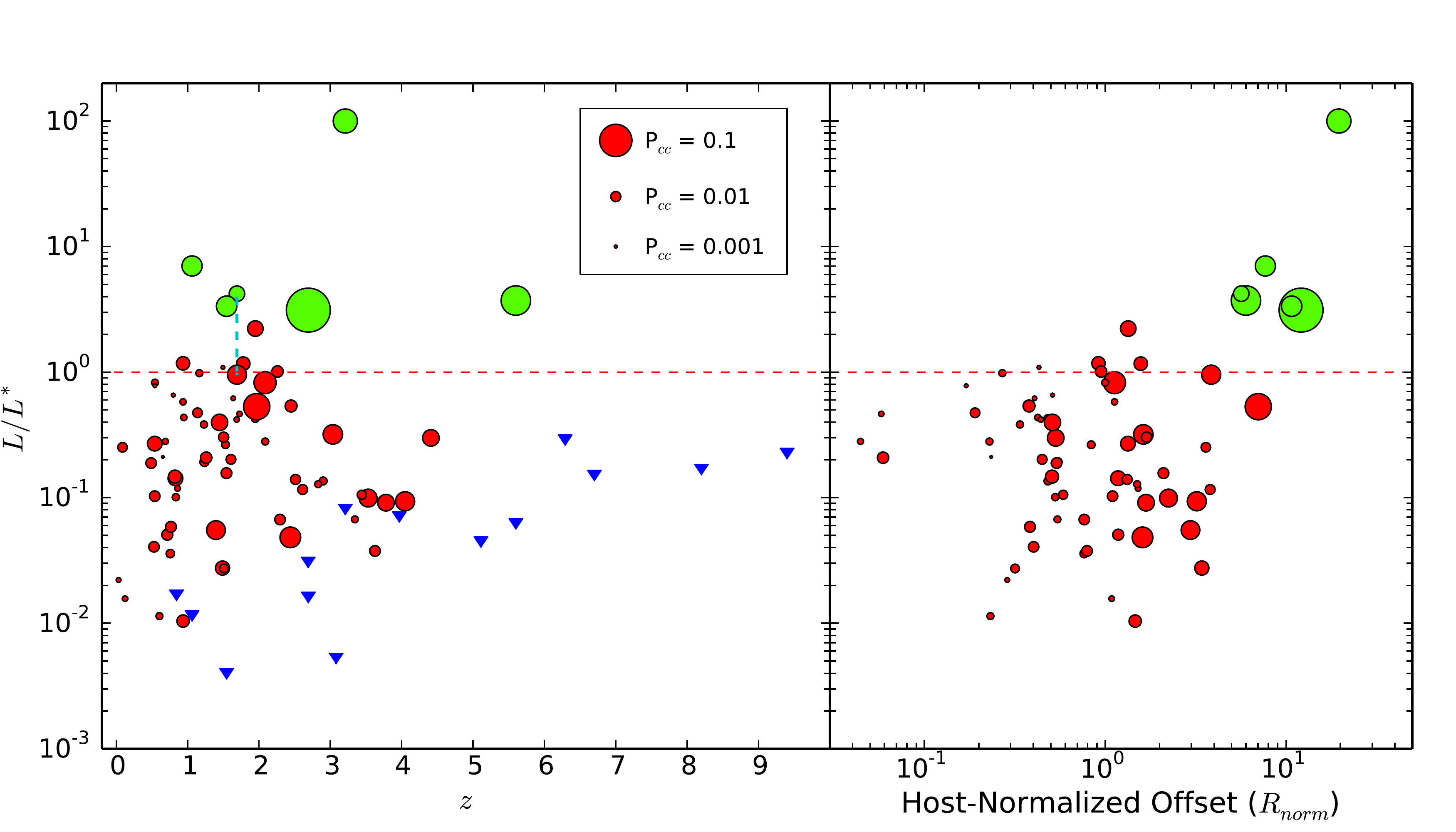}
\end{center}
\caption{\textit{Left:} Host galaxy rest-frame luminosity in units of $L^{*}$ as a function of redshift for assigned host galaxies (red),  rejected host candidates (green), and non-detected hosts (blue).  The data point area is proportional to $P_{\rm cc}$ where a smaller point means a smaller $P_{\rm cc}$.  The red dotted line indicates a luminosity of $L^{*}$, and the dotted cyan line connects the two possible host candidates for GRB 080928 (see Section \ref{sec:lum}).  We do not include bursts with unknown redshift.  \textit{Right:} Host galaxy luminosity versus host-normalized offset.  As expected this is a scatter plot for the robust host associations.  The correlated tail caused by the rejected candidates is an indication that they are unrelated galaxies.  }
\label{fig:lum}
\end{figure*}

We plot the luminosities as a function of redshift for the full sample, including coincident and non-coincident candidates, in the left panel of Figure \ref{fig:lum}.  The right panel shows the luminosities as a function of host-normalized offset.  The luminosities of the robust host galaxy associations span $\sim (0.01 - 2) L^{*}$, whereas some host candidates at relatively large offset and high $P_{\rm cc}$ have luminosities ranging from $\sim (3 - 100) L^{*}$, inconsistent with the distribution for bursts with coincident host galaxies and indicating that they are unrelated galaxies.  The luminosity of the host candidate for GRB 060526 is $\sim 100L^{*}$ assuming it is at the redshift of the GRB, much larger than expected for a GRB host.  We therefore reject this association.  We also reject the candidates for GRBs 060927, 070125, 071031, and 091208B because their luminosities are unreasonably high given the observed distribution for securely assigned hosts.  In the case of GRB 080928, there are two candidates with $P_{\rm cc}$ $\lesssim$ 0.1 where the object with $P_{\rm cc} \approx 0.023$ has a luminosity of $\sim 4L^{*}$ and the object with $P_{\rm cc} \approx 0.035$ has a luminosity of $\sim L^{*}$.  We therefore accept the lower luminosity galaxy as the host.  We also accept as the host the candidate for GRB 081008 due to its reasonable luminosity of $\sim 0.6L^{*}$.  The left panel of Figure \ref{fig:lum} also shows upper limits on the luminosities of the hosts where we are confident the host was not detected due to the lack of any nearby host candidates with $P_{\rm cc}$ $\lesssim$ 0.1.  Upper limits for the cases with rejected host candidates discussed above are also shown.

Furthermore, we note that there is no trend of host luminosity versus host-normalized offset.  There is no reason to suspect from predictions that bursts at large or small offsets occur only in bright or faint galaxies.  As shown in Figure \ref{fig:lum}, the inclusion of the rejected host candidates creates an artificial trend because the only galaxies at higher normalized offset that have $P_{\rm cc}$ $\lesssim$ 0.1 are bright galaxies.  Therefore we are confident these rejected candidates are unrelated galaxies and the true host was not detected.  Indeed, the resulting upper limits on the luminosities of these hosts are not surprising given the upper limits we calculate for unambiguous non-detected hosts.  Upon completing our host assignment procedure we obtain a sample size of 71 LGRBs for which we make offset and fractional flux measurements.

\subsection{Fractional Flux Measurements}
\label{sec:FF}
Studying the locations of LGRBs within their hosts can be expanded beyond an analysis of the offset distribution.  In many cases where the host's morphology is irregular, the location of the host center is unclear and the offset from an empirically defined center may not be meaningful.  Furthermore, the host galaxy size defined by the half-light radius does not capture the complex morphology of many of the hosts.   While measuring offsets is useful for comparing the locations of LGRBs with predicted profiles of star formation, we would like to go one step further and ask whether or not LGRBs are spatially coincident with sites of massive star formation.  This is possible using a method, less sensitive to galaxy morphology than offsets, of comparing the brightness at the LGRB location relative to the entire host light distribution \citep{Fruchter2006}.  We calculate the fraction of the total flux from the host galaxy that is contained in pixels fainter than or equal to the flux at the LGRB location.  The resulting "fractional flux" is a statistic that measures the brightness of the burst site compared to the entire galaxy.  A fractional flux value of 1 means that the burst occurred on the brightest pixel in the galaxy. 

We measure the fractional flux for each LGRB using the following procedure.  First we determine the flux value at the LGRB position.  When the 1$\sigma$ error circle defined by the uncertainty on the LGRB position as determined by the quadrature sum of $\sigma_{\rm tie}$ and $\sigma_{\rm AG}$ spans less than 1 pixel, we simply take the flux value of that pixel.  When the error circle encompasses more than one pixel but is smaller than the PSF, we measure an average flux within the 1$\sigma$ error circle weighted by the fractional area encompassed by a given pixel.  These scenarios are relevant for most bursts where we use afterglows from ground-based optical/NIR telescopes and \textit{HST} (see Section 2).  Using {\tt SExtractor} we extract the host galaxy pixels using a threshold of 1$\sigma$ above the sky background and then calculate the fraction of the total flux in pixels fainter than or equal to the average flux of the LGRB error region (the fractional flux).  

\begin{figure*}[ht!]
\begin{center}
\includegraphics[scale=0.45]{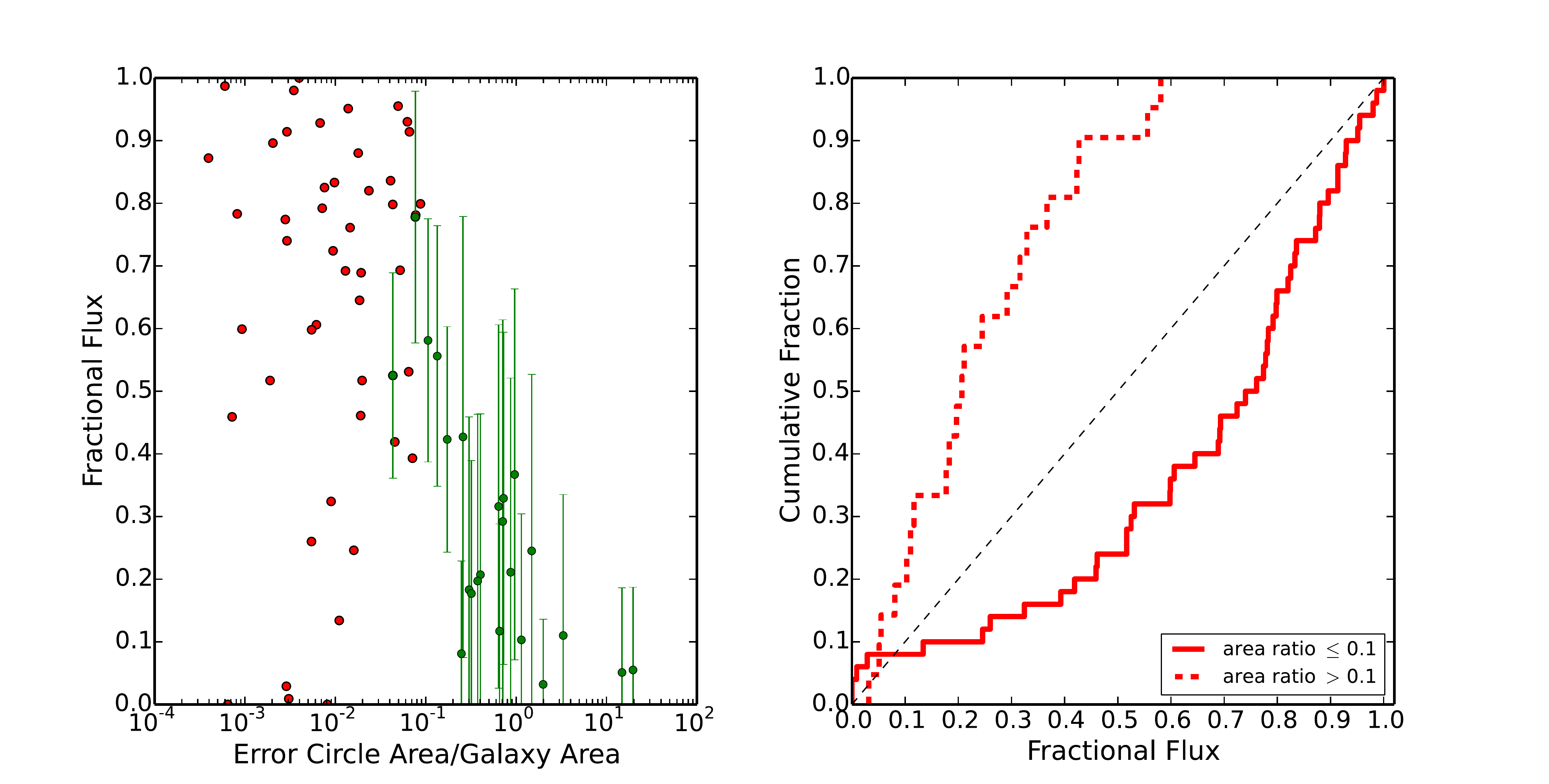}
\end{center}
\caption{\textit{Left:} Fractional flux versus the ratio of error circle area to galaxy area.  Green points with error bars represent bursts with error circles larger than the \textit{HST} PSF.  \textit{Right:} The fractional flux distribution binned into two bins separated at an area ratio of 0.1.  In both plots it is apparent that there is a bias to lower fractional flux values when the error circle covers a substantial fraction of the galaxy area.  In the analysis in Section \ref{sec:FFresults} we only use bursts with a ratio of $\lesssim$ 0.1}
\label{fig:FFdiag}
\end{figure*}

When the error circle is larger than the PSF, the galaxy brightness as a function of position could vary significantly so that the uncertainty in the position of the LGRB may have a large impact on the fractional flux.  This scenario is often encountered when using optical afterglow detections from \textit{Swift}/UVOT, X-ray afterglows from \textit{Chandra}, or radio afterglow detections, where the error regions can encompass tens to hundreds of pixels.  In some cases, these pixels include both the faintest and brightest pixels of the host galaxy.  In this regime, it is unclear whether the fractional flux determined as described above remains a meaningful quantity.  Here we show that for bursts with error circles larger than the PSF, a  different weighting is required to properly assess the fractional flux in these cases. 

To assess this, we employ a different procedure for determining the fractional flux, taking into account the Gaussian error distribution associated with each burst.  As before we extract the galaxy pixels using {\tt SExtractor}.  Each pixel in the image has an associated probability for hosting the burst, governed by the 2D Gaussian distribution that takes into account the astrometric tie and afterglow positional uncertainties.  For all pixels encompassed by the error circle, a corresponding fractional flux value can be calculated.  Pixels not associated with the galaxy will have fractional flux values of zero.  The posterior probability distribution for the fractional flux can then be constructed by weighting each value by its corresponding Gaussian probability.  The fractional flux and corresponding uncertainty are estimated as the mean and standard deviation of this distribution.  

By comparing the fractional flux values for bursts with large error regions with the well-localized bursts, we find that there exists a considerable bias towards lower fractional flux values in the former.  We hypothesize that not only is the absolute size of the error circle important but also the error circle size relative to the galaxy size.  If the galaxy is much larger than the error circle, even a large error circle in an absolute sense can yield a well-constrained fractional flux value.  If a relatively well-localized burst occurs in a small galaxy, the fractional flux value may not be well-constrained.  Figure \ref{fig:FFdiag} illustrates this bias by plotting fractional flux versus the ratio of error circle area to galaxy area (left panel).  The right panel of Figure \ref{fig:FFdiag} shows the cumulative distributions for two subsets of the data divided at an area ratio of 0.1.  Of the bursts with area ratios $\gtrsim 0.1$, there are none with fractional flux values greater than $\sim 0.6$, while bursts with an area ratio of $\lesssim 0.1$ span the full range of fractional flux values. This bias can be understood by noting that as the error region increases relative to the galaxy size, there is greater probability that the burst occurred on background sky pixels.  In general we find that for error regions large in an absolute or relative sense there is a bias towards lower fractional flux values.  To avoid biasing our results we remove from consideration the bursts with area ratios of $\gtrsim 0.1$, resulting in a sample size of 50 LGRBs.  By removing these bursts we are not concerned about selecting against bursts with intrinsically low fractional flux values.  Here we are removing a bias associated with the measurement of fractional flux when burst localization is poor.   

In Table \ref{tab:FF} we list the fractional flux values and area ratios for each LGRB.  Bursts with no fractional flux value are either bursts for which a host was not detected or a sufficient afterglow position was not available.

\section{Results}
\label{sec:results}
We have compiled the largest sample of LGRB offsets and fractional flux measurements presented to date.  Our aim is to use them to understand where within their host galaxies LGRBs occur and what the results from a large sample indicate about the progenitors.  Here we present the results of our offset and fractional flux distribution measurements summarized in Tables \ref{tab:data} and \ref{tab:FF}.    

\subsection{Offset Distribution}
In this section we present the measured physical and host-normalized offset distributions.  We also present the results of an analysis of the offset uncertainties to understand how they affect the offset distributions.  Each offset measurement has an associated uncertainty ($\sigma_{R}$) dependent on $\sigma_{\rm tie}$, $\sigma_{\rm AG}$, and $\sigma_{\rm host}$ (Section \ref{sec:off}).  Due to the inherent non-Gaussian nature of offsets and the non-uniform uncertainties, we employ a Monte Carlo approach to assess the uncertainty on the resulting distributions of offsets.  At each iteration a random offset ($x$) was drawn for each burst from its offset probability distribution defined by the measured offset ($R$) and uncertainty ($\sigma_{R}$).  For bursts with $R/\sigma_{R} \lesssim 5$ we use a Rice distribution to represent the probability distributions, defined by:
\begin{equation}
p(x|R,\sigma_{R}) = \frac{x}{\sigma_{R}^{2}}\text{exp}\left[-\frac{(x^{2} + R^{2})}{2\sigma_{R}^{2}}\right]I_{0}\left(\frac{xR}{\sigma_{R}^{2}}\right)
\label{eqn:rice}
\end{equation}
where $I_{0}$ is the zeroth order modified Bessel function of the first kind.  For $R/\sigma_{R} \gtrsim 5$, the Rice distribution can be approximated as a Gaussian.  We make this approximation for numerical ease.  In all cases discussed in the following sections we use 10,000 iterations.   

\begin{figure*}[!hbt]
\begin{center}
\includegraphics[scale=0.45]{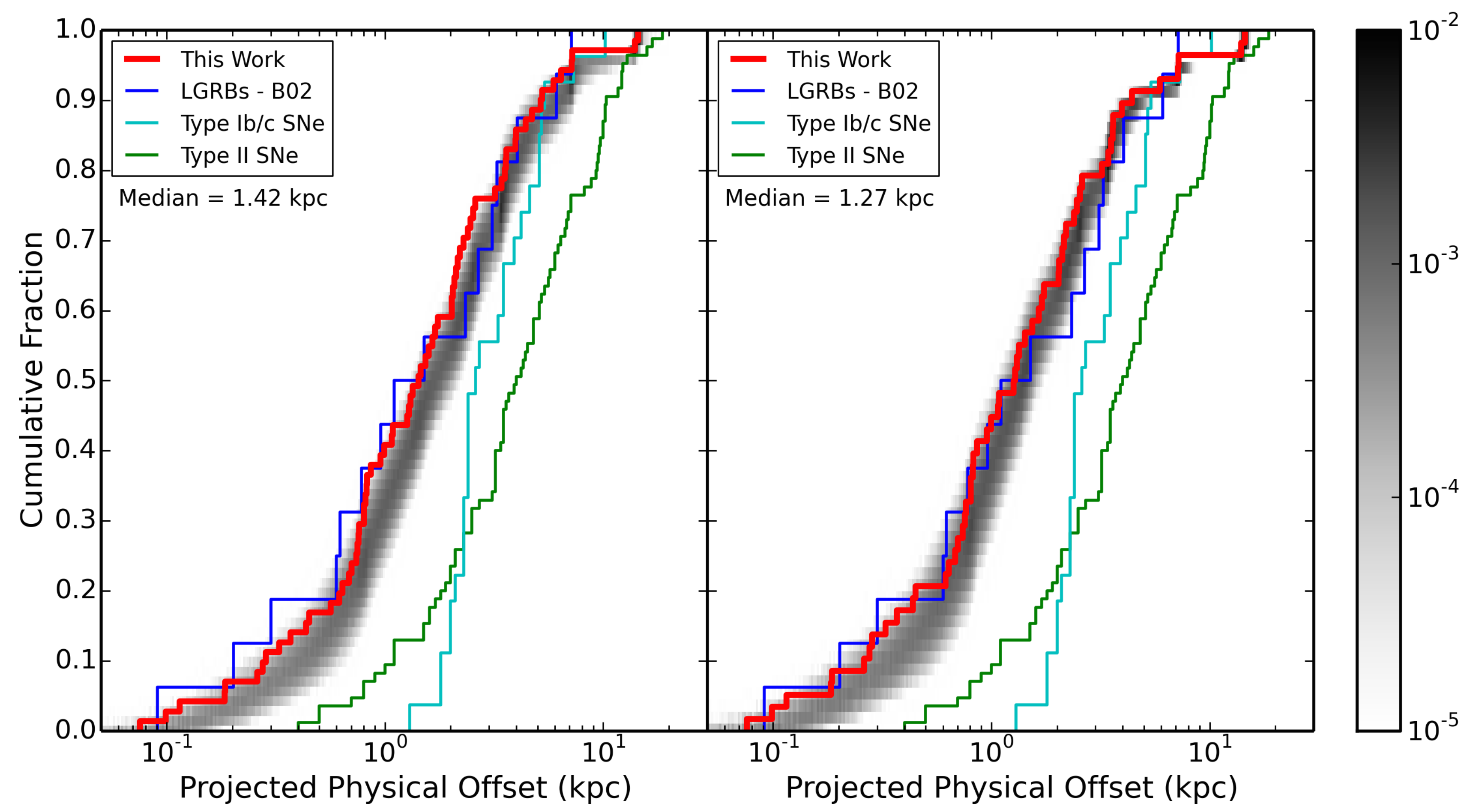}
\end{center}
\caption{\textit{Left:} Cumulative distributions of the projected physical offsets of 71 LGRBs from this work (red) and \citet{Bloom2002} (B02, blue).  We also show the distributions for Type Ib/c (cyan) and Type II (green) SNe from \citet{Prieto2008}.  The shaded region displays the results of our Monte Carlo assessment of the uncertainties on the offsets in the form of a 2D histogram. \textit{Right:} Similar to the left panel, except here we plot the sample of 58 bursts with $\sigma_{R_{\rm phys}} \lesssim 1.0$ kpc to reduce any bias caused by bursts with a large offset uncertainty (See Figure \ref{fig:offvserr}).  The results are essentially unchanged.}
\label{fig:phys}
\end{figure*}

\subsubsection{Physical Offset Distribution}

In Figure \ref{fig:phys} we show the cumulative distribution of projected physical offsets, calculated as described in Section \ref{sec:off}.  The final sample size included in this distribution is 71 LGRBs, the subset of the original sample where offset measurement is possible.  The distribution ranges from offsets of 0.075 to 14.4 kpc with the median at 1.42 kpc.  About 90\% of bursts occur within $\sim$ 5 kpc.  The distribution is remarkably smooth and has bursts representing the full range of offsets.  There are few gaps where we do not find bursts, with one notable gap at $\sim 7 - 14$ kpc.  For comparison we also show the distribution of physical offsets for the small sample of 20 LGRBs measured by \citet{Bloom2002}.  The two distributions agree exceptionally well.  A Kolmogorov-Smirnov (KS) test yields a p-value of 0.94.  We also show in Figure \ref{fig:phys} the results of our Monte Carlo simulation (taking into account the offset uncertainties) by plotting a 2D histogram showing the density of points from the resulting cumulative distributions generated at each iteration.  Darker regions indicate a higher density of points, in other words a high fraction of synthetic distributions pass through that region.  The apparent shift of the distribution from the measured distribution towards higher offset values is indicative of the skew to higher offsets resulting from the fact that an offset is a positive-definite quantity described by a Rice distribution.  The shift is dominated by the $\sim$ 10 LGRBs for which the uncertainty is larger than the offset.  

\begin{figure*}[ht!]
\begin{center}
\includegraphics[scale=0.45]{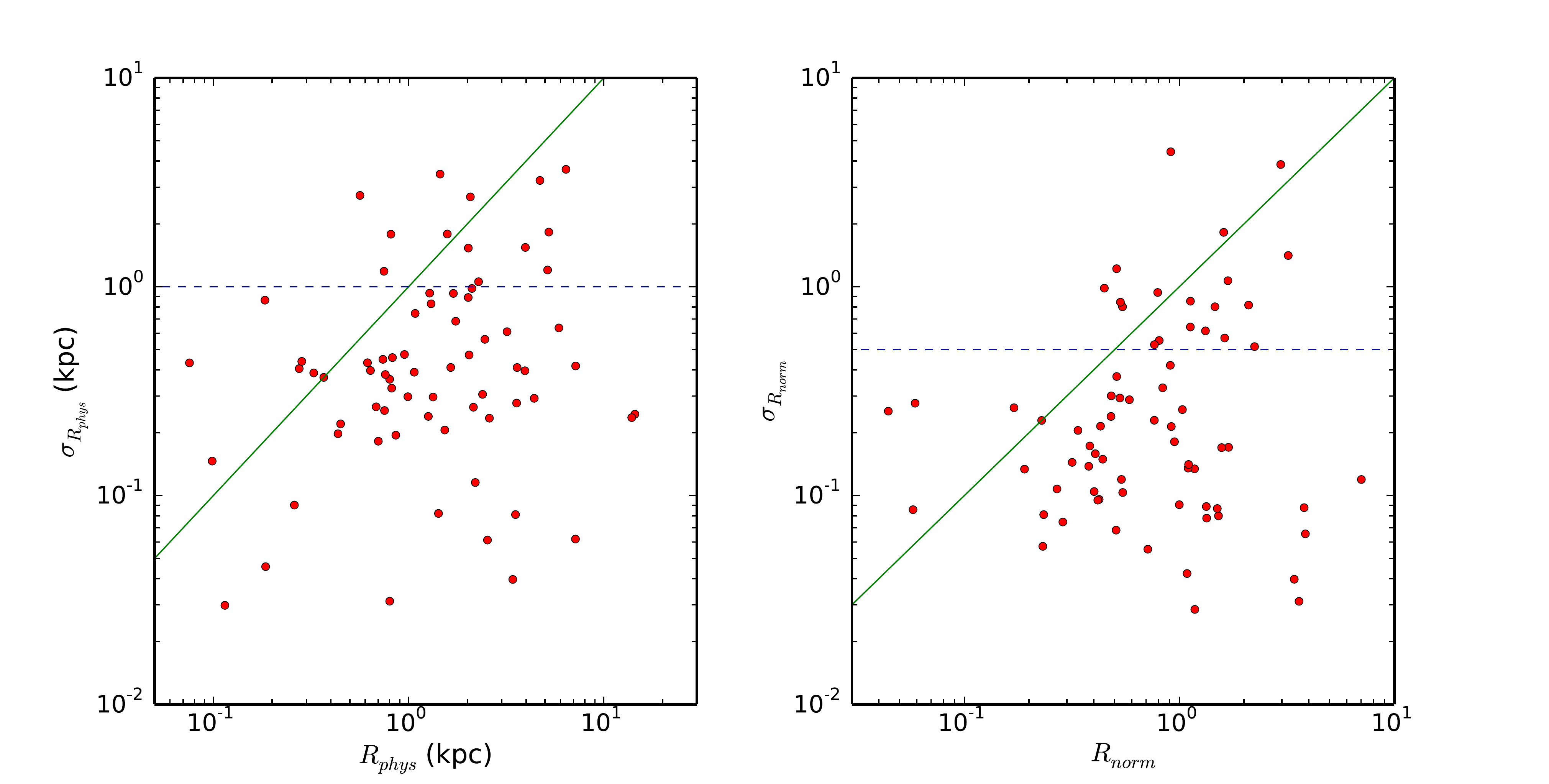}
\end{center}
\caption{\textit{Left:} Uncertainty in physical offset ($\sigma_{R_{\rm phys}}$) versus physical offset ($R_{\rm phys}$).  The green line marks $\sigma_{R_{\rm phys}} = R_{\rm phys}$.  \textit{Right:} Uncertainty in host-normalized offset ($\sigma_{R_{\rm norm}}$) versus host-normalized offset ($R_{\rm norm}$).  Similarly, the green line marks $\sigma_{R_{\rm norm}}$ = $R_{\rm norm}$.    Since $\sigma_{R_{\rm norm}}$ and $\sigma_{R_{\rm phys}}$ are not related to the intrinsic offset, the apparent lack of small offsets with large uncertainty (the gap in the upper left quadrant of both plots) is indicative of a bias to large offset associated with bursts that have poor localizations.  The dashed blue lines at $\sigma_{R_{\rm phys}} = 1$ kpc and $\sigma_{R_{\rm norm}}$ = 0.5 mark the locations of our cuts.}
\label{fig:offvserr}
\end{figure*}

It is important to carefully consider the effects of the uncertainties on the distribution of offsets because offsets with uncertainties large relative to the offset itself have probability distributions skewed to larger offset.  Given a large uncertainty, the measured offset is more likely to be large due to the large area encompassed by the error circle at large offset.  As $\sigma_{R_{\rm phys}}$ should not be correlated with the intrinsic offset, if the true offset distribution extended to small offsets, this effect would cause a bias against finding them when the uncertainties are large.  In Figure \ref{fig:offvserr} we plot $\sigma_{R_{\rm phys}}$ versus $R_{\rm phys}$.  Of the 13 bursts in the sample with $\sigma_{R_{\rm phys}}$ $\gtrsim 1.0$ kpc, all are at measured offsets $\gtrsim 0.5$ kpc, whereas the remaining bursts with $\sigma_{R_{\rm phys}}$ $\lesssim 1.0$ kpc span the full range of offsets from 0.075 to 14 kpc.  Since there is no relationship between the true offset and $\sigma_{R_{\rm phys}}$, the range of offsets found with large uncertainty should not be significantly different from the range found with small uncertainty.  The fact that we only measure large offsets when we have large uncertainty is indicative of the above bias.  We note that even at relatively small uncertainty we may still be missing bursts at very small offsets because any amount of uncertainty prevents the measurement of arbitrarily small offsets.      

To reduce this bias, we make a quality cut on our physical offset sample by also plotting in Figure \ref{fig:phys} the distribution for bursts with $\sigma_{R_{\rm phys}} \lesssim 1.0$ kpc, resulting in a sample of 58 LGRBs.  The resulting distribution has a median at 1.27 kpc.  The mean of the distribution of medians for each iteration of the Monte Carlo simulation is 1.34 kpc with a 90\% confidence interval of 1.17 $-$ 1.51 kpc.  

We also compare the distribution of projected physical offsets to the distributions for Type II and Ib/c SNe.  KS tests with the supernovae samples of \citet{Prieto2008} yield p-values of 1.33 $\times 10^{-6}$ and 1.53 $\times 10^{-7}$ for the Type Ib/c and II SNe, respectively.  LGRBs occur significantly closer to the centers of their host galaxies in a physical sense than SNe.  To disentangle the effects of galaxy size differences between LGRB hosts and SNe hosts, we revisit supernova comparisons in the next section where we analyze the LGRB host-normalized offsets.  

\subsubsection{Host-Normalized Offset Distribution}
\begin{figure*}[ht!]
\begin{center}
\includegraphics[scale=0.45]{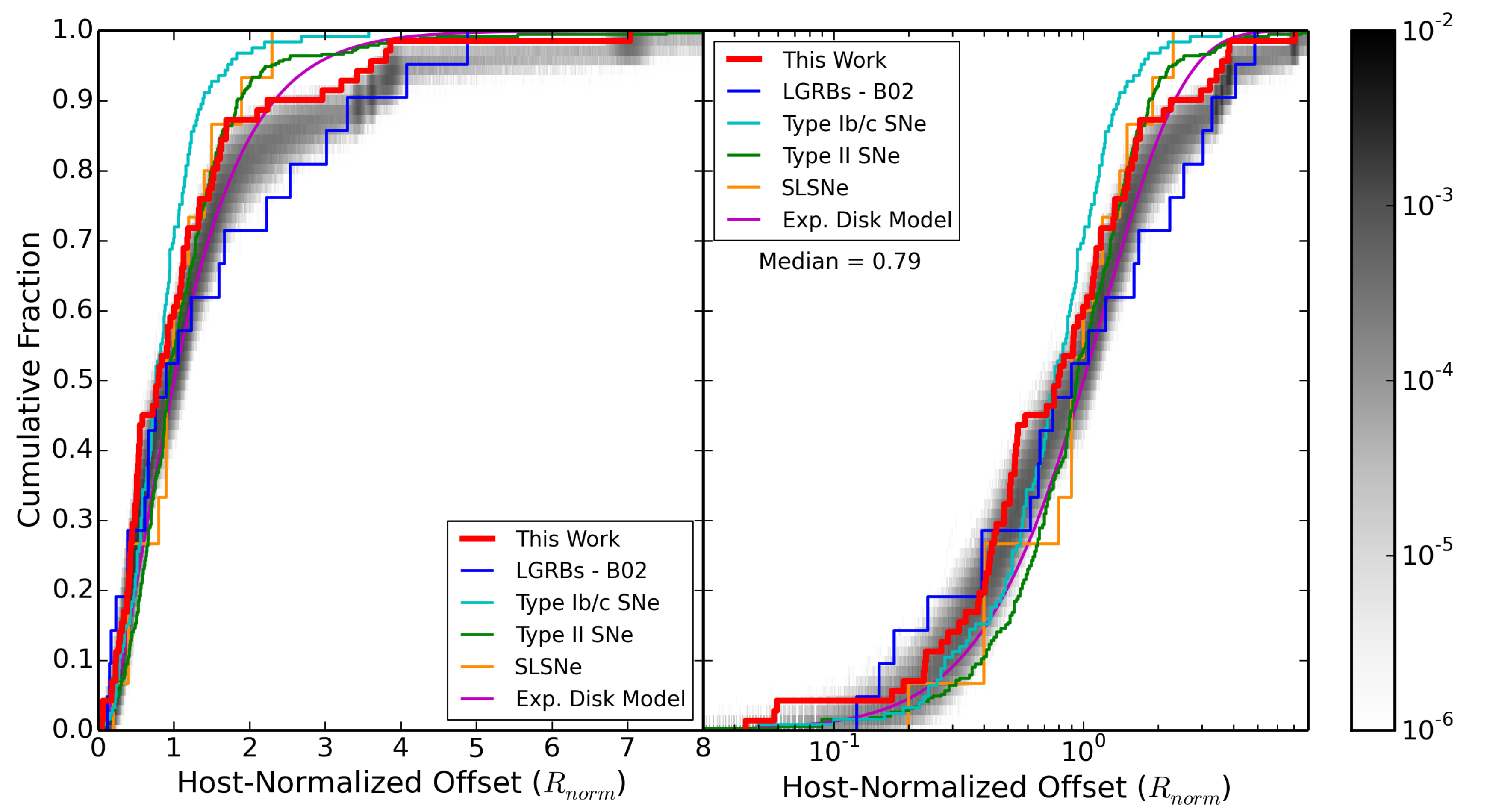}
\end{center}
\caption{Cumulative distributions of LGRB host-normalized offsets for our sample (red) and the sample of \citet{Bloom2002} (B02,  blue) shown on linear (left) and log (right) $x$-axis scales.  Also shown are the distributions for Type Ib/c (cyan) and Type II (green) SNe from \citet{KellyKirshner2012} and SLSNe (orange) from \citet{Lunnan2015}.  We also plot the distribution expected for an exponential disk profile (magenta).  The shaded regions depict the uncertainties on our offset measurements determined using a Monte Carlo simulation.  We find an offset distribution qualitatively similar to an exponential disk profile but statistically inconsistent.}
\label{fig:norm}
\end{figure*}

Normalizing the offsets by the host galaxy sizes enables a fair comparison across the LGRB sample and with the SNe samples.  In Figure \ref{fig:norm} we plot the cumulative distribution of host-normalized offsets ($R/R_{h}$).  The distribution ranges from about 0.04 to 7.0.  The median of the distribution is 0.79 and about 90\% of the bursts occur within a host-normalized offset of $\approx 2.2$.  The distribution is overall reminiscent of an exponential disk profile, the expected surface brightness profile of star forming disk galaxies.  For comparison we also plot the predicted distribution of host-normalized offsets if LGRB locations exactly trace an exponential disk profile.  Although the shapes are overall consistent between our measured distribution and that of an exponential disk, there is a notable shift to lower offsets seen in our distribution.  Again, to compare with the previous offset sample of LGRBs we show the host-normalized offsets from \citet{Bloom2002}.  A KS test between this sample and our sample gives a p-value of 0.50.  As a result of the small sample size, \citet{Bloom2002} concluded that the LGRB host-normalized offsets were consistent with being drawn from the exponential disk profile.  With our larger sample we can rule this out at a significance level of 0.01.  Given the irregular morphologies of LGRB host galaxies this result is not surprising even if LGRBs trace star formation.

\begin{figure*}[!ht]
\begin{center}
\includegraphics[scale=0.45]{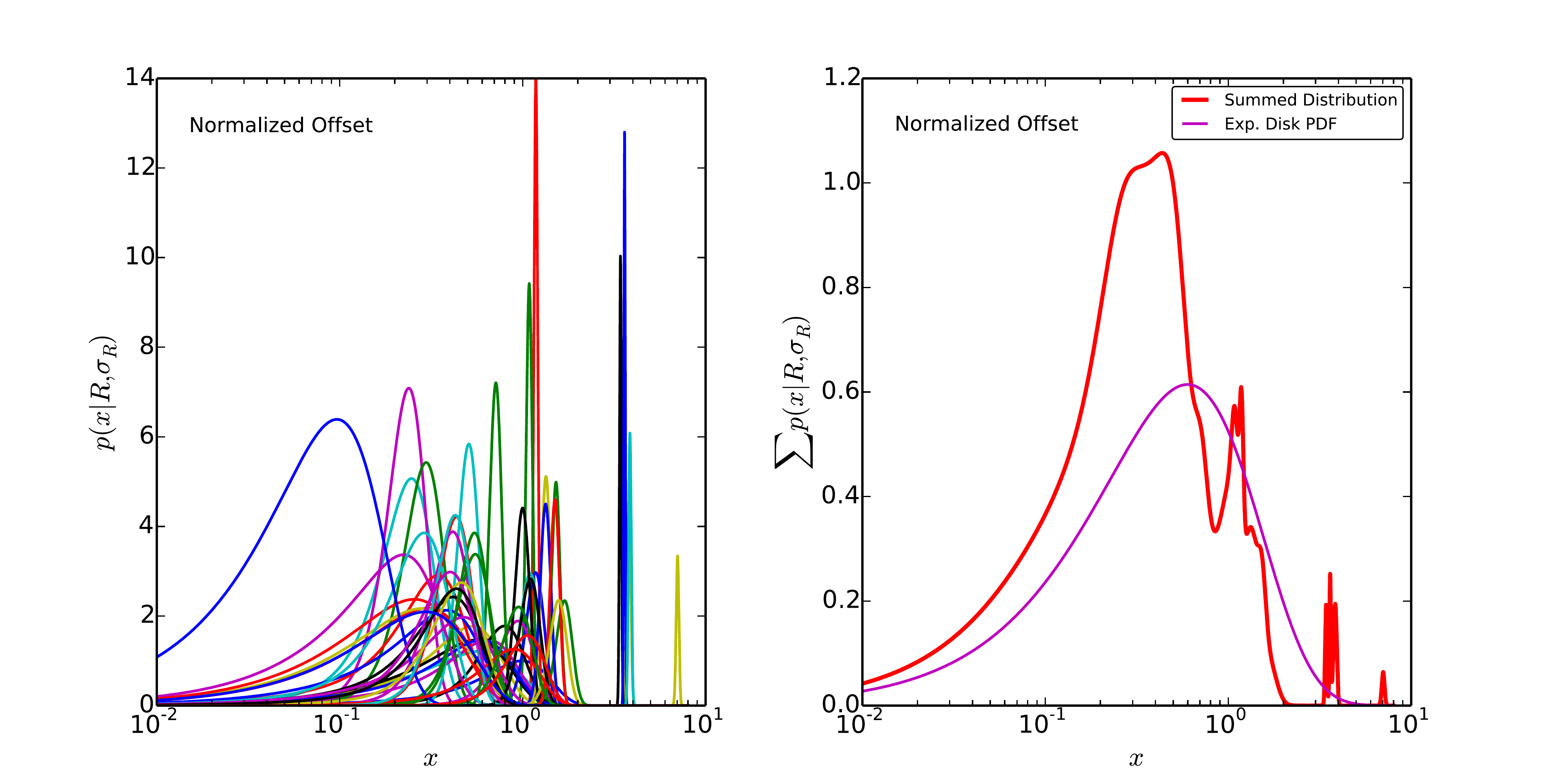}
\end{center}
\caption{\textit{Left:} Probability distributions for each LGRB with $\sigma_{R_{\rm norm}} \lesssim 0.5$ given its measured offset and offset uncertainty calculated using Equation \ref{eqn:rice}.  Narrow peaks indicate well-localized bursts while broader distributions indicate poorer localizations.  \textit{Right:} Sum (red) of the individual probability distributions (left) effectively producing a smoothed histogram of the offsets.  In magenta we show the probability distribution for an exponential disk.  The summed distribution is clearly offset to lower offsets than what would be expected for a population tracing an exponential disk.}
\label{fig:sumdist}
\end{figure*}

\begin{figure*}[ht!]
\begin{center}
\includegraphics[scale=0.45]{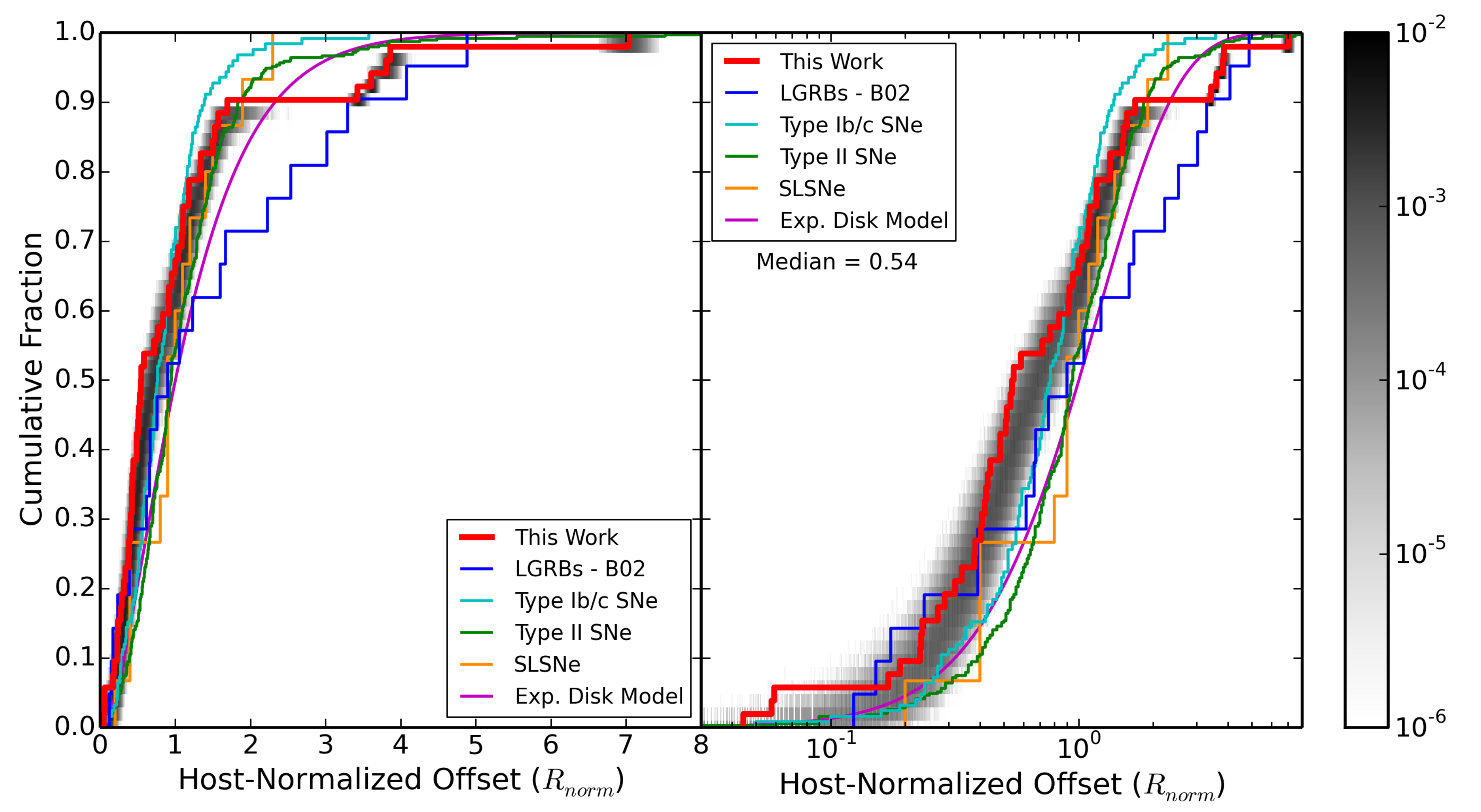}
\end{center}
\caption{Similar to Figure \ref{fig:norm} except here we show the distribution of host-normalized offsets only for bursts with $\sigma_{R_{\rm norm}} \lesssim 0.5$.  The distribution is notably shifted to lower offset with a median of 0.54.  This is because we have removed bursts with large offset uncertainties which bias the distribution to larger offsets (See Figure \ref{fig:offvserr}).}
\label{fig:norm_res}
\end{figure*}

As with the physical offsets, we also show in Figure \ref{fig:norm} the uncertainty region determined from our Monte Carlo procedure. 
Again, we carefully consider the effects of the uncertainties on the distribution of host-normalized offsets to avoid the aforementioned bias associated with the offset measurement of bursts with large offset uncertainties.  In Figure \ref{fig:offvserr} we also plot $\sigma_{R_{\rm norm}}$ versus $R_{\rm norm}$.  Of the 20 bursts in the sample with $\sigma_{R_{\rm norm}} \gtrsim 0.5$, all are at measured offsets $\gtrsim 0.4$, whereas the remaining bursts with $\sigma_{R_{\rm norm}} \lesssim 0.5$ span the full range of offsets from 0.04 to 7.0.  As with the physical offsets, our inability to measure small offsets when we have large uncertainty is indicative of a bias.  Again, it is likely the case that even at relatively small uncertainty we may still be missing bursts at very small offsets.      

As before, we make a quality cut on our host-normalized offset sample by restricting the sample to bursts with $\sigma_{R_{norm}} \lesssim 0.5$, resulting in a sample of 52 LGRBs.  The resulting distribution has a median at 0.54.  In Figure \ref{fig:sumdist} we show the individual probability distributions of each LGRB used in the Monte Carlo simulation.  The mean of the distribution of medians for each iteration of the Monte Carlo simulation is 0.67 with a 90\% confidence interval of 0.58 $-$ 0.76.  We plot the distribution and Monte Carlo results in Figure \ref{fig:norm_res}.  A KS test with the exponential disk model yields a p-value of $2.6 \times 10^{-4}$.  Furthermore, only $\sim$ 4\% of the Monte Carlo synthetic distributions have p-values $> 0.05$ when compared to an exponential disk profile, meaning that we can rule out the hypothesis that LGRB host-normalized offsets are drawn from an exponential disk.  As another way to view these results we also show in Figure \ref{fig:sumdist}  the sum of the individual probability distributions of each LGRB.  The summed distribution effectively produces a smoothed histogram of the offsets and shows a clear shift to smaller offsets from the probability distribution for an exponential disk profile.  LGRBs are apparently highly concentrated in the inner parts of their hosts.  In particular, they are more concentrated than the radial light distributions of their hosts; namely, 50\% of LGRBs occur within 67\% of a half-light radius.  Stated differently, about half of all LGRBs occur within a region that contains only 33\% of the underlying distribution of light in their host galaxies.  The possible bias caused by the inability to find bursts at very small offset mentioned above would act in the opposite direction of the striking result found here, indicating that the distribution may be even more concentrated than we find here.    

A striking feature of the distribution in Figure \ref{fig:norm_res} is the apparent tail to large offset extending from a gap at $R_{\rm norm} \sim 1.6 - 3.3$.  While it is not unexpected, given the rough similarity of our distribution to an exponential disk, to find bursts at large offset, it is somewhat surprising that $\sim$10\% of LGRBs appear to be located at $R_{\rm norm} \gtrsim 3$.  These bursts include GRBs 050820, 081008, 080928, 060418, and 060505.  We revisit the properties of these bursts and their assigned hosts.  The host of GRB 050820 has a morphology consisting of a bright core with a diffuse tail with the GRB located on this diffuse emission, leading to a host-normalized offset of 3.8 from the bright core.  The burst's location on the underlying galaxy emission and the $P_{\rm cc}$ of 0.01 do not give reason to doubt the host association, the large offset being due to the unusual morphology.  As discussed in Section \ref{sec:lum} we assigned as the host of GRB 080928 a non-coincident galaxy with a luminosity consistent with the observed distribution of LGRB host luminosities.  However, we cannot rule out the possibility that the host was not detected and that the assigned candidate is instead the source of an intervening absorption system observed in the afterglow \citep{Vrees2008}, casting doubt on the large inferred offset.  GRB 081008, another burst with a host assigned based on our luminosity analysis, has a large normalized offset of $\approx 7$ and $P_{\rm cc} \approx 0.07$.  While the host assignment matches our criteria, we acknowledge the possibility that the assignment is not correct.  For GRB 060418 we assign as the host the brighter of the two nearby host components identified by \citet{Pollack2009}.  Although these components are distinct from each other (see Figure \ref{fig:mosaic}), the assignment of one or the other has little effect on the offset and no effect on the fractional flux.  GRB 060505 is a controversial $z = 0.089$ burst that did not have an associated supernova as expected for a low-$z$ LGRB leading some to argue in favor of a short GRB scenario \citep{Ofek2007}.  However, given its location in a bright, low-metallicity HII region of its host, others argue in favor of 060505 as a LGRB \citep{Thone2014}.  In this case, 060505 is an example of a LGRB occurring at large normalized offset (3.6) and at high fractional flux (0.99) in the outskirts of its host.  It is interesting to speculate that the other bursts at large offset may be higher redshift versions of 060505 where underlying HII regions are too faint to be detected.  For the reasons discussed here, the large offset tail in the distribution of host-normalized offsets may not be real.  However, it does not affect the overall conclusion that LGRBs as a population are more centrally concentrated than the underlying light distributions of their hosts.     

We compare our distribution of LGRB host-normalized offsets with those measured for supernovae by \citet{KellyKirshner2012}, also shown in Figure \ref{fig:norm_res}.  KS tests with the Type Ib/c and Type II SNe yield p-values of $1.3 \times 10^{-2}$ and $5.2 \times 10^{-5}$, respectively.  While the measured distribution for Type Ib/c and our sample of LGRBs are statistically inconsistent, about 80\% of Monte Carlo synthetic distributions have p-values $> 0.05$ when compared to Type Ib/c SNe.  This percentage is $\sim$1\% when compared to Type II SNe.  When taking into account the uncertainties, the distributions for LGRBs and Type Ib/c SNe are therefore consistent.  They both are much more centrally concentrated than Type II SNe and the exponential disk profile.  Comparing our LGRBs to the distribution measured for super-luminous supernovae \citep[SLSNe;][]{Lunnan2015} we find a KS test p-value of 0.17.  This comparison, though, is limited by the small SLSNe sample size.

\subsubsection{Offset Distribution Summary}
We find that our sample of LGRBs spans the range 0.075 to 14.4 kpc (mean of Monte Carlo distribution of medians = 1.34 kpc) in physical offset from the centers of their host galaxies.  When appropriately accounting for uncertainties and removing bursts with $\sigma_{R_{\rm norm}} \gtrsim 0.5$, we find that LGRB host-normalized offsets are considerably smaller and more centrally concentrated (mean of Monte Carlo distribution of medians = 0.67, with 90\% confidence interval of 0.58 - 0.76) than what would be expected if LGRBs traced an exponential disk profile.  We can rule out at high significance that LGRB offsets are drawn from an exponential disk distribution and their distribution is inconsistent with Type II SNe.  Accounting for uncertainties, the LGRB offset distribution we find is consistent with that for Type Ib/c SNe.  LGRBs prefer the central locations in their hosts, being more concentrated than the underlying host light distributions.  Table \ref{tab:sum} provides a summary of the KS p-values we find for comparisons between our sample of LGRB host-normalized offsets and other types of stellar explosions and in Table \ref{tab:medians} we show the medians.

\subsection{Host Light Distribution} 
\label{sec:FFresults}

\begin{figure*}[ht!]
\begin{center}
\includegraphics[scale=0.45]{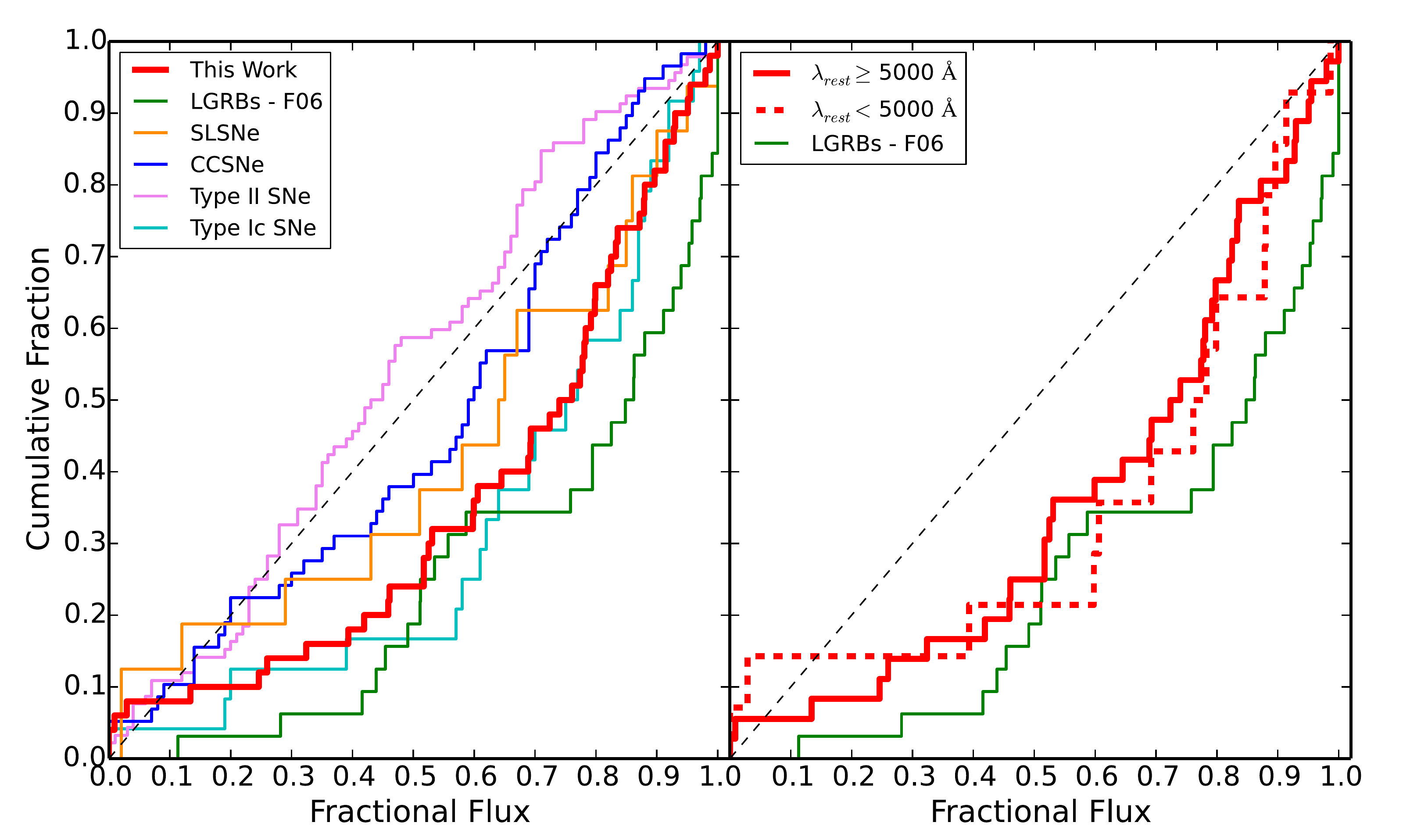}
\end{center}
\caption{\textit{Left:} Cumulative distributions of fractional flux for our sample (red), the LGRB sample of \citet{Fruchter2006} (F06, green), CCSNe from \citet{Svensson2010} (blue), SLSNe from \citet{Lunnan2015} (orange), and Type Ic (cyan) and II (violet) SNe from \citet{Kelly2008}.  We only include bursts in our sample with a ratio of error circle to galaxy area of $\lesssim 0.1$ (Figure \ref{fig:FFdiag}).  The diagonal dashed line shows the expected fractional flux distribution for a population uniformly tracing the underlying light of their host galaxies.  \textit{Right:} Cumulative distributions of fractional flux for two subsets of the sample divided at a rest-frame wavelength of 5000 \AA.  The agreement between the two means that we can use the rest-frame optical observations as a proxy for rest-frame UV.}
\label{fig:FF}
\end{figure*}

In Figure \ref{fig:FF} we show the cumulative distribution of fractional flux for the 49 bursts in our sample with error circle to galaxy area ratios of $\lesssim 0.1$ (left panel).  Ideally, we would like to assess whether or not LGRBs are spatially coincident with bright rest-frame UV regions of their hosts, which trace young, massive stars.  As shown in Figure \ref{fig:red} our observations primarily probe the rest-frame optical emission, which may probe older stellar populations.  Here we compare the fractional flux distribution for the $\sim 20\%$ of the sample that probes rest-frame UV to the distribution that probes rest-frame optical.  The right panel of Figure \ref{fig:FF} shows the resulting distributions when dividing the sample by wavelength into two bins separated at $\lambda_{\rm rest} = 5000$ \AA.  The rest-frame optical and blue/UV distributions agree well, which we interpret as indicating that there is no bias in using rest-frame optical observations as a proxy for rest-frame UV.  If there was a separate older stellar population, we would not expect to find LGRB positions correlated so strongly with both the optical and UV light.  In other words, the optical and UV light are probing the same population of stars.  We can then interpret our fractional flux measurements as reflecting the relationship between location and the host UV emission.  

As shown in Figure \ref{fig:FF} the fractional flux distribution for our sample spans the full range from 0 to 1 with 90\% of values falling in the range 0.03 to 0.98.  The median is 0.75 and nearly 80\% of bursts have fractional flux values greater than 0.5.  This strong preference towards higher fractional flux values indicates that LGRBs are typically located on some of the brightest regions of their hosts, and lends further support to the idea that LGRBs originate from very massive stars.  This is in broad agreement with previous claims \citep{Fruchter2006, Svensson2010}.  We also show in Figure \ref{fig:FF} the distribution of fractional flux for the sample of 30 LGRBs from \citet{Fruchter2006}.  A KS test between the two samples yields a p-value of 0.09.  This indicates that while our sample and the \citet{Fruchter2006} sample are marginally inconsistent, we cannot rule out that they are drawn from the same distribution.  Nevertheless, the sample of \citet{Fruchter2006} has a higher median and general inclination towards higher fractional flux values.  This slight discrepancy is probably not caused by effects intrinsic to the LGRB population but rather by systematic differences between the two samples.  We also show the \citet{Fruchter2006} sample with the wavelength binned distributions in Figure \ref{fig:FF}.  The slight shift of the rest-frame blue/UV distribution towards the distribution of \citet{Fruchter2006}, mostly comprised of rest-frame UV observations, may indicate the discrepancy is a bandpass effect.  We address another potential effect in Section \ref{sec:disc}.       

We also compare our measured distribution to the corresponding distributions measured for CCSNe \citep{Svensson2010}, SLSNe \citep{Lunnan2015}, and Type Ic and II SNe \citep{Kelly2008}.  KS tests between our sample, the SLSNe sample, the Type Ic SNe sample, the Type II SNe sample, and the CCSNe sample yield p-values of 0.52, 0.79, 5.4$\times10^{-5}$, and 0.03, respectively.  We cannot rule out that our sample of LGRBs and the SLSNe sample are drawn from the same distribution.  Interestingly, our sample overlaps the SLSNe distribution at fractional flux values of $\gtrsim 0.8$.  The very good agreement between the distributions for our sample and the Ic SNe sample is consistent with the fact that Ic SNe, particularly Ic-BL, are the only types of SNe observed to be associated with LGRBs.  The distributions of fractional flux for the CCSNe, mostly consisting of Type II SNe, and the Type II SNe sample are clearly distinct from our LGRB distribution, strong evidence that Type II SNe and LGRBs have different progenitors.  We summarize the KS p-values for our comparisons in Table \ref{tab:sum}.

\subsection{Galaxy Sizes}
\begin{figure}[ht!]
\begin{center}
\includegraphics[scale=0.45]{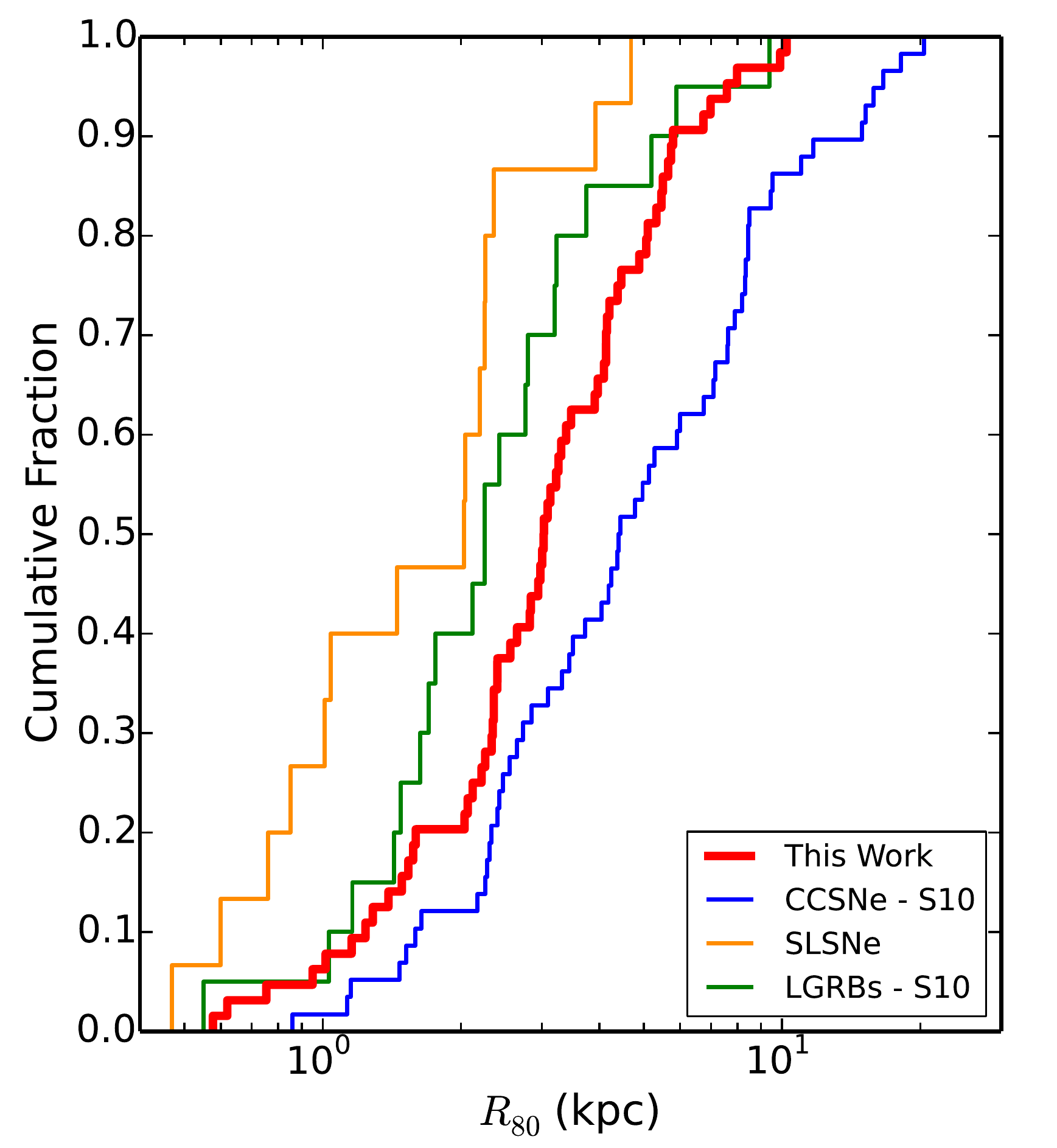}
\end{center}
\caption{Cumulative distributions of $R_{80}$ for our sample of LGRBs (red), the LGRB (green) and CCSNe (blue) samples from \citet{Svensson2010} (S10), and the SLSNe sample from \citet{Lunnan2015} (orange).  At high significance we find that LGRB host galaxy sizes are larger than SLSNe host sizes but smaller than CCSNe host sizes.}
\label{fig:sizes80}
\end{figure}

As it pertains to our knowledge of the host galaxies and therefore implications for the progenitors, we plot in Figure \ref{fig:sizes80} the cumulative distribution of 80\% light-radii ($R_{80}$) of our host galaxies.  The median size for the LGRB sample is 3.0 kpc and the distribution spans from $R_{80} \sim 0.6$ kpc to $R_{80} \sim 10$ kpc.  For comparison, we show the distributions for the samples of LGRBs and CCSNe studied by \citet{Svensson2010}.  A KS test reveals that our distribution, though shifted to slightly higher values of $R_{80}$, is consistent with their distribution of LGRB sizes.  A KS test between our distribution of LGRB sizes and the CCSNe distribution yields a p-value of $3 \times 10^{-3}$ indicating our larger sample of LGRBs is in full agreement with the conclusion of \citet{Svensson2010} that LGRB host galaxies are on average smaller than CCSNe host galaxies.  We also show in Figure \ref{fig:sizes80} the sample of SLSNe host galaxies studied by \citet{Lunnan2015}.  With the previous small sample of LGRBs, the SLSNe and LGRB host size distributions were statistically consistent.  Now with our larger sample we find that the apparent larger sizes of LGRB hosts is statistically robust (KS p-value = $1.4 \times 10^{-3}$).            

\subsection{Durations and Energies}

\begin{figure*}[!h]
\begin{center}
\includegraphics[scale=0.5]{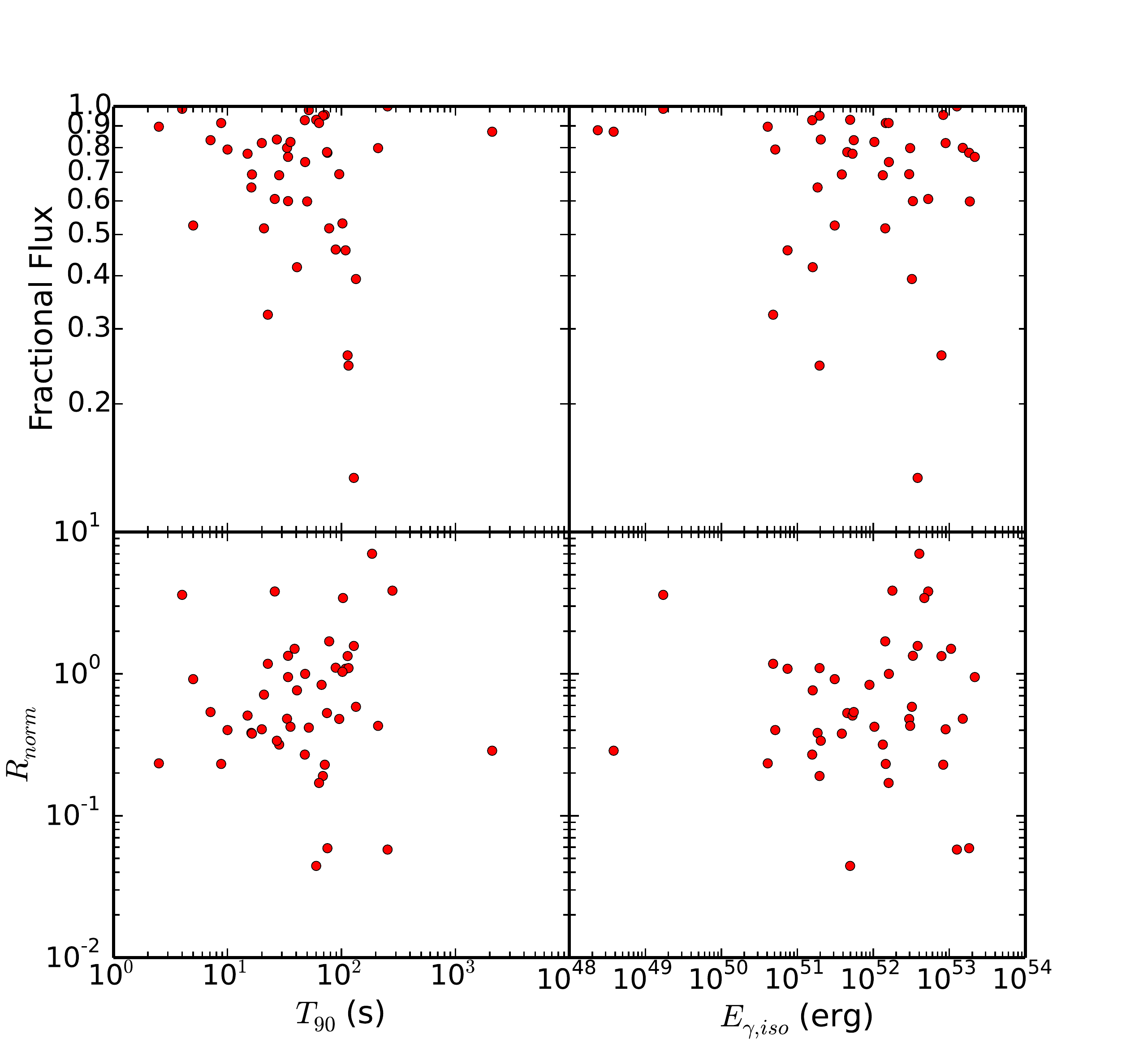}
\end{center}
\caption{Scatter plots of fractional flux and host-normalized offset ($R_{\rm norm}$) versus duration ($T_{90}$) and equivalent isotropic energy ($E_{\gamma,iso}$).  We find no significant correlations between these quantities.}
\label{fig:T90}
\end{figure*}

The durations (as measured by $T_{90}$) and isotropic equivalent energies ($E_{\gamma,iso}$) of LGRBs each vary by over several orders of magnitude.  Whether or not environmental factors influence these variations in burst properties is an open question that we can address with our sample of offsets and fractional fluxes.  In Figure \ref{fig:T90} we plot fractional flux and host-normalized offset versus $T_{90}$ and $E_{\gamma,iso}$.  We find no significant correlations between fractional flux or offset with $T_{90}$ or $E_{\gamma,iso}$, indicating that the locations of LGRBs within their host galaxies do not affect the properties of the bursts themselves.  Our findings do not support the claim of a possible trend between host-normalized offset and isotropic equivalent energy reported by \citet{RamirezRuiz2002} based on a much smaller sample of 16 pre-\textit{Swift} events.       

\section{Discussion}
\label{sec:disc}
The distributions of offsets and fractional flux presented in Section \ref{sec:results} provide the most in-depth look at the locations and environments of LGRBs presented to date.  Furthermore, our sample is sufficiently large and wide in redshift to allow investigation of trends with cosmic time.  Here we discuss possible redshift evolution, the relationship between fractional flux and host-normalized offset, and the implications for the progenitors of LGRBs provided by this new view of their locations.

\subsection{Redshift Trends}
\begin{figure*}[ht!]
\begin{center}
\includegraphics[scale=0.45]{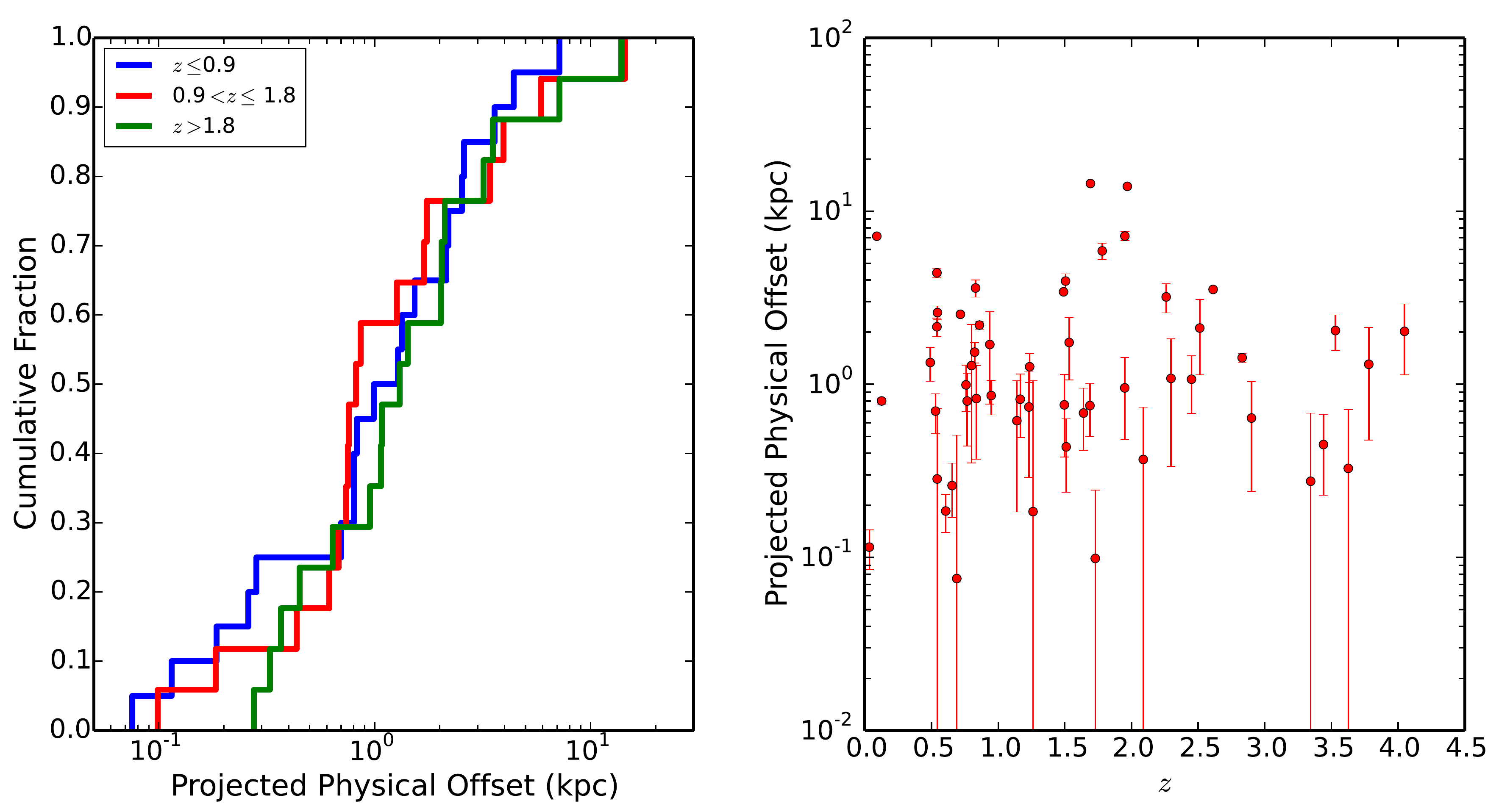}
\end{center}
\caption{\textit{Left:} Cumulative distribution of physical offsets for bursts with $\sigma_{R_{\rm phys}} \lesssim 1.0$ kpc binned into three equal-numbered bins of redshift: $z \leq 0.9$ (blue), $0.9 < z \leq 1.8$ (red), and $z > 1.8$ (green).  \textit{Right:}  Physical offsets with their associated uncertainties as a function of redshift.  In both plots we do not include bursts with unknown redshift.  We do not find a statistically significant trend in physical offset as a function of redshift.}
\label{fig:phys_vs_z}
\end{figure*}

In Figure \ref{fig:phys_vs_z} we show the physical offsets for bursts with $\sigma_{R_{\rm phys}} \lesssim 1.0$ kpc as a function of redshift (right panel) and the cumulative distribution of physical offsets binned into three equally populated bins of redshift (left panel): $z \leq 0.9$, $0.9 < z \leq 1.8$, and $z > 1.8$.  We find that there is no statistically significant trend of the physical offsets as a function of redshift.  KS tests between the three redshift bins yield p-values of 0.97, 0.77, and 0.45 for the low-$z$ and mid-$z$ bins, low-$z$ and high-$z$ bins, and mid-$z$ and high-$z$ bins, respectively.

\begin{figure*}[ht!]
\begin{center}
\includegraphics[scale=0.45]{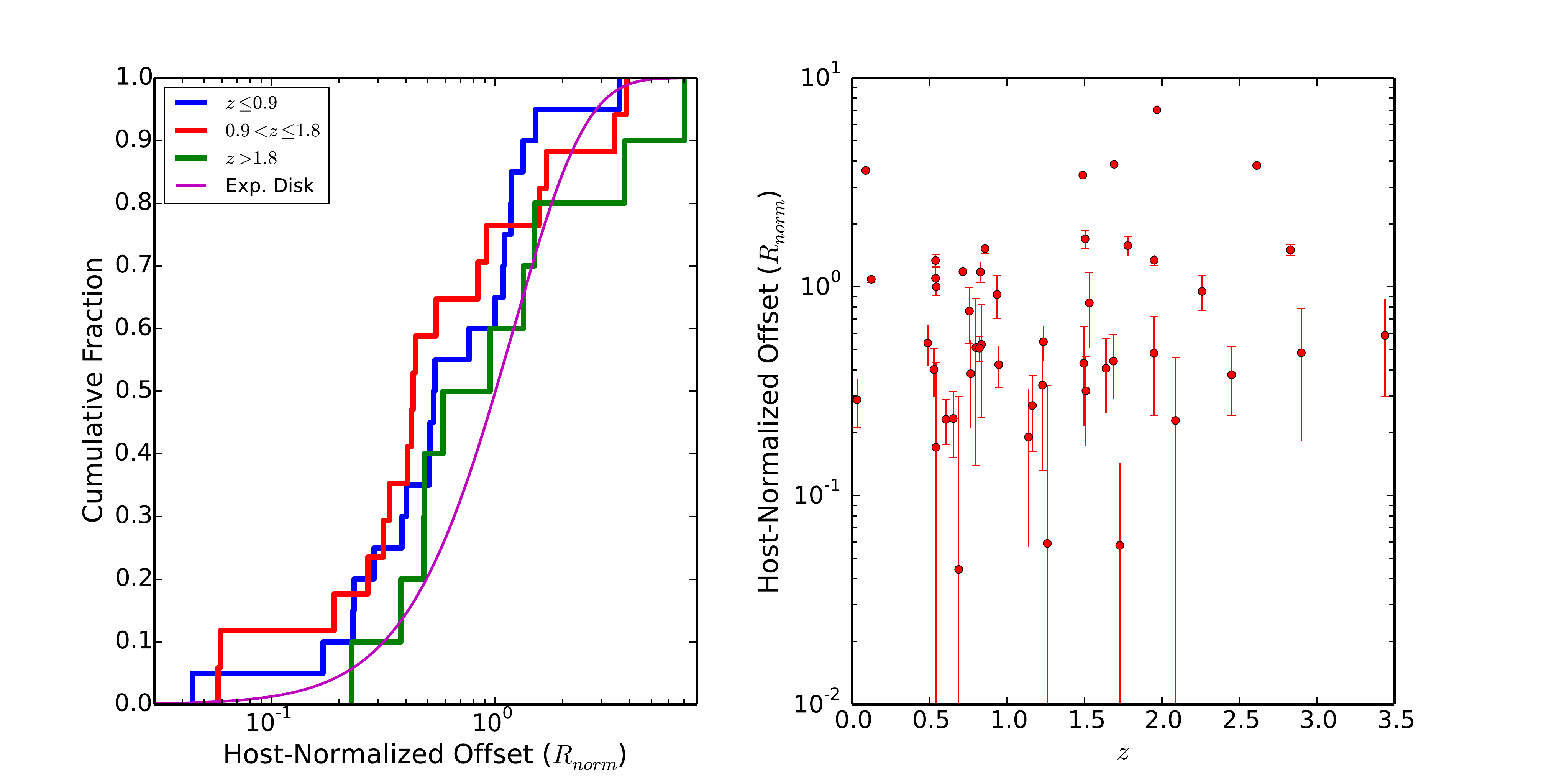}
\end{center}
\caption{Same as Figure \ref{fig:phys_vs_z} but for the host-normalized offsets of bursts with $\sigma_{R_{\rm norm}} \lesssim 0.5$.  We do not find a statistically significant trend in host-normalized offset as a function of redshift.  The slight inclination towards larger offsets at higher redshift may be the result of surface brightness dimming.}
\label{fig:norm_vs_z}
\end{figure*}

\begin{figure}[ht]
\begin{center}
\includegraphics[scale=0.45]{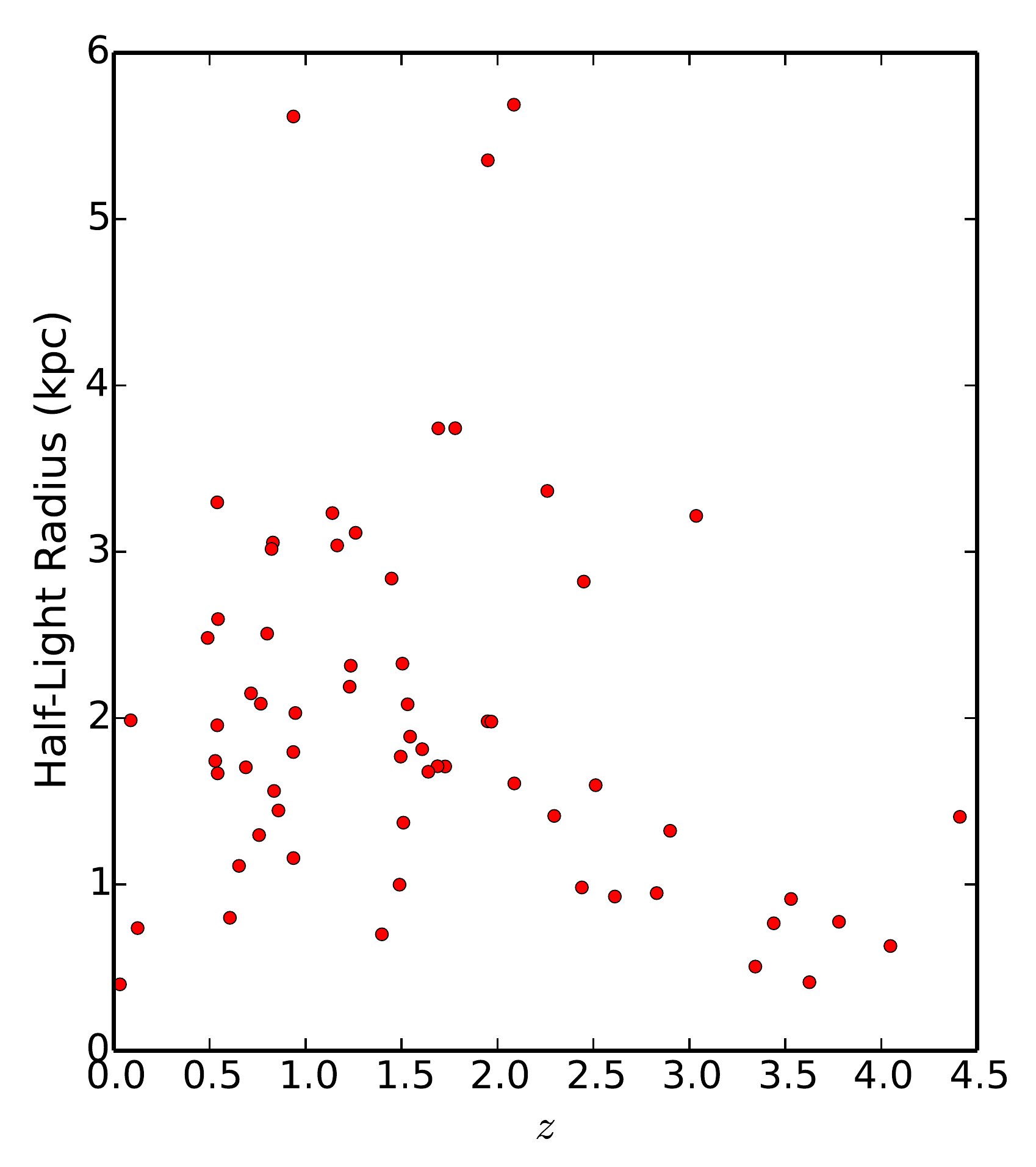}
\end{center}
\caption{Half-light radii as a function of redshift showing a clear trend that may explain the slight shift to larger host-normalized offsets at $z \gtrsim 2$.  Whether this trend reflects real galaxy growth or the effects of surface brightness dimming is less clear.}
\label{fig:sizes}
\end{figure}

In Figure \ref{fig:norm_vs_z} we show the host-normalized offsets of bursts with $\sigma_{R_{\rm norm}} \lesssim 0.5$ as a function of redshift (right panel) and the cumulative distribution of host-normalized offsets binned by redshift (left panel).  Overall we find no significant trend of the host-normalized offsets with redshift.  KS tests between the three redshift bins yield p-values of 0.61, 0.51, and 0.23 for the low-$z$ and mid-$z$ bins, low-$z$ and high-$z$ bins, and mid-$z$ and high-$z$ bins, respectively.  Despite the statistical insignificance, there appears to be a slight shift of the $z > 1.8$ bin toward larger offsets.  If this reflects an actual trend, coupled with the lack of a trend in the distribution of physical offsets, it would indicate an effect related to the sizes of the galaxies.  In Figure \ref{fig:sizes} we show the measured half-light radii of the host galaxies as a function of redshift.  There is clearly a trend of smaller galaxy sizes with increasing redshift.  At $z < 2$ the median galaxy half-light radius is 1.98 kpc and at $z > 2$ the median radius is 1.15 kpc.  It is possible that surface brightness dimming, a strongly redshift dependent effect ($\propto (1 + z)^{-4}$), may be limiting our ability to measure the sizes of the host galaxies at higher redshift.  At higher redshifts the faint outer regions of galaxies become much harder to detect resulting in an underestimate of galaxy size and corresponding overestimate of the host normalized offset.  However, \textit{HST} observations of Lyman-break galaxies indicate that high redshift galaxies are intrinsically compact and show a size evolution of approximately $(1 + z)^{-1}$ \citep{Bouwens2004,Calvi2014}.  If low-mass galaxies such as LGRB hosts also lack diffuse outer components at high redshift, then it is possible the observed size trend in Figure \ref{fig:sizes} reflects actual galaxy growth.       
      
\begin{figure*}[ht!]
\begin{center}
\includegraphics[scale=0.45]{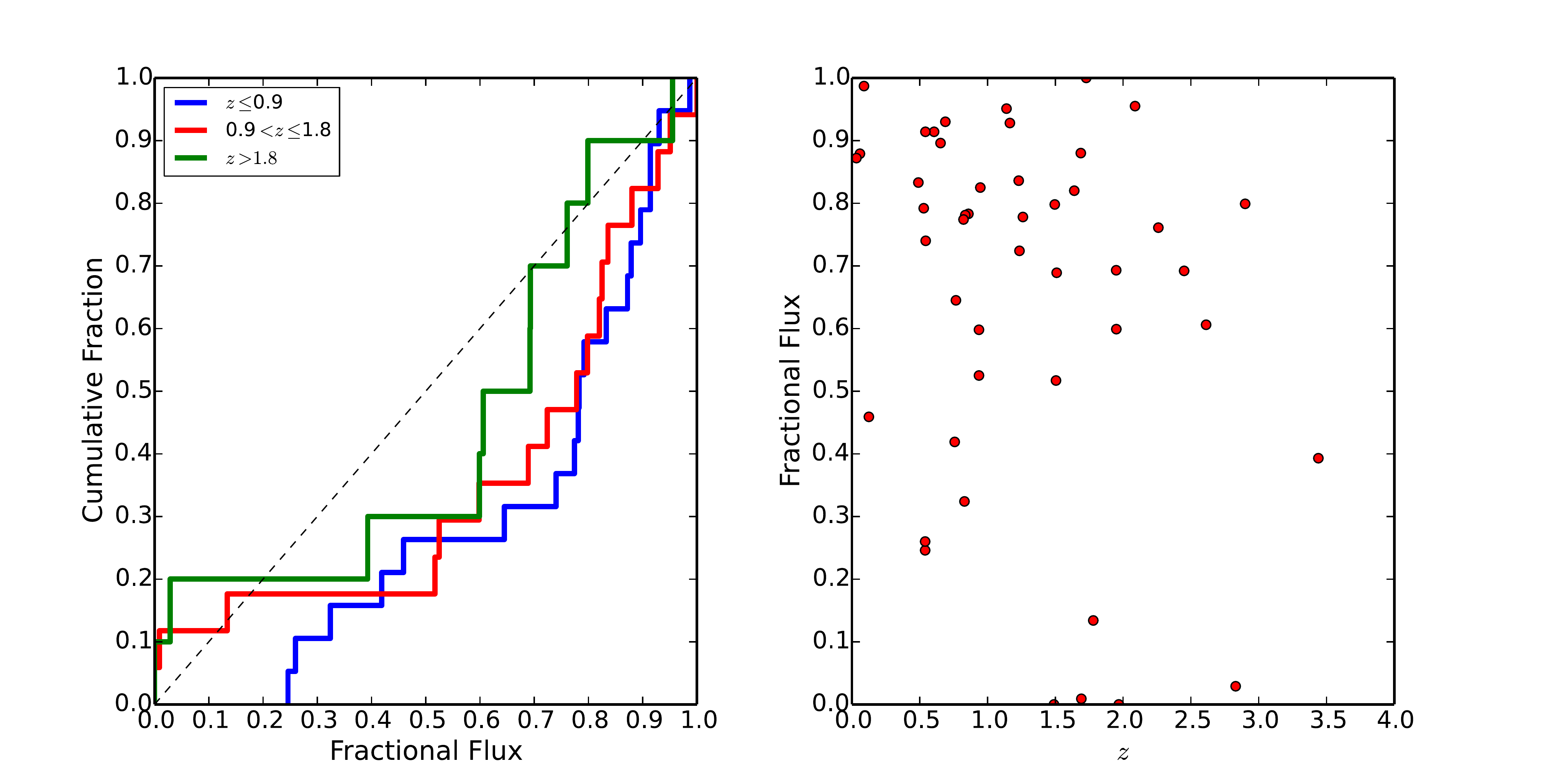}
\end{center}
\caption{Same as Figure \ref{fig:phys_vs_z} but for fractional flux.  We do not find a statistically significant trend in fractional flux as a function of redshift.}
\label{fig:FFzbin}
\end{figure*}

In Figure \ref{fig:FFzbin} we show the fractional flux distribution binned into the same three redshift bins.  KS tests between the three redshift bins yield p-values of 0.92, 0.13, and 0.42 for the low-$z$ and mid-$z$ bins, low-$z$ and high-$z$ bins, and mid-$z$ and high-$z$ bins, respectively.  While the three binned distributions are statistically consistent, the lowest fractional flux values ($\lesssim 0.2$) only occur at $z > 1.5$.  It is possible that surface brightness dimming is also playing a role here. 

Overall we find no statistically significant trends with redshift in the offset or fractional flux distributions.  The slight shift towards larger offsets and lower fractional flux in the highest redshift bin of $z > 1.8$ may indicate we are missing some of the diffuse outer parts of the hosts at high redshift.  

\subsection{The Fractional Flux - Offset Relationship}
\begin{figure*}[ht!]
\begin{center}
\includegraphics[scale=0.45]{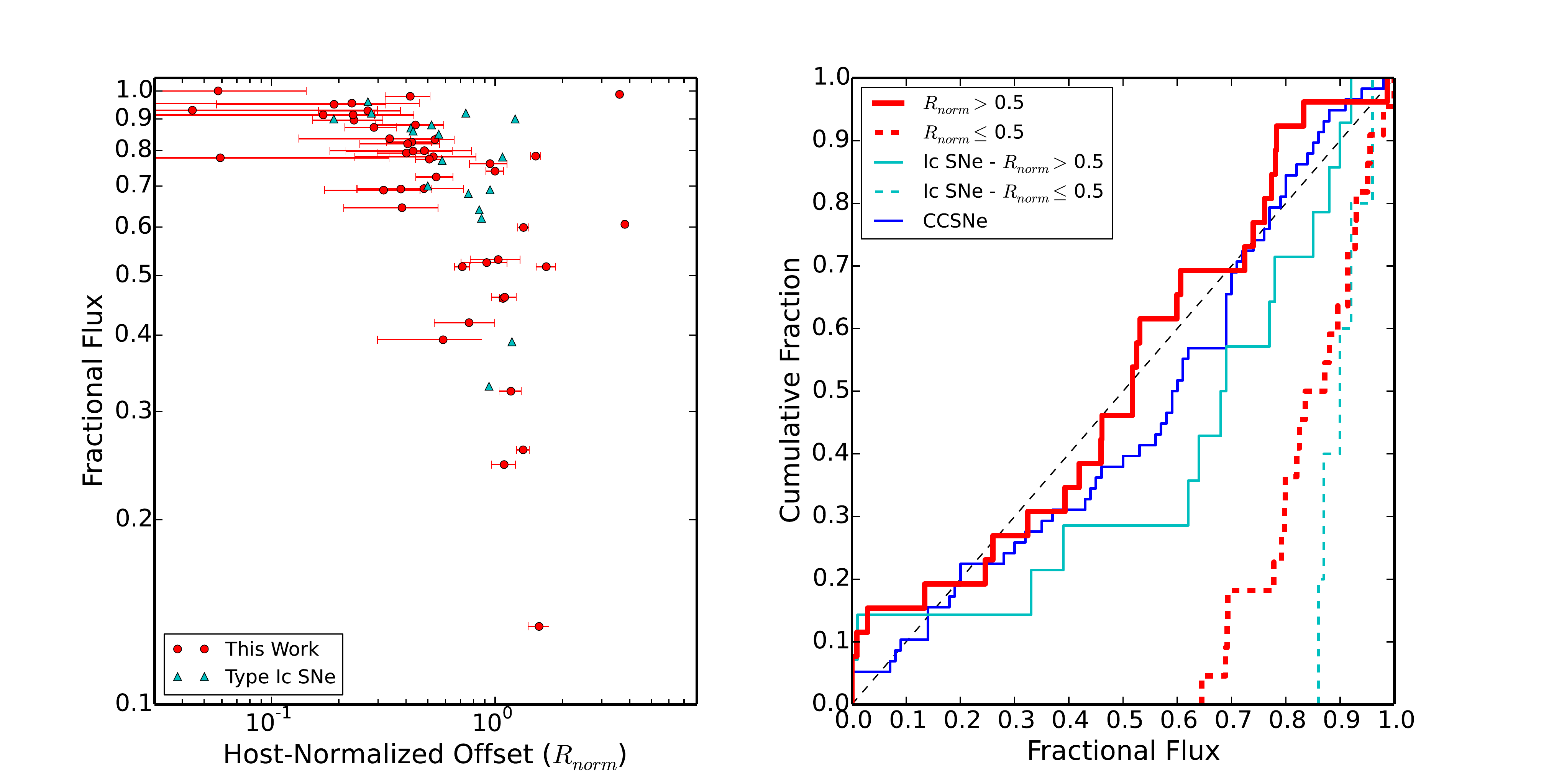}
\end{center}
\caption{\textit{Left:} Fractional flux versus host-normalized offset for LGRBs (red) with $\sigma_{R_{\rm norm}} \lesssim 0.5$ and error circle to galaxy area ratio $\lesssim 0.1$ and Type Ic/Ic-BL SNe (cyan) from \citet{Kelly2008} and \citet{KellyKirshner2012}.  \textit{Right:}  Cumulative distributions of fractional flux for two subsets of the LGRB sample divided at $R_{\rm norm} = 0.5$.  There is a clear trend relating these two quantities where LGRBs at small host-normalized offset, near the bright cores of their hosts, are exclusively at high fractional flux.  At $R_{\rm norm} \gtrsim 0.5$, the bursts uniformly trace the underlying light of their host galaxies similar to CCSNe as also shown in the right panel.  Therefore, the shift to high fractional flux seen in the full sample distribution (Figure \ref{fig:FF}) is due to bursts occurring in the inner parts of their host galaxies.  A similar trend is seen in the Ic SNe sample.}
\label{fig:FFvsnorm}
\end{figure*}

In Figure \ref{fig:FFvsnorm} we plot the fractional flux versus host-normalized offset for bursts satisfying error circle area to galaxy area ratio of $\lesssim 0.1$ and $\sigma_{R_{\rm norm}} \lesssim 0.5$.  We find a clear correlation between fractional flux and host-normalized offset, where bursts at smaller offsets have high fractional flux values and bursts at larger offsets have on average lower fractional flux values.  The scatter in the fractional flux values increases with increasing $R_{\rm norm}$.  All bursts in our sample with $R_{\rm norm} \lesssim 0.5$ have fractional flux values greater than 0.6, yielding a notable lack of bursts with small offsets and low fractional flux values.  Beyond $R_{\rm norm} \approx 0.5$ the fractional flux values are on average lower but also show considerable scatter.  Bursts with $R_{\rm norm} \gtrsim 1$ have fractional flux values spanning the full range from $0 - 1$.  

While it is not surprising that a trend should exist between fractional flux and offset given that regions close to the centers of galaxies are brighter, the coupling of fractional flux and offset information explains the results seen in the fractional flux distribution.  To illustrate this, we also show in Figure \ref{fig:FFvsnorm} the cumulative distributions of fractional flux for the subsamples with $R_{\rm norm} \leq 0.5$ and $R_{\rm norm} > 0.5$.  Bursts with $R_{\rm norm} \leq 0.5$ exclusively occur on regions of high fractional flux and bursts with $R_{\rm norm} > 0.5$ uniformly trace the light of their hosts, showing no preference for unusually bright regions.  This means that the shift to high fractional flux we find in the total distribution is entirely due to LGRBs at small offsets and not to bursts at large offset that happen to reside in bright star forming regions.  Therefore, it is not simply the presence of a bright star forming region that affects LGRB production, but rather another factor playing a role near the centers of galaxies must be at play.  

Previous studies indicated that the agreement between LGRB offsets and the exponential disk was inconsistent with the observation that LGRBs were highly correlated with the brightest regions of their hosts \citep{Bloom2002,Fruchter2006}.  The offset distribution we measure coupled with the cumulative distribution of fractional flux showing a preference for bright regions does not imply a discrepancy.  Indeed, the fact that we find LGRBs concentrated in the inner regions of their galaxies with corresponding high fractional fluxes can be interpreted as a relief of the tension between the conclusions of \citet{Bloom2002} and \citet{Fruchter2006}. 
Ultimately, the slight difference between our distribution of fractional flux and the distribution of \citet{Fruchter2006} may be explained by a difference in the offset distributions of the two samples, and may simply be the result of small number statistics where our sample more closely approximates the true underlying distribution.              

In Figure \ref{fig:FFvsnorm} we also plot the normalized offsets and fractional fluxes of Type Ic and Ic-BL SNe for objects common to both the sample with fractional fluxes from \citet{Kelly2008} and the sample with host-normalized offsets from \citet{KellyKirshner2012}.  We find a similar trend between the two quantities that indicates the shift to high fractional flux in the Ic SNe distribution can be attributed to SNe that occurred at small offset.

\subsection{Progenitor Implications}
We find that LGRBs are on average located at smaller host-normalized offset than expected for an exponential disk profile of star formation and are correspondingly more concentrated on brighter regions of their hosts than expected for a population simply tracing the underlying light distributions.  The evidence presented here suggests that the progenitors of LGRBs prefer the central regions of their hosts, indicating that the properties of star formation towards the inner regions of LGRB host galaxies are favorable for LGRB production.  As demonstrated in Figure \ref{fig:FFvsnorm}, the bursts responsible for the strong preference towards higher fractional flux values are dominated by bursts at small offsets.  The fractional flux distribution of bursts at offsets $R_{\rm norm} \gtrsim 0.5$ uniformly tracks the host light.  In effect the high fractional flux values are due to the fact that the bursts occur at small offsets where the host is brighter.

How the suspected metallicity bias of LGRB production fits into this new view of their preferred locations is an important question.  Star forming disk galaxies typically have metallicity gradients with higher metallicities towards the cores \citep{VilaCostas1992,Zaritsky1994}.  One might then expect, if metallicity gradients are also present in the hosts of LGRBs, to more frequently find LGRBs in the outskirts of their hosts contrary to what we find here.  Recent resolved host studies of low redshift hosts indicate weak metallicity gradients \citep{Levesque2011,Thone2014} and that the global host metallicity is often a sufficient proxy for the metallicity of the burst site.  Though the metallicity typically shows only slight variation across the host, the burst site often has the lowest metallicity \citep{Christensen2008, Levesque2011,Thone2014}.  In future work, we plan to test for the presence of metallicity gradients, in an average sense, in high redshift hosts by combing ISM metallicity measurements with our measured offsets.  While it is clear that metallicity is a factor affecting the production of LGRBs, our results indicate that other factors relating to the mode of star formation may also play a dominant role.  The conditions in the central regions of LGRB hosts, perhaps a consequence of their formation histories, may be such that the IMF is different, producing more massive stars and thus more potential LGRB progenitors.  It has been shown that massive stars are often found in binary systems \citep{Mason2009,Sana2012} and that binary interactions may be important for LGRB production \citep{Fryer1999,BrommLoeb2006}.  There may be increased numbers of massive star binaries in the central regions of LGRB hosts.          

In both the offset and fractional flux distributions we find that LGRBs and Type Ic SNe prefer similar locations, indicating a common link between their progenitors.  While LGRBs and Type Ic SNe broadly prefer similar locations, the exclusive association of LGRBs with Ic-BL SNe indicates that some factor must determine whether a star explodes as a normal Typc Ic, a Type Ic-BL without a LGRB, or a Type Ic-BL with a LGRB.    Previous observations suggest that the environments where Type Ic-BL SNe occur are more metal poor than those for normal Type Ic SNe, and that the environments of Type Ic-BL SNe associated with LGRBs are even more metal poor \citep{Modjaz2008,Modjaz2011,KellyKirshner2012}.  Recent work suggests that metallicity is likely not the only factor influencing the production of Ic-BL SNe and LGRBs.  \citet{Kelly2014} found that Ic-BL SNe and LGRBs occur in host galaxies that have higher stellar mass and star formation rate densities than a control sample of SDSS star forming galaxies.  They show that metallicity cannot explain this preference and instead suggest that another environmental factor such as increased numbers of young, bound clusters where tight binaries form or a top-heavy IMF may be at play in Ic-BL SNe and LGRB hosts.  The results we find, namely that LGRBs prefer the central regions of their hosts, may be a sub-galactic manifestation of the same environmental factor responsible for the preference for high global stellar mass and star formation rate densities seen by \citet{Kelly2014}.  Whatever this factor is, it seems to influence primarily where LGRBs occur but not the properties of the explosions themselves as we do not find correlations between fractional flux or offset with burst duration or isotropic equivalent energy.         

Although the SLSN sample size is much smaller than the LGRB sample size, the statistical consistency between their distributions of fractional flux may be hinting that SLSNe and LGRBs share a common mass progenitor.  It could be that in the case of SLSNe, the core is not quite massive enough to form a black hole and instead collapses into a magnetar, while in the case of LGRBs, the core collapses into a black hole \citep{Lunnan2014}.  However, the difference in the size distributions of SLSN and LGRB hosts indicates other factors also influence the production of their progenitors.  

\section{Conclusions}
We have carried out a comprehensive study of the locations of LGRBs within their host galaxies using \textit{HST} observations of  \textit{Swift} bursts compiled over the last decade.  Using astrometry from ground- and space-based detections of optical, NIR, radio, and X-ray afterglows, we measure the projected physical and host-normalized offsets of 71 LGRBs from their host centers.  In addition, we measure the fractional flux, the brightness of the burst site relative to the total host light distribution.  Upon restricting the sample to avoid biases associated with measurement uncertainty we obtain sample sizes of 58, 52, and 50 LGRBs for the physical offsets, host-normalized offsets, and fractional fluxes, respectively.  These measurements have enabled us to study the offset and fractional flux distributions in great detail, providing an in-depth view into the preferred locations of LGRBs.  Our results are as follows:  

\begin{itemize}
\item In agreement with previous work \citep{Wainwright2007}, we find that LGRB hosts typically show irregular morphologies and are on average smaller (median = 3.0 kpc) than the host galaxies of CCSNe, though not as small as the hosts of SLSNe.  
\item We find that the physical offsets of LGRBs have a mean of the Monte Carlo medians of 1.34 kpc with a 90\% confidence interval of 1.17 $-$ 1.51 kpc. 
\item The host-normalized offsets span a wide range of 0.04 to 7.0 but are concentrated at small values with a mean of the Monte Carlo medians of 0.67, with a 90\% confidence interval of 0.58 $-$ 0.76.  This distribution of LGRB offsets is considerably more concentrated than expected if LGRBs traced an exponential disk (p-value = $2.6 \times 10^{-4}$).  LGRBs are more concentrated than their host galaxies' own light distributions, such that 50\% of bursts reside within a region containing only 33\% of the underlying host light.  
\item Compared to Type Ib/c SNe, LGRBs have consistent host-normalized offset distributions with 80\% of synthetic Monte Carlo distributions yielding p-values $> 0.05$.  On the other hand, LGRB offsets are inconsistent with those of Type II SNe.  
\item The fractional flux distribution of our sample of LGRBs has a median at 0.75 and $\sim$ 80\% of bursts have fractional flux values $\gtrsim 0.5$, indicating a strong preference towards high fractional flux though not as strong as previous studies of smaller samples indicated.
\item The host-normalized offsets and fractional fluxes are correlated, such that bursts at small offset ($\lesssim$ 0.5) occur exclusively on regions of high fractional flux ($\gtrsim$ 0.6) and bursts at larger offset uniformly track the underlying host light distributions, indicating that the preference for high fractional flux present in the full sample is entirely due to LGRBs at small offset.  
\item The offset and fractional flux distributions show no statistically significant trends with redshift, though a slight shift towards larger offsets and lower fractional flux values at $z \gtrsim 1.8$ may indicate the effects of surface brightness dimming.
\end{itemize}
  
From our study it appears that the star formation occurring within $\sim R_{h}$ is the most favorable for LGRB production.  The distributions of offset and fractional flux for our sample of LGRBs agree well with those for Type Ic SNe, a scenario consistent with the well-known connection between LGRBs and Type Ic-BL SNe.  Both types of transients occur in special types of galaxies and in special environments within their hosts, where metallicity is likely not the only important factor.  In the central regions of galaxies it is possible that changes in the IMF or increased massive star binary fractions may be responsible for the sub-galactic preferences of LGRBs and Ic-BL SNe.  This new view of LGRB locations we have presented is unlikely to change in the near future because the sample size is unlikely to increase significantly.  Progress toward understanding the precise factors at play in the central regions of LGRB host galaxies will likely come through studies of LGRB progenitor analogs in the local universe whose environments can be studied in great detail.   

\acknowledgments
The Berger GRB group at Harvard is supported in part by the NSF under grant AST-1411763 and by NASA under grant NNX15AE50G.  This paper is based upon work supported by the National Science Foundation Graduate Research Fellowship Program under Grant No. DGE1144152.  W.F. acknowledges support provided by NASA through Einstein Postdoctoral Fellowship grant number PF4-150121 issued by the \textit{Chandra} X-ray Observatory Center, which is operated by the Smithsonian Astrophysical Observatory for NASA under contract NAS8-03060.  We thank Ragnhild Lunnan, Maria Drout, Josh Grindlay, and Avi Loeb for helpful discussions.  The \textit{HST} data presented here and the \textit{Swift}/UVOT data used in this work were obtained from the Mikulski Archive for Space Telescopes (MAST). STScI is operated by the Association of Universities for Research in Astronomy, Inc., under NASA contract NAS5-26555.  Support for MAST for non-\textit{HST} data is provided by the NASA Office of Space Science via grant NNX09AF08G and by other grants and contracts.  Based in part on observations obtained at the Gemini Observatory, acquired through the Gemini Science Archive and processed using the Gemini IRAF package, which is operated by the Association of Universities for Research in Astronomy, Inc., under a cooperative agreement with the NSF on behalf of the Gemini partnership: the National Science Foundation (United States), the National Research Council (Canada), CONICYT (Chile), the Australian Research Council (Australia), Minist\'{e}rio da Ci\^{e}ncia, Tecnologia e Inova\c{c}\~{a}o (Brazil) and Ministerio de Ciencia, Tecnolog\'{i}a e Innovaci\'{o}n Productiva (Argentina).  Based in part on data obtained from the ESO Science Archive Facility.  This paper makes use of data gathered with the 6.5 meter Magellan Telescopes located at Las Campanas Observatory, Chile.  This research has made use of data obtained from the \textit{Chandra} Data Archive and software provided by the \textit{Chandra} X-ray Center (CXC) in the application package CIAO.  Based in part on data collected at the Subaru Telescope and obtained from SMOKA, which is operated by the Astronomy Data Center, National Astronomical Observatory of Japan.  We thank Brad Cenko for providing Palomar 60inch afterglow images for GRBs 060206 and 090618.    

\bibliographystyle{apj}

\newpage
\LongTables
\begin{deluxetable}{cc|ccccc|cc}
\tablecolumns{9}
\tabcolsep0.1in\footnotesize
\tablewidth{0pc}
\tablecaption{\textit{HST} and Afterglow Observations  
\label{tab:obs}}
\tablehead {
\colhead {}   &
\colhead {}   &
\multicolumn{5}{c}{\textit{HST} Image}   &
\multicolumn{2}{c}{Afterglow Image}   \\
\cmidrule{3-7}\cmidrule{8-9}
\colhead{GRB}              &
\colhead {$z$} &
\colhead {Instrument}             &
\colhead{Filter}                &
\colhead {Exp. Time}       &
\colhead {Obs. Date}     &
\colhead {Program ID}    &
\colhead {Telescope}     &
\colhead {Filter}     \\
\colhead {}      &       
\colhead {}      &
\colhead {}      &
\colhead {}      &
\colhead {(s)}      &
\colhead {(UT)}     &
\colhead {}      &
\colhead {}      &
\colhead {}                               
}
\startdata
040701  & \nodata      & ACS/WFC1              & F775W  &  1940 & 2004 Aug 09 &  10135 & \textit{Chandra}        & \nodata     \\
              &          &                                  & F625W & 1820 &       &     &                      &              \\
 040812  & \nodata      & ACS/WFC1              & F775W  &  2120 & 2004 Oct 04 &  10135 & \textit{Chandra}        & \nodata        \\
              &          &                                  & F625W & 2000   &         &   &                      &              \\
 040916  & \nodata     & ACS/WFC1              & F775W  & 2056 & 2004 Nov 30 &   10135 & \nodata   & \nodata       \\
              &          &                                  & F625W & 1928  &       &     &                      &              \\
 040924  & 0.859  & ACS/WFC1              & F775W  &  3932 & 2005 Feb 18 &  10135 & \textit{HST}/ACS/WFC1   & F775W       \\
              &          &                                  & F850LP &  3932   & 2005 Feb 19  &     &                      &              \\
 041006  & 0.716  & ACS/WFC1              & F775W  &  4224 & 2005 Feb 10 &   10135 & \textit{HST}/ACS/WFC1   & F775W       \\
              &          &                                  & F850LP & 4224  & 2005 Feb 11  &    &                      &              \\
 050315  & 1.949  & WFC3/IR                & F160W  &  1209 & 2011 Jul 20 & 12307 & Magellan/LDSS3 & r           \\
 050401  & 2.9    & WFC3/IR                & F160W  &  1612 & 2010 Oct 1 & 12307 & VLT/FORS2      & R   \\
 050406  & 2.44   & WFC3/IR                & F160W  &  1612 & 2011 Feb 23 & 12307 & Magellan/LDSS3 & r           \\
 050408  & 1.236 & WFC3/IR                & F110W  &  1012 & 2013 Mar 24 & 12949 & Gemini-N/GMOS  & i     \\
               &          &                                  & F160W &  1209        &    &          &                      &              \\
 050416A & 0.654 & ACS/WFC1              & F775W  &  4224 & 2005 Nov 21 & 10135 & \textit{HST}/ACS/WFC1   & F775W       \\
               &          &                                  & F850LP &          &      &      &                      &              \\
 050525  & 0.606  & ACS/WFC1              & F625W  &  4268 & 2006 Mar 10 & 10135 & \textit{HST}/ACS/WFC1   & F625W       \\
               &          &                                  & F775W &          &        &    &                      &              \\
 050730  & 3.967  & ACS/WFC1              & F775W  &  7844 & 2010 Jun 10 & 11734 & LCO/Swope      & WashT1      \\
 050803  & \nodata      & WFC3/IR                & F160W   &  906 & 2011 Sep 3 & 12307 & \textit{Swift}/XRT      & \nodata       \\
 050820  & 2.612  & ACS/WFC1              & F850LP & 14280 & 2006 Jun 8+11  &  10551 & \textit{HST}/ACS/WFC1   & F850LP      \\
               &          &                                  & F625W & 2238   & 2006 Jun 5  &    &                      &              \\ 
               &          &                                  & F775W & 4404  &   &    &                      &              \\
 050824  & 0.83   & WFC3/IR                & F160W  &   906 & 2011 Jan 18 & 12307 & VLT/FORS2      & R   \\
 050904  & 6.29   & ACS/WFC1         & F850LP   &     4216 & 2005 Sep 26 & 10616 & \textit{HST}/NICMOS           & F160W         \\
               &            & NICMOS    & F160W   & 15359     & 2006 Jul 23 &    &             &          \\
 050908  & 3.344  & ACS/WFC1              & F775W  &  7892 & 2010 Oct 31 & 11734 & Gemini-S/GMOS  & r     \\
 051016B & 0.936 & WFC3/IR                & F160W  &   906 & 2011 Jan 16 & 12307 & \textit{Swift}/UVOT     & U+B+V       \\
 051022  & 0.8    & ACS/WFC1           & F606W  &  1560 & 2009 Aug 21 & 11343 & \textit{Chandra}        & \nodata       \\
               &          & WFC3/IR       &  F160W  & 2397  & 2009 Oct 12   &          &             &         \\
 060115  & 3.53   & ACS/WFC1              & F814W  &  7910 & 2010 Aug 27 & 11734 & VLT/FORS1      & I      \\
 060116  & \nodata      & NICMOS         & F160W  &  5120 & 2006 Dec 12 & 10633 & \textit{HST}/NICMOS     & F160W       \\
              &         &                                & F110W   & 5120  & 2006 Dec 11 &            &                              &                    \\
              &          & ACS/WFC1             & F775W  & 4400   & 2006 Oct 10  &          &                    &            \\
              &          &                                & F850LP & 1650 & 2006 Oct 11  &            &                    &           \\
 060124  & 2.296  & WFC3/IR                & F160W  &  1612 & 2010 Sep 28 & 12307 & \textit{Swift}/UVOT     & V           \\
 060206  & 4.048  & ACS/WFC1              & F814W  &  9886 & 2006 Nov 25 & 10817 & Palomar 60inch  & I           \\
 060218  & 0.033  & WFC3/IR           & F160W  &   906 & 2010 Oct 12 & 12307 & \textit{HST}/ACS/WFC1   & F625W       \\
               &            & ACS/WFC1          & F625W & 2040 & 2006 Oct 26 & 10551  &           &      \\   
               &             &                             & F814W & 2040 &                   &          &        &       \\
               &            &                               & F435W & 1530 & 2006 Nov 04  &      &      &       \\
               &           &                                & F555W & 1020 &                       &       &     &         \\                  
 060223  & 4.41   & WFC3/IR                & F110W  &  8395 & 2010 Sep 13 & 11734 & \textit{Swift}/UVOT     & V           \\
 060418\tablenotemark{a}  & 1.49 & ACS/WFC1 & F625W  &  4220 & 2006 May 09 & 10551 & Self-\textit{HST} & \nodata  \\
              &            &                                  & F775W & 4220  &      &      &                               &                   \\
                &            &                                  & F555W & 4386  & 2006 Jul 13 &       &                               &                   \\
 060502A & 1.51   & WFC3/IR                & F160W  &  1209 & 2010 Oct 11 & 12307 & Gemini-N/GMOS  & r     \\
 060505\tablenotemark{b}  & 0.089  & ACS/WFC1 & F475W  & 27774 & 2006 May+Jun & 10551 & Gemini-S/GMOS  & g     \\
 	       &           &                               & F814W & 6840   & 2006 Jun 6+7   &        &                             &    \\
               &           & WFC3/IR                & F160W & 906   & 2011 Aug 3  & 12307  &         &       \\
 060512\tablenotemark{c}  & 0.443 & WFC3/IR                & F160W & 453 & 2011 Feb 23 & 12307 & VLT/FORS1      & \nodata         \\
 060522  & 5.11   & WFC3/IR                & F110W  &  8395 & 2010 Oct 17 & 11734 & \textit{Swift}/UVOT     & White       \\
 060526  & 3.21   & ACS/WFC1              & F775W  &  7844 & 2009 Aug 9 & 11734 & VLT/FORS1      & R      \\
 060602A & 0.787  & WFC3/IR                & F160W  &   906 & 2010 Dec 5 & 12307 & \textit{Swift}/XRT      & \nodata        \\
 060605  & 3.78   & ACS/WFC1              & F775W  &  7862 & 2010 Oct 6 & 11734 & \textit{Swift}/UVOT     & W1+V        \\
 060607  & 3.082  & ACS/WFC1              & F775W  &  7910 & 2010 Sep 17 & 11734 & Gemini-N/GMOS  & r     \\
 060614  & 0.125  & ACS/WFC1       & F606W  &  3600 & 2006 Sep 8 & 10917 & \textit{HST}/WFPC2      & F606W       \\
               &            &                         &  F435W  & 2152   &                     &           &               &    \\
               &            &                         &  F814W  & 4840   & 2006 Oct 31 &           &               &    \\
               &            & WFC3/IR            &  F160W   & 906   & 2010 Oct 8  & 12307  &       &     \\
 060719  & 1.532  & WFC3/IR                & F160W  &  1209 & 2013 Feb 24 & 12949 & VLT/FORS2      & R   \\
              &              &                                & F125W &  1059 &                    &             &             &    \\
 060729  & 0.54   & WFC3/IR                & F160W  &   906 & 2010 Sep 15 & 12307 & Gemini-S/GMOS  & i    \\
 060912A & 0.937  & WFC3/IR                & F160W  &   906 & 2011 Sep 23 & 12307 & \textit{Swift}/UVOT     & White       \\
 060923A & \nodata      & WFC3/IR                & F160W  &  1198 & 2013 Feb 10 & 12949 & Gemini-N/NIRI  & K    \\
                &          &                                & F125W   & 1348 &                     &           &                          &   \\
 060927  & 5.6    & WFC3/IR           & F110W  & 13992 & 2010 Sep 25 & 11734 & VLT/FORS2      & I      \\
               &          & NICMOS  & F160W  & 10240 & 2007 Jun 29  & 10926 &       &     \\
 061007  & 1.261  & WFC3/IR                & F160W  &  1209 & 2011 Jul 8 & 12307 & VLT/FORS1      & R      \\
 061110A & 0.758  & WFC3/IR                & F160W  &   906 & 2010 Sep 30 & 12307 & VLT/FORS1      & R      \\
 061110B & 3.44   & ACS/WFC1              & F775W  &  7862 & 2010 Sep 23 & 11734 & VLT/FORS1      & R      \\
 061222A & 2.088  & NICMOS              & F160W  &  7680 & 2007 Jun 16 & 10908 & Gemini-N/NIRI  & K$_{s}$          \\
 070125  & 1.547  & WFC3/IR          & F110W  &  2812 & 2010 Apr 24 & 11717 & Gemini-N/GMOS  & r     \\
               &            & WFC3/UVIS             & F336W  & 2520  &   &       &       &     \\
 070208  & 1.165  & WFC3/IR                & F160W  &  1209 & 2013 Jan 02 & 12949 & Gemini-N/GMOS  & r     \\
 		&          &                                & F110W & 1012 &                    &            &       &  \\
 070306  & 1.496 & WFC3/IR                & F160W  &  1198 & 2012 Nov 15 & 12949 & VLT/ISAAC      & K$_{s}$          \\
 		&        &                                   & F125W & 1348 &           &       &     &        \\
 070318  & 0.836  & WFC3/IR                & F160W  &   906 & 2010 Dec 31 & 12307 & VLT/FORS1      & R      \\
 070508  & \nodata      & WFC3/IR                & F160W  &   906 & 2011 Feb 22 & 12307 & Magellan/MAGIC & I           \\
 070521  & 1.35   & WFC3/IR                & F160W   &  906 & 2011 Aug 02 & 12307 & \textit{Swift}/XRT      & \nodata       \\
 070721B & 3.626  & ACS/WFC1              & F775W  &  7844 & 2010 Nov 13 & 11734 & VLT/FORS2      & R   \\
 070802  & 2.45   & WFC3/IR                & F160W  &  1209 & 2013 Feb 22 & 12949 & VLT/FORS2      & I      \\
 		&        &                                  & F105W & 1059 &                    &       &       &   \\
 071010A & 0.98   & WFC3/IR                & F160W  &  906 & 2010 Oct 29 & 12307 & \nodata           & \nodata        \\
 071010B & 0.947  & WFC3/IR                & F160W  &   906 & 2010 Nov 25 & 12307 & Gemini-N/GMOS  & r     \\
 071031  & 2.692  & WFC3/IR                & F160W  &  1612 & 2010 Nov 20 & 12307 & VLT/FORS2      & V      \\
 071112C & 0.823  & WFC3/IR                & F160W  &   906 & 2010 Oct 08 & 12307 & Gemini-N/GMOS  & r     \\
 071122  & 1.14   & WFC3/IR                & F160W  &  1209 & 2010 Dec 21 & 12307 & Gemini-N/GMOS  & i     \\
 080207  & 2.086 & WFC3/IR             & F110W  &  2397 & 2009 Dec 09 & 11343 & \textit{Chandra}        & \nodata       \\
               &             & WFPC2           & F606W  & 1600 & 2008 Mar 18  &      &      &       \\ 
               &             &                         &  F814W  & 3320 & 2009 Mar 20 &      &       &     \\
               &             &                                 & F702W  & 3600 & 2009 Mar 21 &      &     &  \\
               &             & NICMOS       &  F160W & 2560 & 2008 Apr 05  &      &       &     \\
 080319B\tablenotemark{d} & 0.937  & WFPC2 & F606W & 3200 & 2008 Jul 03 & 11513 & \textit{HST}/WFPC2 & \nodata \\
 		&            &                             & F814W &           & 2008 Jul 05 &              &                       &              \\
 080319C & 1.95   & WFC3/IR                & F160W  &  1209 & 2010 Sep 19 & 12307 & Gemini-N/GMOS  & r     \\
 080325  & 1.78   & WFC3/IR                & F160W  &  1209 & 2012 Dec 31 & 12949 & Subaru/MOIRCS  & K$_{s}$          \\
 		&         &                                & F125W & 1359 &               &        &       &      \\
 080430  & 0.767  & WFC3/IR                & F160W  &   906 & 2011 Jun 21 & 12307 & \textit{Swift}/UVOT     & U+White     \\
 080520  & 1.545  & WFC3/IR                & F160W  &  1209 & 2011 Feb 08 & 12307 & \textit{Swift}/UVOT     & White       \\
 080603A & 1.688  & WFC3/IR                & F125W  &  1397 & 2013 Apr 10 & 12949 & Gemini-N/GMOS  & r     \\
 		&           &                                & F160W & 1348 &                 &            &           &    \\
 080603B & 2.69   & WFC3/IR                & F160W  &  1612 & 2011 Aug 06 & 12307 & \textit{Swift}/UVOT     & U+B+V+W1    \\
 080605\tablenotemark{b} & 1.640 & WFC3/IR & F160W  &  2418 & 2012+2013  & 12307+12949 & VLT/FORS2 & R   \\
 		&        &                                 & F125W  & 1209 & 2013 Mar 15 & 12949 &      &  \\
 080607  & 3.036  & WFC3/IR            & F160W  & 10791 & 2010 Jul 25 & 12005 & \textit{Swift}/UVOT     & White+V+W2  \\
 080707  & 1.23   & WFC3/IR                & F160W  &  1209 & 2010 Oct 31 & 12307 & VLT/FORS1      & R      \\
 080710  & 0.845  & WFC3/IR                & F160W  &   906 & 2011 Feb 12 & 12307 & Gemini-S/GMOS  & r     \\
 080805  & 1.505  & WFC3/IR                & F160W  &  1209 & 2011 Oct 01 & 12307 & VLT/FORS2      & I      \\
 080913  & 6.7    & WFC3/IR                & F160W  &  7818 & 2009 Nov 30 & 11189 & VLT/FORS2      & z      \\
 080916A & 0.689  & WFC3/IR                & F160W  &   906 & 2011 Mar 19 & 12307 & VLT/FORS1      & R      \\
 080928  & 1.692  & WFC3/IR                & F160W  &  1209 & 2010 Sep 18 & 12307 & Gemini-S/GMOS  & r     \\
 081007  & 0.530 & WFC3/IR                & F160W  &   906 & 2010 Nov 30 & 12307 & Gemini-S/GMOS  & r     \\
 081008  & 1.969 & WFC3/IR                & F160W  &  1209 & 2011 Sep 04 & 12307 & Gemini-S/GMOS  & r     \\
 081109  & 0.979 & WFC3/IR                & F160W  &  1359 & 2013 Mar 09 & 12949 &  \nodata          & \nodata       \\
 	&                  &                               & F110W &  1312  &                      &         &       &   \\
 081121  & 2.512  & WFC3/IR                & F160W  &  1612 & 2011 Jan 04 & 12307 & \textit{Swift}/UVOT     & U+B+V+White \\
 081221  &  2.26   & WFC3/IR               & F160W  & 1209   & 2013 Jun 17  & 12949 & Gemini-N/NIRI & K            \\
 		&         &                               & F105W & 1209 &                    &                &              &            \\
 090113  & 1.749 & WFC3/IR           & F160W  &  2612 & 2009 Oct 17 & 11840 & \nodata          & \nodata        \\
               &             & ACS/WFC1         & F606W  & 2208 & 2009 Oct 15 &       &       &     \\
 090404  & ...      & WFC3/IR           & F160W  &  2612 & 2010 Jan 09 & 11840 & PdBI           & 108 GHz      \\
 		&        &                            & F105W  & 1359 & 2013 Jul 06 & 12949 &        &     \\
		&       &                             & F125W & 1209 &        &           &          &      \\
               &          & ACS/WFC1           & F606W & 2208 & 2010 Sep 02 & 11840  &       &     \\
 090407  & 1.449 & WFC3/IR           & F160W  &  1209 & 2010 Sep 15 & 11840 & \textit{Chandra}        & \nodata     \\
               &            & WFC3/UVIS      & F606W & 740  &      &      &       &     \\
 090417B & 0.345  & WFC3/IR           & F160W  &  2612 & 2009 Oct 17 & 11840 & \textit{Swift}/XRT      & \nodata       \\
               &             & ACS/WFC1         & F606W & 1656 & 2011 Jan 22  &      &       &     \\
 090418A & 1.608  & WFC3/IR                & F160W  &  1209 & 2010 Oct 02 & 12307 & \textit{Swift}/UVOT     & U           \\
 090423\tablenotemark{b}  & 8.2    & WFC3/IR                & F160W  & 42997 & 2010+2011 & 11189 & Gemini-N/NIRI  & H           \\
 		&        &                                 & F125W & 52117 &  2010             &           &         &      \\
 090424  & 0.544  & WFC3/IR                & F160W  &   906 & 2011 May 03 & 12307 & Gemini-S/GMOS  & r     \\
 090429B\tablenotemark{b} & 9.4    & WFC3/IR & F160W  & 4823 & 2010 Jan+Feb & 11189 & Gemini-N/NIRI  & ...         \\
 		&        &                            & F105W & 4823 & 2010 Feb 24+28 &             &       &      \\
               &            & ACS/WFC1        & F606W  & 2100 & 2010 Jan 03 &       &       &     \\ 
 090618  & 0.54   & WFC3/IR                & F160W  &   906 & 2011 Jan 09 & 12307 & Palomar 60inch  & I           \\
 090709A & \nodata     & WFC3/IR                & F160W  &   1359 & 2012 Oct 15 & 12949 & Subaru/MOIRCS  & K$_{s}$         \\
 		&          &                                & F125W & 1397 &                      &          &       &      \\
 091127  & 0.49   & WFC3/IR                & F160W  &   906 & 2010 Dec 16 & 12307 & Gemini-S/GMOS  & i     \\
 091208B & 1.063  & WFC3/IR                & F160W  &  1209 & 2010 Oct 10 & 12307 & Gemini-N/GMOS  & r   \\
 100205A & \nodata      & WFC3/IR           & F160W  &  1209 & 2010 Dec 06 & 11840 & Gemini-N/NIRI  & K     \\
                &           & WFC3/UVIS                 & F606W  & 1140 &        &           &       &     \\
 100316D & 0.059  & WFC3/UVIS          & F336W  &  1680 & 2011 Aug 03 & 12323 & \textit{HST}/WFC3/UVIS  & F336W       \\
 		&           &                               & F225W & 2520 & 2011 Aug 02  &              &         &           \\
		&          &                                 & F438W & 1230 &              &              &         &           \\
		&       &                                  & F814W & 1260  &              &              &         &           \\  
		&        &                                  & F275W & 2520 & 2011 Aug 03 &              &         &           \\
		&        &                                  & F555W & 1680  &              &              &         &           \\
		&         &                                 & F625W & 720 & 2011 Aug 06 &              &         &           \\
		&        &                                  & F763M & 1720  &              &              &         &           \\
                &           & WFC3/IR                  &  F125W & 1312  & 2011 Aug 02  &          &       &     \\
                &           &                                   & F160W & 1312 &                      &            &        &    \\
                &             & ACS/WFC1           & FR716N & 2400 & 2011 Aug 06 &             &       &    \\
 100413A & 3.9    & WFC3/IR           & F160W  & 1209 & 2010 Aug 31 & 11840 & EVLA           & 8.5 GHz      \\
                &          & WFC3/UVIS        & F606W & 752   &      &      &       &     \\
 100526A  & \nodata  & WFC3/IR       & F125W & 1348 & 2013 Jun 30 & 12949 & Gemini-N/NIRI & K      \\
 		&              &                        & F160W & 1209 &                    &            &                   &        \\            
 100615A & 1.398  & WFC3/UVIS           & F606W  &  1128 & 2010 Dec 16 & 11840 & \textit{Chandra}        & \nodata        \\
                &            & WFC3/IR             & F160W  & 1209  &       &       &       &     \\
 100621A & 0.542  & WFC3/IR                & F160W  &   909 & 2013 Mar 09 & 12949 & VLT/XSHOOTER   & I           \\
 		&          &                                  & F105W & 748 &                  &          &         &  \\
		&         &                                  & F125W &  898 &                  &         &      &   \\
 100905A & \nodata      & WFC3/IR                & F140W  &  5440 & 2013 Jul 10 & 12247 & \nodata           & \nodata        \\
 110312A & \nodata      & WFC3/IR           & F160W   &  1209 & 2011 Nov 18 & 12378 & \textit{Swift}/XRT      & \nodata       \\
               &           & WFC3/UVIS      &  F606W & 1110   & 2011 Nov 17     &       &       &     \\
 110709B & \nodata     & WFC3/UVIS            & F606W  &  2480 & 2011 Nov 08 & 12378 & \textit{Chandra}        & \nodata       \\
               &          & WFC3/IR         & F160W & 2612 & 2011 Nov 12   &      &       &     \\
 110731A & 2.83   & WFC3/UVIS             & F625W  &  5200 & 2012 Aug 22 & 12370 & \textit{HST}/WFC3/UVIS  & F625W       \\
 120119A & 1.728  & WFC3/IR                & F160W  &  1209 & 2012 Oct 08 & 12949 & Gemini-S/GMOS  & r  \\
 		&           &                                & F125W & 1209 &           &        &        &         
 \enddata
 \tablenotetext{a}{The afterglow and what we believe to be the host galaxy \citep{Pollack2009} are both detected in the same epoch and offset from each other such that centroid measurements are not biased.  Therefore the offset (see Section \ref{sec:off}) can be measured using this single early epoch image.  To measure the fractional flux (see Section \ref{sec:FF}) we used a late epoch image where the afterglow has faded away.}  
 \tablenotetext{b}{We combined multiple epochs in the same filter to obtain deep images.  For GRB 060505, the afterglow was not detected in early epoch \textit{HST} observations so we combined these observations with a later epoch.}
 \tablenotetext{c}{The \textit{HST} data for GRB 060512 is of poor quality, preventing further analysis.}
 \tablenotetext{d}{We combined the F606W and F814W filters to obtain a higher S/N detection of the host.  The afterglow image is also a stack of these two filters at an earlier epoch.}
 
 \tablecomments{The table is organized such that the first image listed for each burst is the image for which the corresponding offset and fractional flux measurements were used to compile the distribution (Section \ref{sec:results}).  For bursts with multiple observations we group the observations by instrument.  Within an instrument group, the observations are listed by date.  Only the last epoch of observations are shown for bursts with multiple epochs, unless an earlier epoch was chosen as the best host image.  Blank entries in the table indicate inheritance of the entry in the previous row.}
 \end{deluxetable}
 
 \newpage
 \LongTables
 \begin{deluxetable}{ccccccccccc}
\tablecolumns{11}
\tabcolsep0.1in\footnotesize
\tablewidth{0pc}
\tablecaption{Offsets, Galaxy Sizes, Magnitudes, and $P_{\rm cc}$  
\label{tab:data}}
\tablehead {
\colhead{GRB}              &
\colhead {\# Tie Objects} &
\colhead {$\sigma_{\rm tie}$}             &
\colhead{$\sigma_{\rm AG}$}                &
\colhead {$\sigma_{\rm host}$}       &
\colhead {$R$}     &
\colhead {$R_{\rm phys}$}    &
\colhead {$R_{h}$}     &
\colhead {$R_{\rm norm}$}    &
\colhead {AB Mag\tablenotemark{a}} &
\colhead{$P_{\rm cc}$} \\
\colhead {}       &
\colhead {}       &
\colhead {(\arcsec)}      &       
\colhead {(\arcsec)}      &
\colhead {(\arcsec)}      &
\colhead {(\arcsec)}      &
\colhead {(kpc)}      &
\colhead {(\arcsec)}     &
\colhead {}      &
\colhead {}      &
\colhead {}                               
}
\startdata   
040701  &   4 &   0.367 &  0.475 &   \nodata & \nodata     & \nodata    & \nodata                        & \nodata           & \nodata            & \nodata     \\
 040812  &   5 &   0.115 &  0.046 &   0.019 & 0.270 $\pm$ 0.125      & 2.283 $\pm$ 1.057 &  0.297        & 0.908 $\pm$ 0.420 & 19.450 $\pm$ 0.010 &  <0.001  \\
 040916  &   \nodata &   \nodata     &  0.150  & \nodata & \nodata & \nodata & \nodata                   & \nodata           & >27.1              & \nodata     \\
 040924  &  41 &   0.005 &  0.002 &   0.014 & 0.286 $\pm$ 0.015     & 2.197 $\pm$ 0.116  &  0.188      & 1.521 $\pm$ 0.080   & 23.954 $\pm$ 0.024 &  0.003 \\
 041006\tablenotemark{b}  &  24 &   0.004 &  <0.001     &   0.007 & 0.352 $\pm$ 0.008     & 2.536 $\pm$ 0.061 &  0.298  & 1.180 $\pm$ 0.029   & 24.668 $\pm$ 0.057 &  0.012 \\
 050315  &  17 &   0.054 &  0.003 &   0.017 & 0.114 $\pm$ 0.056     & 0.953 $\pm$ 0.474      &  0.236  & 0.481 $\pm$ 0.239   & 23.808 $\pm$ 0.029 &  0.006 \\
 050401  &  24 &   0.044 &  0.025 &   0.009 & 0.082 $\pm$ 0.051     & 0.638 $\pm$ 0.397     &  0.170   & 0.483 $\pm$ 0.300   & 25.217 $\pm$ 0.077 &  0.006 \\
 050406 &   7 &   0.038 &  0.217 &   0.006 & 0.195 $\pm$ 0.220      & 1.579 $\pm$ 1.788      &  0.121   & 1.609 $\pm$ 1.822  & 26.602 $\pm$ 0.140 &  0.041 \\
 050408  &  14 &   0.027 &  0.002 &   0.010  & 0.152 $\pm$ 0.029     & 1.263 $\pm$ 0.239     &  0.278    & 0.545 $\pm$ 0.103 & 23.700 $\pm$ 0.022 &  0.008 \\
 050416A &  24 &   0.007 &  <0.001     &   0.011 & 0.037 $\pm$ 0.013     & 0.260 $\pm$ 0.090 &  0.160  & 0.234 $\pm$ 0.081   & 23.001 $\pm$ 0.009 &  0.001 \\
 050525  &  74 &   0.006 &  0.001 &   0.002 & 0.028 $\pm$ 0.007     & 0.185 $\pm$ 0.046      &  0.119  & 0.232 $\pm$ 0.057   & 26.067 $\pm$ 0.064 &  0.005 \\
 050730  &   9 &   0.038 &  0.014 & \nodata & \nodata & \nodata  & \nodata                              & \nodata            & >27.9              & \nodata     \\
 050803  &   \nodata& \nodata &  0.850  & \nodata & \nodata & \nodata & \nodata                         & \nodata            & \nodata            & \nodata     \\
 050820  &  28 &   0.009 &  0.003 &   0.004 & 0.442 $\pm$ 0.010      & 3.530 $\pm$ 0.081     &  0.116   & 3.808 $\pm$ 0.087  & 25.196 $\pm$ 0.050 &  0.010  \\
 050824  &   6 &   0.036 &  0.012 &   0.038 & 0.473 $\pm$ 0.054     & 3.597 $\pm$ 0.410      &  0.402  & 1.177 $\pm$ 0.134   & 23.716 $\pm$ 0.042 &  0.021 \\
 050904  &   9 &   0.027 &  0.020  & \nodata & \nodata & \nodata  & \nodata                            & \nodata             & >27.2              & \nodata     \\
 050908  &  15 &   0.054 &  0.001 &   0.003 & 0.037 $\pm$ 0.055     & 0.275 $\pm$ 0.405      &  0.068  & 0.544 $\pm$ 0.802   & 27.669 $\pm$ 0.136 &  0.005 \\
 051016B &   4 &   0.062 &  0.184 &   0.013 & 0.257 $\pm$ 0.194     & 2.023 $\pm$ 1.532      &  0.228  & 1.127 $\pm$ 0.853   & 22.314 $\pm$ 0.011 &  0.004 \\
 051022  &   5 &   0.121 &  0.015 &   0.021 & 0.171 $\pm$ 0.124     & 1.283 $\pm$ 0.932      &  0.334   & 0.512 $\pm$ 0.372  & 22.020 $\pm$ 0.006 &  0.002 \\
 060115  &  18 &   0.045 &  0.043 &   0.017 & 0.280 $\pm$ 0.065      & 2.042 $\pm$ 0.471     &  0.125  & 2.239 $\pm$ 0.517   & 27.342 $\pm$ 0.137 &  0.030  \\
 060116  &  10 &   0.024 &  0.029 &   \nodata & \nodata   & \nodata     & \nodata                      & \nodata             & >27.1              & \nodata     \\
 060124  &   7 &   0.058 &  0.065 &   0.025 & 0.132 $\pm$ 0.091     & 1.081 $\pm$ 0.745      &  0.172  & 0.766 $\pm$ 0.528   & 26.133 $\pm$ 0.126 &  0.011 \\
 060206  &   6 &   0.127 &  0.021 &   0.005 & 0.292 $\pm$ 0.128     & 2.021 $\pm$ 0.888      &  0.091  & 3.211 $\pm$ 1.411   & 27.647 $\pm$ 0.094 &  0.035 \\
 060218  &  13 &   0.012 &  <0.001     &   0.044 & 0.174 $\pm$ 0.045     & 0.115 $\pm$ 0.030 &  0.606  & 0.287 $\pm$ 0.075   & 19.635 $\pm$ 0.003 &  0.002 \\
 060223  &   9 &   0.173 &  0.041 &   0.009 & 0.112 $\pm$ 0.178     & 0.749 $\pm$ 1.186      &  0.211  & 0.532 $\pm$ 0.844   & 26.534 $\pm$ 0.069 &  0.026 \\
 060418  &   \nodata &   \nodata &  0.003 &   0.004 & 0.404 $\pm$ 0.005  & 3.419 $\pm$ 0.040 &  0.118  & 3.425 $\pm$ 0.040   & 25.855 $\pm$ 0.067 &  0.019 \\
 060502A &   8 &   0.021 &  0.009 &   0.006 & 0.051 $\pm$ 0.023     & 0.435 $\pm$ 0.198      &  0.162   & 0.317 $\pm$ 0.144  & 25.864 $\pm$ 0.110 &  0.007 \\
 060505  &  20 &   0.029 &  0.004 &   0.023 & 4.307 $\pm$ 0.037     & 7.160 $\pm$ 0.062      &  1.195    & 3.605 $\pm$ 0.031 & 19.101 $\pm$ 0.001 &  0.009 \\
 060522  &   4 &   0.089 &  0.146 &   \nodata & \nodata   & \nodata & \nodata                            & \nodata           & >28.9              & \nodata     \\
 060526  &  23 &   0.040  &  0.006 & \nodata & \nodata & \nodata  & \nodata                              & \nodata           & >27.4              & \nodata     \\
 060602A &   \nodata & \nodata &  2.550  & \nodata & \nodata & \nodata  & \nodata                        & \nodata           & \nodata            & \nodata     \\
 060605  &  12 &   0.115 &  0.020  &   0.002 & 0.183 $\pm$ 0.117     & 1.304 $\pm$ 0.829     &  0.109  & 1.683 $\pm$ 1.069   & 27.555 $\pm$ 0.148 &  0.026 \\
 060607  &  33 &   0.037 &  0.003 & \nodata & \nodata & \nodata  & \nodata                               & \nodata           & >28.9              & \nodata     \\
 060614  &  21 &   0.009 &  0.002 &   0.011 & 0.357 $\pm$ 0.014     & 0.801 $\pm$ 0.031      &  0.329  & 1.086 $\pm$ 0.042   & 22.821 $\pm$ 0.007 &  0.003 \\
 060719  &  12 &   0.051 &  0.062 &   0.011 & 0.206 $\pm$ 0.081     & 1.743 $\pm$ 0.684      &  0.246    & 0.837 $\pm$ 0.328 & 23.400 $\pm$ 0.023 &  0.006 \\
 060729  &  17 &   0.039 &  0.001 &   0.015 & 0.338 $\pm$ 0.042     & 2.149 $\pm$ 0.265      &  0.308  & 1.098 $\pm$ 0.135   & 23.509 $\pm$ 0.028 &  0.011 \\
 060912A &  11 &   0.148 &  0.005 &   0.036 & 0.654 $\pm$ 0.153     & 5.155 $\pm$ 1.203      &  0.713  & 0.918 $\pm$ 0.214   & 21.547 $\pm$ 0.013 &  0.017 \\
 060923A &  38 &   0.039 &  0.013 &   0.052 & 0.291 $\pm$ 0.066     & 2.460 $\pm$ 0.560      &  0.696  & 0.418 $\pm$ 0.095   & 23.278 $\pm$ 0.038 &  0.038 \\
 060927  &  18 &   0.026 &  0.111 & \nodata & \nodata & \nodata  & \nodata                             & \nodata             & >28.6              & \nodata     \\
 061007  &   5 &   0.026 &  0.100   &   <0.001     & 0.022 $\pm$ 0.103    & 0.184 $\pm$ 0.862 & 0.373  & 0.059 $\pm$ 0.277   & 23.621 $\pm$ 0.033 &  0.013 \\
 061110A &  11 &   0.035 &  0.014 &   0.015 & 0.135 $\pm$ 0.040      & 0.992 $\pm$ 0.297     &  0.176  & 0.765 $\pm$ 0.229   & 25.114 $\pm$ 0.083 &  0.007 \\
 061110B &  49 &   0.025 &  0.011 &   0.011 & 0.061 $\pm$ 0.030      & 0.449 $\pm$ 0.220     &  0.104   & 0.586 $\pm$ 0.288  & 27.230 $\pm$ 0.105 &  0.008 \\
 061222A &  15 &   0.026 &  0.034 &   0.011 & 0.044 $\pm$ 0.044     & 0.368 $\pm$ 0.368      &  0.193   & 0.229 $\pm$ 0.229  & 24.399 $\pm$ 0.050 &  0.005 \\
 070125  &   5 &   0.033 &  0.007 & \nodata & \nodata & \nodata  & \nodata                             & \nodata             & >28.0              & \nodata     \\
 070208  &   6 &   0.030  &  0.006 &   0.025 & 0.099 $\pm$ 0.040      & 0.820 $\pm$ 0.327    &  0.368  & 0.270 $\pm$ 0.108   & 21.902 $\pm$ 0.009 &  0.005 \\
 070306  &  14 &   0.043 &  0.005 &   0.011 & 0.090 $\pm$ 0.045      & 0.760 $\pm$ 0.380     &  0.209   & 0.430 $\pm$ 0.215  & 21.860 $\pm$ 0.007 &  0.002 \\
 070318  &   8 &   0.057 &  0.007 &   0.019 & 0.109 $\pm$ 0.060      & 0.827 $\pm$ 0.458     &  0.205    & 0.529 $\pm$ 0.293 & 24.100 $\pm$ 0.034 &  0.005 \\
 070508  &  16 &   0.025 &  0.004 &   0.020  & 0.423 $\pm$ 0.033     & 3.577 $\pm$ 0.277     &  0.592  & 0.714 $\pm$ 0.055   & 22.871 $\pm$ 0.031 &  0.027 \\
 070521  &   \nodata & \nodata &  0.850  & \nodata & \nodata & \nodata  & \nodata                      & \nodata             & \nodata            & \nodata     \\
 070721B &  17 &   0.042 &  0.032 &   0.006 & 0.045 $\pm$ 0.054     & 0.327 $\pm$ 0.387      &  0.057  & 0.793 $\pm$ 0.939   & 28.442 $\pm$ 0.165 &  0.011 \\
 070802  &  10 &   0.039 &  0.007 &   0.027 & 0.132 $\pm$ 0.048     & 1.069 $\pm$ 0.390      &  0.348  & 0.379 $\pm$ 0.138   & 23.994 $\pm$ 0.043 &  0.014 \\
 071010A &   11 &   0.127 &  0.300   & \nodata & \nodata & \nodata  & \nodata                          & \nodata             & \nodata            & \nodata     \\
 071010B &   7 &   0.022 &  0.001 &   0.010  & 0.109 $\pm$ 0.025     & 0.860 $\pm$ 0.195     &  0.257  & 0.424 $\pm$ 0.096   & 22.635 $\pm$ 0.016 &  0.004 \\
 071031  &   7 &   0.068 &  <0.001     & \nodata & \nodata & \nodata  & \nodata                        & \nodata             & >27.4              & \nodata     \\
 071112C &  10 &   0.020  &  0.007 &   0.017 & 0.202 $\pm$ 0.027     & 1.533 $\pm$ 0.206     &  0.398  & 0.508 $\pm$ 0.068   & 23.674 $\pm$ 0.046 &  0.016 \\
 071122  &  10 &   0.032 &  0.033 &   0.025 & 0.075 $\pm$ 0.053     & 0.616 $\pm$ 0.433      &  0.393  & 0.191 $\pm$ 0.134   & 22.678 $\pm$ 0.017 &  0.009 \\
 080207  &   2 &   0.381 &  0.206 &   0.066 & 0.769 $\pm$ 0.438     & 6.402 $\pm$ 3.650      &  0.683  & 1.126 $\pm$ 0.642   & 23.225 $\pm$ 0.025 &  0.047 \\
 080319B &  18 &   0.011 &  0.001 &   0.117 & 0.215 $\pm$ 0.118     & 1.697 $\pm$ 0.929      &  0.147    & 1.465 $\pm$ 0.802 & 26.679 $\pm$ 0.200 &  0.015 \\
 080319C &  10 &   0.018 &  0.008 &   0.046 & 0.855 $\pm$ 0.050      & 7.177 $\pm$ 0.417     &  0.638  & 1.341 $\pm$ 0.078   & 22.018 $\pm$ 0.011 &  0.023 \\
 080325  &  60 &   0.022 &  0.041 &   0.059 & 0.697 $\pm$ 0.075     & 5.890 $\pm$ 0.635      &  0.443  & 1.574 $\pm$ 0.170   & 22.537 $\pm$ 0.016 &  0.017 \\
 080430  &   4 &   0.035 &  0.015 &   0.030  & 0.108 $\pm$ 0.049     & 0.800 $\pm$ 0.361     &  0.282  & 0.384 $\pm$ 0.173   & 24.595 $\pm$ 0.071 &  0.011 \\
 080520  &  17 &   0.106 &  0.146 &   0.024 & 0.468 $\pm$ 0.182     & 3.966 $\pm$ 1.543      &  0.223  & 2.100 $\pm$ 0.817   & 23.965 $\pm$ 0.032 &  0.011 \\
 080603A &  22 &   0.027 &  <0.001     &   0.013 & 0.089 $\pm$ 0.030    & 0.753 $\pm$ 0.255  &  0.202  & 0.440 $\pm$ 0.149   & 22.884 $\pm$ 0.011 &  0.003 \\
 080603B &   6 &   0.117 &  0.034 & \nodata & \nodata & \nodata  & \nodata                             & \nodata             & >26.7              & \nodata     \\
 080605  &  38 &   0.029 &  0.009 &   0.008 & 0.081 $\pm$ 0.031     & 0.682 $\pm$ 0.266      &  0.198  & 0.407 $\pm$ 0.159   & 22.467 $\pm$ 0.008 &  0.002 \\
 080607  &   5 &   0.230  &  0.056 &   0.024 & 0.681 $\pm$ 0.238     & 5.229 $\pm$ 1.827     &  0.419  & 1.626 $\pm$ 0.568   & 24.372 $\pm$ 0.027 &  0.036 \\
 080707  &   6 &   0.047 &  0.024 &   0.009 & 0.089 $\pm$ 0.054     & 0.739 $\pm$ 0.449      &  0.263   & 0.338 $\pm$ 0.205  & 22.950 $\pm$ 0.017 &  0.005 \\
 080710  &   6 &   0.035 &  <0.001     & \nodata & \nodata & \nodata  & \nodata                         & \nodata            & >26.1              & \nodata     \\
 080805  &  15 &   0.038 &  0.006 &   0.026 & 0.466 $\pm$ 0.047     & 3.944 $\pm$ 0.396      &  0.275   & 1.695 $\pm$ 0.170  & 23.249 $\pm$ 0.020 &  0.010  \\
 080913  &  13 &   0.035 &  0.044 & \nodata & \nodata & \nodata  & \nodata                              & \nodata            & >28.0              & \nodata     \\
 080916A &  12 &   0.056 &  0.022 &   0.010  & 0.011 $\pm$ 0.061     & 0.075 $\pm$ 0.432    &  0.240   & 0.044 $\pm$ 0.254   & 22.762 $\pm$ 0.016 &  0.004 \\
 080928  &  16 &   0.021 &  0.012 &   0.016 & 1.705 $\pm$ 0.029     & 14.437 $\pm$ 0.245     &  0.442    & 3.858 $\pm$ 0.066 & 21.990 $\pm$ 0.010 &  0.035 \\
 081007  &  11 &   0.023 &  <0.001     &   0.017 & 0.111 $\pm$ 0.029     & 0.700 $\pm$ 0.182 &  0.277  & 0.402 $\pm$ 0.105   & 24.489 $\pm$ 0.059 &  0.011 \\
 081008  &  28 &   0.021 &  0.002 &   0.018 & 1.658 $\pm$ 0.028     & 13.902 $\pm$ 0.236     &  0.236  & 7.027 $\pm$ 0.119   & 23.591 $\pm$ 0.028 &  0.066 \\
 081109  &   \nodata & \nodata &  0.850  & \nodata & \nodata & \nodata  & \nodata                      & \nodata             & \nodata            & \nodata     \\
 081121  &   6 &   0.12  &  0.014 &   0.011 & 0.262 $\pm$ 0.122     & 2.112 $\pm$ 0.981      &  0.198  & 1.323 $\pm$ 0.615   & 24.923 $\pm$ 0.054 &  0.010  \\
 081221  &  11 &   0.047 &  0.016 &   0.055 & 0.388 $\pm$ 0.074     & 3.195 $\pm$ 0.609      &  0.409  & 0.949 $\pm$ 0.181   & 23.156 $\pm$ 0.025 &  0.013 \\
 090113  &   \nodata & \nodata &  0.910  & \nodata & \nodata & \nodata  & \nodata                      & \nodata             & \nodata            & \nodata     \\
 090404  &   4 &   0.115 &  0.361 &   0.044 & 0.556 $\pm$ 0.381     & 4.709 $\pm$ 3.229     &  0.690    & 0.806 $\pm$ 0.553  & 23.370 $\pm$ 0.018 &  0.045 \\
 090407  &   5 &   0.134 &  0.387 &   0.021 & 0.172 $\pm$ 0.410      & 1.451 $\pm$ 3.463     &  0.336  & 0.511 $\pm$ 1.220   & 22.953 $\pm$ 0.019 &  0.026 \\
 090417B &   \nodata & \nodata &  0.850  & \nodata & \nodata & \nodata  & \nodata                       & \nodata            & \nodata            & \nodata     \\
 090418A &   5 &   0.209 &  0.014 &   0.017 & 0.096 $\pm$ 0.211     & 0.812 $\pm$ 1.785      &  0.214  & 0.448 $\pm$ 0.984   & 23.687 $\pm$ 0.025 &  0.009 \\
 090423  &   6 &   0.031 &  0.005 & \nodata & \nodata & \nodata  & \nodata                              & \nodata            & >28.2              & \nodata     \\
 090424  &   9 &   0.021 &  0.005 &   0.030  & 0.407 $\pm$ 0.037     & 2.594 $\pm$ 0.235     &  0.407   & 1.000 $\pm$ 0.090  & 21.260 $\pm$ 0.007 &  0.005 \\
 090429B &  10 &   0.053 &  0.012 & \nodata & \nodata & \nodata  & \nodata                              & \nodata            & >28.1              & \nodata     \\
 090618  &  12 &   0.038 &  0.001 &   0.025 & 0.693 $\pm$ 0.046     & 4.402 $\pm$ 0.292      &  0.519   & 1.335 $\pm$ 0.089  & 22.464 $\pm$ 0.020 &  0.020  \\
 090709A &  25 &   0.028 &  0.021 &   0.007 & 0.283 $\pm$ 0.036     & 2.393 $\pm$ 0.305      &  0.256  & 1.105 $\pm$ 0.141   & 24.213 $\pm$ 0.038 &  0.010  \\
 091127  &  16 &   0.040  &  0.005 &   0.028 & 0.221 $\pm$ 0.049     & 1.335 $\pm$ 0.296     &  0.411    & 0.538 $\pm$ 0.119 & 22.700 $\pm$ 0.021 &  0.011 \\
 091208B &   7 &   0.019 &  0.001 & \nodata & \nodata & \nodata  & \nodata                             & \nodata             & >26.7              & \nodata     \\
 100205A &  12 &   0.034 &  0.028 & \nodata & \nodata & \nodata  & \nodata                             & \nodata             & >26.7              & \nodata     \\
 100316D\tablenotemark{c} &  15 &   0.007 &  <0.001     &   <0.001     & \nodata    & \nodata     & \nodata             & \nodata             & \nodata            & \nodata     \\
 100413A &   8 &   0.098 &  0.600   &   \nodata & \nodata    & \nodata     & \nodata                   & \nodata             & \nodata            & \nodata     \\
 100526A &  10 &   0.047 &  0.008 &   0.008 & 0.194 $\pm$ 0.049     & 1.645 $\pm$ 0.411      &  0.188  & 1.034 $\pm$ 0.258   & 24.766 $\pm$ 0.033 &  0.007 \\
 100615A &   3 &   0.318 &  0.028 &   0.004 & 0.246 $\pm$ 0.320      & 2.073 $\pm$ 2.694     &  0.083  & 2.963 $\pm$ 3.850   & 25.092 $\pm$ 0.033 &  0.034 \\
 100621A &   6 &   0.062 &  0.025 &   0.015 & 0.045 $\pm$ 0.069     & 0.284 $\pm$ 0.439      &  0.262  & 0.17 $\pm$ 0.263    & 21.316 $\pm$ 0.006 &  0.002 \\
 100905A &   \nodata & \nodata &  0.910  & \nodata & \nodata & \nodata  & \nodata                      & \nodata             & \nodata            & \nodata     \\
 110312A &   \nodata & \nodata &  0.850  & \nodata & \nodata & \nodata  & \nodata                      & \nodata             & \nodata            & \nodata     \\
 110709B &   1 &   0.318 &  0.057 &   0.005 & 0.067 $\pm$ 0.323     & 0.564 $\pm$ 2.737      &  0.073   & 0.912 $\pm$ 4.430  & 26.714 $\pm$ 0.080 &  0.111 \\
 110731A & 211 &   0.006 &  0.001 &   0.008 & 0.182 $\pm$ 0.010      & 1.423 $\pm$ 0.082     &  0.121  & 1.502 $\pm$ 0.087   & 25.236 $\pm$ 0.043 &  0.005 \\
 120119A &  49 &   0.013 &  <0.001     &   0.012 & 0.012 $\pm$ 0.017     & 0.099 $\pm$ 0.146 &  0.202   & 0.058 $\pm$ 0.086  & 23.480 $\pm$ 0.022 &  0.003         
\enddata
\tablenotetext{a}{Magnitudes are corrected for Galactic extinction \citep{SF2011}}
\tablenotetext{b}{The SN associated with GRB 041006 is detected in the final epoch.  The SN and host center are sufficiently offset from each other such that centroid measurements are not biased.  However, measurement of fractional flux is not possible.}
\tablenotetext{c}{GRB 100316D occurred in a low-redshift galaxy with irregular morphology.  The complexity of the host system prevents a meaningful definition of the host center.  We therefore do not make an offset measurement for this burst.} 
\tablecomments{For bursts with multiple \textit{HST} images of the host galaxy, the image used to make the measurements listed here corresponds to the first image listed for such bursts in Table \ref{tab:obs}.  We show the 3$\sigma$ upper limits for bursts with no host detection.  Bursts with no listed offset or upper limit correspond to those bursts for which we were not able to obtain a suitable afterglow image.}
\end{deluxetable}

\newpage
\begin{deluxetable}{ccc|ccc}

\tablecolumns{6}
\tabcolsep0.1in\footnotesize
\tablewidth{0pc}
\tablecaption{Fractional Flux and Ratio of Error Circle to Galaxy Area   
\label{tab:FF}}
\tablehead {
\colhead {GRB}   &
\colhead {Fractional Flux} &  
\colhead {Ratio}  &
\colhead {GRB}   &
\colhead {Fractional Flux}  &
\colhead {Ratio}                              
}
\startdata   
040701 & \nodata & \nodata &                  070802 & 0.692 & $1.30 \times 10^{-2}$ \\
040812 & 0.423 & $1.74 \times 10^{-1}$  &      071010A & \nodata & \nodata \\          
040916 & \nodata & \nodata &                  071010B & 0.825 & $7.34 \times 10^{-3}$ \\
040924 & 0.783 & $8.21 \times 10^{-4}$ &      071031 & \nodata & \nodata \\           
041006 & \nodata & $1.80 \times 10^{-4}$ &    071112C & 0.774 & $2.83 \times 10^{-3}$ \\
050315 & 0.693 & $5.25 \times 10^{-2}$ &      071122 & 0.951 & $1.37 \times 10^{-2}$ \\
050401 & 0.799 & $8.86 \times 10^{-2}$ &      080207 & 0.207 & $4.02 \times 10^{-1}$ \\
050406 & 0.110 &   3.31                          &       080319B & 0.598 & $5.65 \times 10^{-3}$ \\
050408 & 0.724 & $9.48 \times 10^{-3}$ &      080319C & 0.599 & $9.53 \times 10^{-4}$ \\
050416A & 0.896 & $1.91 \times 10^{-3}$ &     080325 & 0.134 & $1.10 \times 10^{-2}$ \\
050525 & 0.914 & $2.61 \times 10^{-3}$ &      080430 & 0.645 & $1.82 \times 10^{-2}$ \\
050730 & \nodata & \nodata &                  080520 & 0.117 & $6.55 \times 10^{-1}$ \\
050803 & \nodata & \nodata &                  080603A & 0.880 & $1.79 \times 10^{-2}$ \\
050820 & 0.606 & $6.69 \times 10^{-3}$ &      080603B & \nodata & \nodata \\          
050824 & 0.324 & $8.91 \times 10^{-3}$ &      080605 & 0.820 & $2.35 \times 10^{-2}$ \\
050904 & \nodata & \nodata &                  080607 & 0.177 & $3.19 \times 10^{-1}$ \\
050908 & 0.316 & $6.31 \times 10^{-1}$ &      080707 & 0.836 & $4.03 \times 10^{-2}$ \\
051016B & 0.329 & $7.25 \times 10^{-1}$ &     080710 & \nodata & \nodata \\           
051022 & 0.556 & $1.33 \times 10^{-1}$ &      080805 & 0.517 & $1.96 \times 10^{-2}$ \\
060115 & 0.081 & $2.48 \times 10^{-1}$ &      080913 & \nodata & \nodata \\           
060116 & \nodata & \nodata &                 080916A & 0.930 & $6.28 \times 10^{-2}$ \\
060124 & 0.427 & $2.57 \times 10^{-1}$ &      080928 & 0.009 & $2.99 \times 10^{-3}$ \\
060206 & 0.032 &   2.00                         &      081007 & 0.792 & $6.89 \times 10^{-3}$ \\
060218 & 0.872 & $3.92 \times 10^{-4}$ &      081008 & 0.000 & $7.99 \times 10^{-3}$ \\ 
060223 & 0.292 & $7.10 \times 10^{-1}$ &      081109 & \nodata & \nodata \\           
060418 & 0.000 & $6.46 \times 10^{-4}$ &    081121 & 0.197 & $3.72 \times 10^{-1}$ \\
060502A & 0.689 & $1.99 \times 10^{-2}$ &     081221 & 0.761 & $1.47 \times 10^{-2}$ \\
060505 & 0.987 & $6.00 \times 10^{-4}$ &      090113 & \nodata & \nodata \\           
060522 & \nodata & \nodata &                  090404 & 0.183 & $3.02 \times 10^{-1}$ \\
060526 & \nodata & \nodata &                  090407 & 0.245 &  1.49  \\
060602A & \nodata & \nodata &                 090417B & \nodata & \nodata \\          
060605 & 0.103 &   1.15                   &       090418A & 0.367 & $9.58 \times 10^{-1}$ \\
060607 & \nodata & \nodata &                  090423 & \nodata & \nodata \\           
060614 & 0.459 & $7.85 \times 10^{-4}$ &      090424 & 0.740 & $2.81 \times 10^{-3}$ \\
060719 & 0.581 & $1.07 \times 10^{-1}$ &      090429B & \nodata & \nodata \\          
060729 & 0.246 & $1.60 \times 10^{-2}$ &      090618 & 0.260 & $5.36 \times 10^{-3}$ \\
060912A & 0.525 & $4.31 \times 10^{-2}$ &     090709A & 0.461 & $1.87 \times 10^{-2}$ \\
060923A & 0.980 & $3.49 \times 10^{-3}$ &      091127 & 0.833 & $9.62 \times 10^{-3}$ \\
060927 & \nodata & \nodata &                  091208B & \nodata & \nodata \\          
061007 & 0.778 & $7.67 \times 10^{-2}$ &      100205A & \nodata & \nodata \\          
061110A & 0.419 & $4.59 \times 10^{-2}$ &     100316D & 0.879 & \nodata \\            
061110B & 0.393 & $6.90 \times 10^{-2}$ &     100413A & \nodata & \nodata \\          
061222A & 0.955 & $4.92 \times 10^{-2}$ &     100526A & 0.531 & $6.43 \times 10^{-2}$ \\
070125 & \nodata & \nodata &                  100615A & 0.051 &  14.8  \\
070208 & 0.928 & $6.91 \times 10^{-3}$ &      100621A & 0.914 & $6.51 \times 10^{-2}$ \\
070306 & 0.798 & $4.29 \times 10^{-2}$ &      100905A & \nodata & \nodata \\          
070318 & 0.781 & $7.85 \times 10^{-2}$ &      110312A & \nodata & \nodata \\          
070508 & 0.517 & $1.83 \times 10^{-3}$ &      110709B & 0.055 &  19.6  \\
070521 & \nodata & \nodata &                  110731A & 0.029 & $2.53 \times 10^{-3}$ \\
070721B & 0.211 & $8.58 \times 10^{-1}$ &     120119A & 1.000 & $4.14 \times 10^{-3}$ 
  \enddata
 \end{deluxetable}

\newpage
\begin{deluxetable*}{c|cccccccc}[!h]
\tablecolumns{9}
\tabcolsep0.1in\footnotesize
\tablewidth{0pc}
\tablecaption{Summary of KS Test p-values   
\label{tab:sum}}
\tablehead {
\colhead {LGRBs - This Work  } &
\colhead{Exp. Disk}  &
\colhead {Type Ib/c SNe\tablenotemark{a}}   &
\colhead {Type Ic SNe}   &
\colhead {Type II SNe} &
\colhead {CCSNe}       &
\colhead {SLSNe}             &
\colhead {LGRBs - F06}     &
\colhead {LGRBs - B02}    
}   
\startdata
Physical Offsets             & \nodata & 1.33$\times10^{-6}$  & \nodata  & 1.53$\times10^{-7}$  & \nodata & \nodata & \nodata & 0.84   \\
				  &         &         &        &        &        &       &            &        \\
Host-Normalized Offsets & 2.6$\times10^{-4}$ & 1.3$\times10^{-2}$ & \nodata & 5.2$\times10^{-5}$ & \nodata & 0.17 & \nodata & 0.21  \\
                                    &         &         &        &        &        &       &            &        \\
Fractional Fluxes             & \nodata & \nodata & 0.79  & 5.4$\times10^{-5}$ & 3.0$\times10^{-2}$ & 0.52  & 0.09  & \nodata
\enddata
\tablenotetext{a}{For the host-normalized offsets, 80\% of LGRB Monte Carlo synthetic distributions have p-values $> 0.05$ when compared to Type Ib/c SNe.} 
\tablecomments{The values reported here for the physical offsets, host-normalized offsets, and fractional fluxes include only bursts with $\sigma_{R_{\rm phys}} \lesssim 1.0$ kpc, $\sigma_{R_{\rm norm}} \lesssim 0.5$, and error circle to galaxy area ratio $\lesssim 0.1$, respectively.  Comparison samples of Type Ib/c and II SNe (physical offsets), Type Ib/c and II SNe (host-normalized offsets), Type Ic and II SNe (fractional fluxes), CCSNe, SLSNe, LGRBs - F06, and LGRBs - B02 are from \citet{Prieto2008}, \citet{KellyKirshner2012}, \citet{Kelly2008}, \citet{Svensson2010}, \citet{Lunnan2015}, \citet{Fruchter2006}, and \citet{Bloom2002}, respectively.}  
\end{deluxetable*}

\begin{deluxetable*}{c|ccc}
\tablecolumns{4}
\tabcolsep0.1in\footnotesize
\tablewidth{0pc}
\tablecaption{Summary of LGRB Offset and Fractional Flux Medians  
\label{tab:medians}}
\tablehead {
\colhead {                          } &
\colhead {Sample Median}  &
\colhead {Mean of Monte Carlo Medians} &
\colhead {90\% Confidence Interval}
}
\startdata
Physical Offsets              & 1.27 kpc & 1.34 kpc & 1.17 $-$ 1.51 kpc \\
				    &        &          &     \\
Host-Normalized Offsets & 0.54  & 0.67  & 0.58 $-$ 0.76   \\
                                      &        &          &     \\
Fractional Fluxes              & 0.75    &  \nodata        &  \nodata
\enddata
\tablecomments{The values reported here for the physical offsets, host-normalized offsets, and fractional fluxes include only bursts with $\sigma_{R_{\rm phys}} \lesssim 1.0$ kpc, $\sigma_{R_{\rm norm}} \lesssim 0.5$, and error circle to galaxy area ratio $\lesssim 0.1$, respectively.}
\end{deluxetable*}

\end{document}